\pdfoutput=1
\documentclass[american,english,structabstract]{aa}
\usepackage{mathptmx}
\usepackage[T1]{fontenc}
\usepackage[utf8]{inputenc}
\usepackage{color}
\usepackage{array}
\usepackage{multirow}
\usepackage{amsmath}
\usepackage{amssymb}
\usepackage{graphicx}
\usepackage{natbib}

\makeatletter

\providecommand{\tabularnewline}{\\}

\makeatother

\usepackage{babel}
\begin{document}

\title{Efficient ortho-para conversion of $\mathrm{H}_{2}$ on interstellar
grain surfaces}

\author{Emeric Bron\inst{1}\and Franck Le Petit\inst{1}\and Jacques Le
Bourlot\inst{1,2}}

\institute{\inst{1}LERMA, Observatoire de Paris, PSL Research University, CNRS,
Sorbonne Universités, UPMC Univ. Paris 06, F-92190, Meudon, France\\
\inst{2}Université Paris Diderot, Sorbonne Paris Cité, F-75013, Paris,
France}

\offprints{emeric.bron@obspm.fr}

\date{Received 3 December 2015; Accepted 16 January 2016}

\abstract{Fast surface conversion between ortho- and para-H$_{2}$ has been
observed in laboratory studies, and this mechanism has been proposed
to play a role in the control of the ortho-para ratio in the interstellar
medium. Observations of rotational lines of H$_{2}$ in Photo-Dissociation
Regions (PDRs) have indeed found significantly lower ortho-para ratios
than expected at equilibrium. The mechanisms controlling the balance
of the ortho-para ratio in the interstellar medium thus remain incompletely
understood, while this ratio can affect the thermodynamical properties
of the gas (equation of state, cooling function).} {We aim to build
an accurate model of ortho-para conversion on dust surfaces based
on the most recent experimental and theoretical results, and to validate
it by comparison to observations of H$_{2}$ rotational lines in PDRs.
} {We propose a statistical model of ortho-para conversion on dust
grains with fluctuating dust temperatures, based on a master equation
approach. This computation is then coupled to full PDR models and
compared to PDR observations.} {We show that the observations of
rotational H$_{2}$ lines indicate a high conversion efficiency on
dust grains, and that this high efficiency can be accounted for if
taking dust temperature fluctuations into account with our statistical
model of surface conversion. Simpler models neglecting the dust temperature
fluctuations do not reach the high efficiency deduced from the observations.
Moreover, this high efficiency induced by dust temperature fluctuations
is quite insensitive to the values of microphysical parameters of
the model.} {Ortho-para conversion on grains is thus an efficient
mechanism in most astrophysical conditions that can play a significant
role in controlling the ortho-para ratio.}

\keywords{astrochemistry - ISM: molecules - molecular processes - ISM: dust,
extinction - ISM: photon-dominated region (PDR) - methods: numerical}

\maketitle

\section{Introduction\label{sec:Intro}}

H$_{2}$ is the main constituent of molecular clouds. It can be observed
in absorption in diffuse gas, and in emission in warm gas and/or high
UV field conditions (mainly photodissociation regions, hereafter PDRs,
and shocks). Its low rotational levels are collisionaly excited and
trace the gas temperature in the emitting material, while its vibrational
levels are either pumped by the UV radiation field or collisionaly
excited in hot shocked material. Excitation at formation has also
been proposed to contribute to vibrational excitation.

H$_{2}$ in the diffuse ISM has thus mainly been observed in absorption
\citep{Savage77,Rachford02,Tumlinson02,Gry02,Richter03,Lacour05,Gillmon06,Rachford09},
with the only observation in emission being \citet{Falgarone05}.
The excitation temperature of the first two rotational levels ($T_{01}$)
is commonly used as a measure of the gas temperature in diffuse clouds,
although it could stop being a meaningful measure at low $N(\mathrm{H_{2}})$
\citep{Srianand05,Roy06}. Higher rotational levels appear suprathermaly
excited and could trace a small fraction of warm gas heated by the
dissipation of interstellar turbulence in shocks or vortices \citep{Gredel02,Godard14,These}.

Observed in emission in brighter PDRs \citep{Fuente99,Moutou99,Habart03,Habart04,Allers05,Thi09,Fleming10,Habart11,Sheffer11},
it traces the surface layer of warm molecular gas close to the H/H$_{2}$
transition, and can be used as a diagnostics of the gas temperature
and UV radiation field. It can also help in constraining processes
such as H$_{2}$ formation or photoelectric heating \citep{Habart04,Habart11}.

H$_{2}$ has also been observed in extragalactic environments. Observations
of rotational emission in other galaxies has shown that the overall
H$_{2}$ emission could be explained by PDRs (\citealp{Naslim15}
in the LMC, \citealp{Roussel07} in the SINGS galaxy sample, \citealp{Higdon06}
in ULIRGs) except in Seyfert galaxies where a significant shock contribution
might be present \citep{Rigopoulou02,Pereira14}. In addition, H$_{2}$
has been detected in absorption in DLAs \citep{Ledoux03,Noterdaeme07,Muzahid15}.

H$_{2}$ exists as two spin isomers : para-H$_{2}$ with the spins
of its two nucleus in opposite directions, and ortho-H$_{2}$ with
parallel nuclear spins. In the ground electronic state, ortho-H$_{2}$
can only have an odd rotational number $J$, while para-H$_{2}$ only
takes even rotational numbers. As conversion between the two spin
isomers is forbidden for an isolated molecule (e.g., \citealp{Pachucki08}:
radiative transition rate of $6\times10^{-14}\,\mathrm{yr}^{-1}$)
and only reactive collisions can induce conversion, the ortho-to-para
ratio (hereafter OPR) can be out of local thermal equilibrium (hereafter
LTE). At LTE, the OPR is close to 3 at high temperatures ($>200\,\mathrm{K}$)
and goes to zero at low temperatures. The ratio between the successive
rotational lines of H$_{2}$ can thus be affected by an out-of-equilibrium
OPR, and the interpretation of H$_{2}$ emission requires a good understanding
of the OPR. Moreover, the excitation temperature $T_{01}$ used in
absorption studies only traces the gas temperature if the OPR is thermalized.

Several observations have derived out-of-equilibrium OPR values from
the pure rotational lines of H$_{2}$ in PDRs \citep{Fuente99,Moutou99,Habart03,Fleming10,Habart11},
with OPR values $\sim1$, significantly lower than the value of $\sim3$
expected from the excitation temperature of the low-$J$ rotational
lines. Out-of-equilibrium rotational OPR values have also been reported
in the SINGS galaxy sample by \citet{Roussel07}. Note that this is
a different problem than the low values of the OPR derived from vibrational
lines of H$_{2}$. Low OPR values in the vibrational levels can be
caused by preferential UV pumping of para-H$_{2}$ due to preferential
self-shielding of ortho-H$_{2}$ even when the true OPR (dominated
by the $v=0$ levels) is $3$, as described in great detail in \citet{Sternberg99}. 

In dark dense clouds, H$_{2}$ cannot be directly observed, and the
OPR can only be determined indirectly. For instance, \citet{Troscompt09}
deduce the OPR from the anomalous absorption of H$_{2}$CO owing to
the different collisional rates with ortho-H$_{2}$ and para-H$_{2}$.
\citet{Maret07} use DCO$^{+}$, as the fractionation reaction $\mathrm{H}_{3}^{+}+\mathrm{HD}\rightleftharpoons\mathrm{H}_{2}\mathrm{D}^{+}+\mathrm{H}_{2}$
is very sensitive to the OPR of H$_{2}$. \citet{Pagani09} similarly
use the influence of the OPR on the deuterium chemistry and deduce
the OPR from observations of N$_{2}$D$^{+}$, N$_{2}$H$^{+}$ and
H$_{2}$D$^{+}$. All find an OPR higher than the LTE value in cold
gas. Non dissociative shocks, in which the quickly heated gas makes
H$_{2}$ observable in emission but does not have time to significantly
change its OPR, can also offer a way to estimate the OPR in the preshock
dark molecular gas \citep{Neufeld06,Yuan11}.

The OPR of H$_{2}$ plays several important roles in the physico-chemistry
of the interstellar medium. First, it can affect significantly the
dynamics of core formation through gravitational collapse in star
forming clouds by modifying the heat capacity and the equation of
state of the gas, as was shown by \citet{Vaytet14} who compared numerical
simulations with different prescriptions corresponding to LTE OPR
or fixed OPR of 3. Second, it controls large parts of the chemistry
in dense clouds, such as the nitrogen chemistry \citep{Dislaire12,Faure13}
through the reaction $\mathrm{N}^{+}+\mathrm{H}_{2}\rightleftharpoons\mathrm{NH}^{+}+\mathrm{H}$,
or the deuterium chemistry \citep{Flower06} through the fractionation
reaction $\mathrm{H}_{3}^{+}+\mathrm{HD}\rightleftharpoons\mathrm{H}_{2}\mathrm{D}^{+}+\mathrm{H}_{2}$.
\foreignlanguage{american}{Finally}, the slow conversion process between
the two spin isomers has been used as a tool to measure the age of
molecular clouds \citep{Pagani11,Pagani13}.

The OPR is \foreignlanguage{american}{controlled} by several processes,
as first investigated by \citet{Burton92} : 
\begin{itemize}
\item Formation is usually assumed to occur with an OPR of $3$ but this
value is uncertain \citep{Takahashi01,Yabushita08,Gavilan12}.
\item Photo-dissociation can sometimes destroy para-H$_{2}$ \foreignlanguage{american}{preferentially}
due to the faster self-shielding of ortho-H$_{2}$ when the OPR is
$>1$. This can lead to the OPR being locally $>3$ \citep{Abgrall92,Sternberg99}.
\item Reactive collisions can induce conversion between the two spin isomers.
The main conversion reactions are proton exchange with H$^{+}$ \citep{Gerlich90,Honvault11}
and with H$_{3}^{+}$ \citep{Gomez12}, and hydrogen exchange with
H \citep{Schulz65,Truhlar76,Mandy92,Mandy93,Sun94,Lique12}. This
last reaction possesses an activation barrier ($\sim5000\,\mathrm{K}$)
and is thus only efficient in warm gas. Reactive collisions with H$_{2}$
have an even higher activation barrier ($\sim60000\,\mathrm{K}$,
\citealp{Carmona-Novillo07}) and are usually neglected. These reactions
tend to thermalize the OPR to the gas temperature.
\item H$_{2}$ molecules that are adsorbed on the surface of dust grains
can interact with the magnetic fields caused by impurities and surface
defects and convert from one spin isomer to the other \citep[and references therein]{Fukutani13}.
This process is the subject of the present article and tends to thermalize
the OPR to the dust temperature, which is significantly lower than
the gas temperature in PDRs.
\end{itemize}
An approximate treatment of ortho-para conversion on dust was used
in \citet{LeBourlot00} to investigate the effect of this process
in PDRs. The process was found to be only efficient on cold dust grains.
It was then shown that with a few hypotheses favoring high efficiency
(high binding energy, low dust temperatures), the pure rotational
lines of H$_{2}$ were strongly affected. \citet{Sheffer11} found
that PDR models could successfully explain observed OPR values lower
than LTE in the PDR NGC~2023 South when including this process with
the high efficiency hypothesis of \citet{LeBourlot00}.

In this article, we investigate the efficiency of this process using
a detailed model based on the most recent experimental and theoretical
results reviewed by \citet{Fukutani13}. As the surface processes
are highly sensitive to the dust temperature and as small dust grains
are known to have fluctuating temperatures that can significantly
affect the efficiency of surface processes (e.g., \citealp{Bron14}
for H$_{2}$ formation), we build a statistical model of ortho-para
conversion on dust grains with temperature fluctuations, which we
compare to a simpler rate equation model without fluctuations. We
then investigate the effect of this process in PDR models and compare
the predicted observable OPR values to PDR observations.

In Sect. \ref{sec:Processes_rate_equations}, we present the physical
processes at play and define their rates. We also present a simple
rate equation model neglecting the temperature fluctuations for comparison
with the more sophisticated model that we develop in the following
sections. In Sect. \ref{sec:Statistical_method}, we present the statistical
method that we employ to compute the effect of dust temperature fluctuations
on ortho-para conversion on grains. Sect. \ref{sec:Results} presents
the results of this statistical computation and discusses the importance
of the various microphysical parameters. In Sect. \ref{sec:PDR_models_and_obs},
we couple this statistical computation of the ortho-para conversion
rate to the Meudon PDR Code to study the impact of this new computation
on full PDR models, and confront their results to observations of
the ortho-para ratio in PDRs. Finally we give our conclusions in Sect.
\ref{sec:Conclusions}.

\section{Processes and rate equation model\label{sec:Processes_rate_equations}}

We will consider dust grains of sizes above $1\,\mathrm{nm}$. Ortho-para
conversion on Polycyclic Aromatic Hydrocarbons (PAHs) is probably
much less efficient due to the lack of surface defects and impurity
sites which are thought to allow ortho-para conversion on graphite
surfaces. 

We adopt a simple spherical grain model. We note $a$ the grain radius,
and assume uniformly distributed adsorption sites on the surface,
characterized equivalently by the surface density of sites $n_{\mathrm{s}}$,
the typical distance between sites $d_{\mathrm{s}}$ or the total
number of sites $N_{\mathrm{s}}$. Those quantities are related by
\[
N_{\mathrm{s}}=4\,\pi\,a^{2}n_{\mathrm{s}}=\frac{4\,\pi\,a^{2}}{d_{\mathrm{s}}^{2}}.
\]
In addition, we note $T_{\mathrm{d}}$ the grain temperature, $T_{\mathrm{gas}}$
the gas temperature, $n_{\mathrm{o}}$ and $n_{\mathrm{p}}$ respectively
the number of ortho-$\mathrm{H}_{2}$ and para-$\mathrm{H}_{2}$ molecules
physisorbed on the grain surface, and $n(\mathrm{H}_{2}^{(\mathrm{o})})$
and $n(\mathrm{H}_{2}^{(\mathrm{p})})$ respectively the gas phase
densities of ortho- and para-$\mathrm{H}_{2}$.

We now describe the different processes affecting the adsorbed $\mathrm{H}_{2}$
molecules that we include in our model.

\subsection{Adsorption of molecular hydrogen\label{sub:sticking}}

The first step of ortho-para conversion of an H$_{2}$ molecule is
phy\-sisorption on the grain surface, in which the molecule binds
to the grain through van der Waals interactions. A gas $\mathrm{H}_{2}$
molecule hitting the grain on an empty site becomes physisorbed with
a probability $S(T_{\mathrm{gas}})$ called the sticking probability.
We discuss the choice of this sticking function below. We assume rejection
if the molecule hits an occupied site. The rates of adsorption of
ortho-$\mathrm{H}_{2}$ and para-$\mathrm{H}_{2}$ molecules on the
grain (in $\mathrm{s}^{-1}$) are thus 
\begin{equation}
k_{\mathrm{coll}}^{(i)}\,S(T_{\mathrm{gas}})\,\left(1-\frac{n_{\mathrm{o}}+n_{\mathrm{p}}}{N_{\mathrm{s}}}\right)\label{eq:Adsorption_rate_ortho}
\end{equation}
with $k_{\mathrm{coll}}^{(i)}=\pi\,a^{2}\,n(\mathrm{H}_{2}^{(i)})\,\sqrt{{\displaystyle \frac{8\,k_{B}\,T_{\mathrm{gas}}}{\pi\,2\,m_{\mathrm{H}}}}}$,
with $i=\mathrm{o}$ for ortho-H$_{2}$ and $i=\mathrm{p}$ for para-H$_{2}$.
In the following, we will note $k_{\mathrm{ads}}^{(i)}=k_{\mathrm{coll}}^{(i)}\,S(T_{\mathrm{gas}})$
to simplify the notations.

Few measurements of the sticking function for $\mathrm{H}_{2}$ on
dust surfaces have been made. The only full measurement as a function
of the gas temperature is given by \citet{2010JChPh.133j4507M}, who
measured the sticking function on amorphous water ice in the temperature
range $30-350\,\mathrm{K}$. They give the prescription :
\begin{equation}
S(T_{\mathrm{gas}})=S_{0}\frac{1+\beta\frac{T_{\mathrm{gas}}}{T_{0}}}{\left(1+\frac{T_{\mathrm{gas}}}{T_{0}}\right)^{\beta}}\label{eq:Matar_sticking}
\end{equation}
with $S_{0}=0.76$, $T_{0}=87\,\mathrm{K}$ and $\beta=5/2$. This
formula is derived from a statistical model of sticking on amorphous
surfaces, fitted to the experimental results. 

\begin{figure}
\begin{centering}
\includegraphics[width=1\columnwidth]{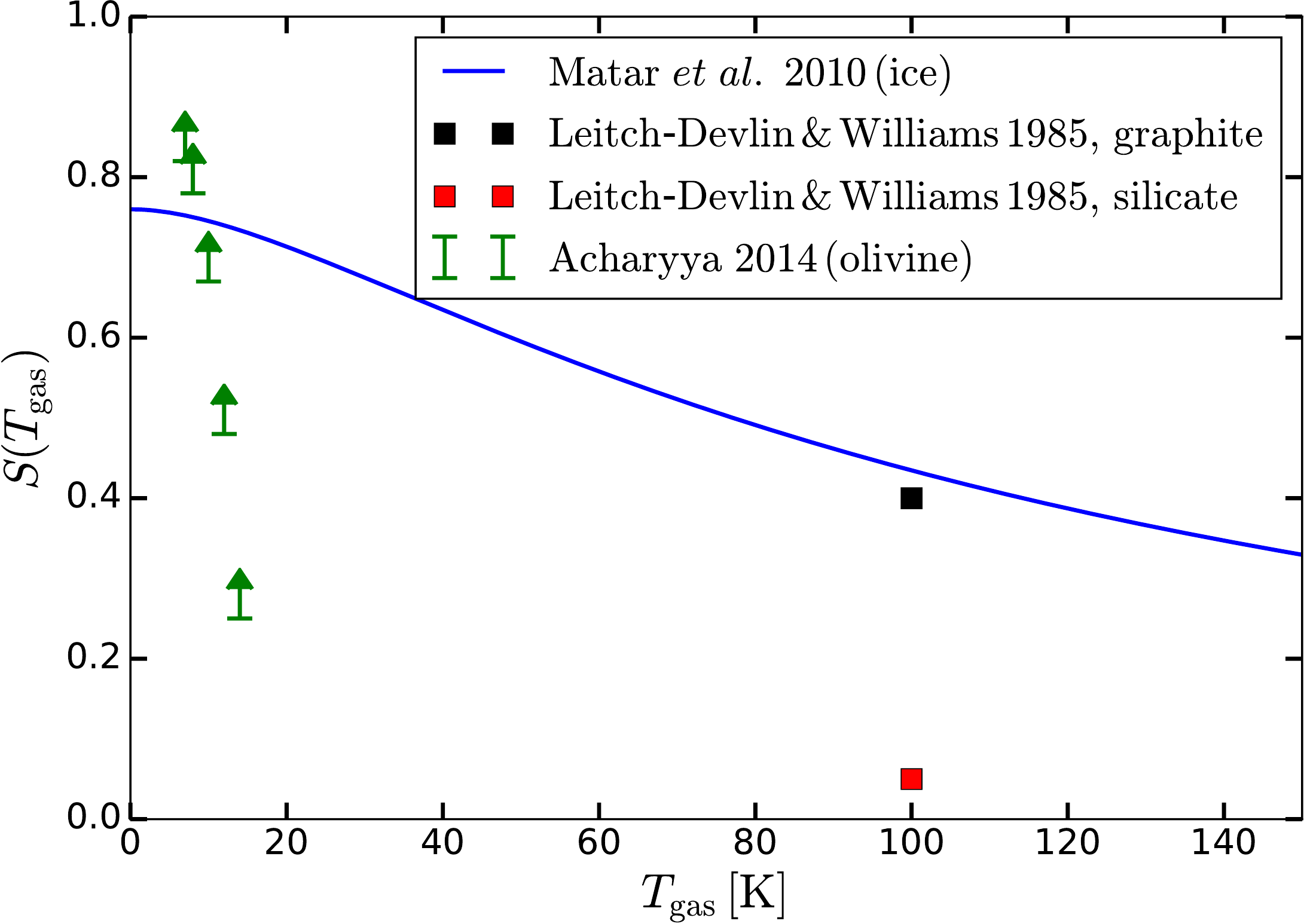}
\par\end{centering}

\protect\caption{Sticking coefficients of H$_{2}$ on different grain surfaces from
the literature.\label{fig:sticking_functions}}
\end{figure}
This sticking function is shown on Fig. \ref{fig:sticking_functions}
(blue curve). Theoretical calculation have also been presented in
\citet{1985MNRAS.213..295L} and are also shown on the graph: $0.4$
for graphite and $0.05$ for silicates at $\sim100\,\mathrm{K}$.
In the present paper, our comparison to observations will focus on
rotational H$_{2}$ emission in PDRs. In the regions of interest,
ice mantles are thus not yet present, and we need sticking functions
on bare carbonaceous and silicate surfaces. As we lack precise determinations
of these sticking functions, and as the \citet{2010JChPh.133j4507M}
sticking function is compatible with the value for graphite from \citet{1985MNRAS.213..295L},
we will use this sticking function (Eq. \ref{eq:Matar_sticking})
for all models in this article. The surface ortho-para conversion
efficiency is directly proportional to the sticking coefficient, so
that a lower sticking coefficient would proportionally reduce the
importance of surface conversion.

In comparison \citet{2014MNRAS.443.1301A} determined lower limits
to the sticking coefficient of H$_{2}$ on olivine as a function of
temperature for a very limited range of gas temperatures (7-14 K).
These lower limits are also presented on Fig. \ref{fig:sticking_functions}
and seem to indicate a steeper decrease with temperature, but the
limited range of temperatures and the fact that they are only lower
limits make conclusions difficult to draw.

\subsection{Ortho-para conversion\label{sub:conversion}}

Once physisorbed, an ortho-H$_{2}$ molecule can convert to a para-H$_{2}$
molecule with rate $k_{\mathrm{o}\rightarrow\mathrm{p}}(T_{\mathrm{d}})$,
and vice versa with rate $k_{\mathrm{p}\rightarrow\mathrm{o}}(T_{\mathrm{d}})$.
Conversion occurs because of electro-magnetic interactions with the
surface allowing spin-transfer through various processes, and often
involves impurity sites or surface defects (see \citealp{Fukutani13}
and \citealp{Ilisca14} for recent overviews of these processes).

Those two rates are related by the fact that for an adsorbed population
of $\mathrm{H}_{2}$ without desorption or arrival, the populations
of ortho-$\mathrm{H}_{2}$ and para-$\mathrm{H}_{2}$ at equilibrium
must yield an equilibrium ortho-para ratio. We thus have 
\begin{equation}
\frac{k_{\mathrm{p}\rightarrow\mathrm{o}}(T_{\mathrm{d}})}{k_{\mathrm{o}\rightarrow\mathrm{p}}(T_{\mathrm{d}})}=OPR^{(\mathrm{eq})}(T_{\mathrm{d}})\label{eq:OPR_grain}
\end{equation}
We will neglect here the fact that the energy levels of phy\-sisorbed
$\mathrm{H}_{2}$ are slightly modified compared to the gas phase
values, and take the gas phase equilibrium OPR function.

Several experiments, which we discuss below, have measured a conversion
timescale $\tau_{\mathrm{conv}}$ to pure para-H$_{2}$ on a cold
surface. In such experiments, $\mathrm{H_{2}}$ is first adsorbed
on the surface with some initial OPR. As the temperature of the surface
is very low, it then progressively converts to para-$\mathrm{H_{2}}$.
Desorption is negligible. In such case, the evolution matrix is 
\begin{equation}
\left(\begin{array}{cc}
-k_{\mathrm{p}\rightarrow\mathrm{o}}(T_{\mathrm{d}}) & k_{\mathrm{o}\rightarrow\mathrm{p}}(T_{\mathrm{d}})\\
k_{\mathrm{p}\rightarrow\mathrm{o}}(T_{\mathrm{d}}) & -k_{\mathrm{o}\rightarrow\mathrm{p}}(T_{\mathrm{d}})
\end{array}\right)
\end{equation}
with eigenvalues $k_{\mathrm{p}\rightarrow\mathrm{o}}(T_{\mathrm{d}})+k_{\mathrm{o}\rightarrow\mathrm{p}}(T_{\mathrm{d}})$
and $0$. The measured characteristic time is thus 
\begin{equation}
\tau_{\mathrm{conv}}=\frac{1}{k_{\mathrm{p}\rightarrow\mathrm{o}}(T_{\mathrm{d}})+k_{\mathrm{o}\rightarrow\mathrm{p}}(T_{\mathrm{d}})}\label{eq:tau_conv}
\end{equation}
We note that this characteristic time scale is related to the half-life
that is often used by $t_{\mathrm{half-life}}\simeq0.69\,\tau_{\mathrm{conv}}$.

From Eq. \ref{eq:OPR_grain} and \ref{eq:tau_conv}, we can deduce
expressions for the conversion rate coefficients :
\begin{equation}
\begin{cases}
{\displaystyle k_{\mathrm{o}\rightarrow\mathrm{p}}(T_{\mathrm{d}})=\frac{1}{\tau_{\mathrm{conv}}}\frac{1}{1+OPR^{(\mathrm{eq})}(T_{\mathrm{d}})}}\\
\\
{\displaystyle k_{\mathrm{p}\rightarrow\mathrm{o}}(T_{\mathrm{d}})=\frac{1}{\tau_{\mathrm{conv}}}\frac{OPR^{(\mathrm{eq})}(T_{\mathrm{d}})}{1+OPR^{(\mathrm{eq})}(T_{\mathrm{d}})}}
\end{cases}\label{eq:conversion_rates}
\end{equation}
Because we lack experimental determinations of the temperature dependance
of $\tau_{\mathrm{conv}}$, we will take it as a constant.

\begin{table*}
\protect\caption{Experimental measurements of the ortho-para conversion timescale on
different surfaces. See the text for details. \label{tab:tau_conv_literature}}

\begin{centering}
\smallskip{}
\begin{tabular}{lllll}
\hline 
\hline Reference & Surface & Temperature & $\tau_{\mathrm{conv}}^{\mathrm{H_{2}}}$ & $\tau_{\mathrm{conv}}^{\mathrm{D_{2}}}$\tabularnewline
 &  & (K) & (s) & (s)\tabularnewline
\hline 
(1) & Graphite & $<12$ & $10^{4}$ & $5\times10^{6}$\tabularnewline
(2) & Graphite & $10$ & $<20$ & -\tabularnewline
(3) & Graphite & $10$ & - & $1.2\times10^{3}$\tabularnewline
(4) & Water Ice & $90$ & $5\times10^{4}$ & -\tabularnewline
(5) & Water Ice & $12$ & $3.6\times10^{3}$ & -\tabularnewline
(6) & Water Ice & $10$  & - & $>2\times10^{4}$\tabularnewline
(7) & Water Ice & $15$ & $5\times10^{3}-1\times10^{4}$ & -\tabularnewline
\multirow{2}{*}{(8)} & Water Ice (co-adsorbed O$_{2}$) & \multirow{2}{*}{$10$} & $2.2\times10^{2}$ & -\tabularnewline
 & Water Ice (clean) &  & $>2\times10^{4}$ & -\tabularnewline
(9) & Water Ice (clean) & $10$ & $2.3\times10^{2}-7.1\times10^{2}$ & -\tabularnewline
\hline 
\end{tabular}
\par\end{centering}

\smallskip{}

\centering{}(1) : \citet{1985CaJPh..63..605K}; (2) : \citet{1987SurSc.179L...1P};
(3) : \citet{1990PhRvB..42..820Y}; (4) : \citet{1954JPhysChem..58..54};
(5) : \citet{1992JChPh..97..753H}; (6) : \citet{2008PhRvL.100e6101A};
(7) : \citet{2010ApJ...714L.233W}; (8) : \citet{2011PCCP...13.2172C};
(9) : \citet{2011NatPh...7..307S}
\end{table*}

Several experiments have measured the ortho-para conversion timescale:

A few experiments have measured the conversion timescale on graphite.
The first measurement was done by \citet{1985CaJPh..63..605K}, who
measured a conversion rate of $0.40\,\%/\mathrm{h}$ (characteristic
time $\tau_{\mathrm{conv}}\sim10^{4}\,\mathrm{s}$) for $\mathrm{H}_{2}$
ortho-para conversion, and of $0.069\,\%/\mathrm{h}$ ($\tau_{\mathrm{conv}}^{(\mathrm{D}_{2})}\sim5\times10^{6}\,\mathrm{s}$)
for $\mathrm{D_{2}}$ para-ortho conversion. Another measurement was
performed by \citet{1987SurSc.179L...1P} at a surface temperature
of $10\,\mathrm{K}$, finding full conversion for H$_{2}$ in less
than 1 min ($\tau_{\mathrm{conv}}<20\,\mathrm{s}$). Finally, \citet{1990PhRvB..42..820Y}
measured the para-ortho conversion of $\mathrm{D_{2}}$ and found
a conversion time of $20\,\mathrm{min}$ at $10\,\mathrm{K}$. Assuming
the same ratio between the $\mathrm{H_{2}}$ ortho-para conversion
rate and the $\mathrm{D}_{2}$ para-ortho conversion rate as found
in \citet{1985CaJPh..63..605K}, this would lead to a conversion timescale
of $2.5\,\mathrm{s}$ for $\mathrm{H_{2}}$, in agreement with \citet{1987SurSc.179L...1P}.
\citet{1990PhRvB..42..820Y} also showed that the conversion timescale
increases with the surface coverage, and that it is roughly constant
with temperature in the range 10-25 K, and increases sharply to $\sim70\,\mathrm{min}$
at $8-10\,\mathrm{K}$.

On amorphous water ice, several measurements have been made. The oldest
estimation is \citet{1954JPhysChem..58..54}, giving a conversion
half-life of $\sim10\,\mathrm{h}$ ($\tau_{\mathrm{conv}}\sim5\times10^{4}\,\mathrm{s}$)
on solid $\mathrm{D_{2}O}$ at $90\,\mathrm{K}$. \citet{1992JChPh..97..753H}
then measured a half-time of $\sim40\,\min$ ($\tau_{\mathrm{conv}}\sim3.6\times10^{3}\,\mathrm{s}$)
on solid D$_{2}$O at 12 K. They proposed the presence of different
kinds of sites, some of which having enhanced conversion. They also
suspect a low level of oxygen contamination. \citet{2008PhRvL.100e6101A}
found no detectable para-ortho conversion of $\mathrm{D_{2}}$ at
10 K on a time of $\sim10^{3}\,\mathrm{s}$ ($\tau_{\mathrm{conv}}^{(\mathrm{D}_{2})}>2\times10^{4}\,\mathrm{s}$).
\citet{2010ApJ...714L.233W} experimental results seem to indicate
a conversion of the order of $10-20\%$ in $20\,\mathrm{min}$ at
15 K ($\tau_{\mathrm{H_{2}}}=5\times10^{3}-1\times10^{4}\,\mathrm{s}$).
\citet{2011PCCP...13.2172C} find that the conversion is helped by
co-adsorbed O$_{2}$. In the presence of O$_{2}$ they find a conversion
time of $\sim220\,s$, while in the absence of O$_{2}$, they find
a rate less than $15\%/\mathrm{h}$ ($\tau_{\mathrm{H_{2}}}>2\times10^{4}\,\mathrm{s}$).
In contrast, \citet{2011NatPh...7..307S} measure a conversion timescale
in the range $230-710\,\mathrm{s}$ on clean ice at $10\,\mathrm{K}$,
and propose a theoretical model predicting a timescale of the order
of $\sim10^{2}\,\mathrm{s}$.

These values are gathered in Table \ref{tab:tau_conv_literature}.
The results are thus quite contradictory, and we will have to explore
values of the conversion timescale in the range $1-10^{4}\,\mathrm{s}$.
The most likely value seems to be $1-10\,\mathrm{s}$ on graphite,
and $10^{2}-10^{4}\,\mathrm{s}$ on ices.

\subsection{Thermal desorption\label{sub:desorption}}

\begin{table*}
\begin{centering}
\protect\caption{Experimental and theoretical determinations of the binding energy
of H$_{2}$ on different surfaces. See the text for details.\label{tab:T_phys_literature}}
\smallskip{}
\begin{tabular*}{2\columnwidth}{@{\extracolsep{\fill}}>{\raggedright}p{0.1\paperwidth}>{\raggedright}p{0.25\paperwidth}>{\raggedright}p{0.2\paperwidth}>{\raggedright}p{0.25\paperwidth}}
\hline 
\hline Reference & Surface & $T_{\mathrm{phys}}$ & Note\tabularnewline
 &  & (K) & \tabularnewline
\hline 
(1) & Crystalline silicate & $314.5$ & -\tabularnewline
(2) & Amorphous silicates & $406$ and $615$ & Binding energy distribution described by two values\tabularnewline
(3) & Amorphous silicates & $662$ ($[522-1044]$) & Binding energy distribution\tabularnewline
(1) & Amorphous carbon & $542$ & -\tabularnewline
(4) & Amorphous water ice & $646\pm177$ & Binding energy difference between ortho-H$_{2}$ and para-H$_{2}$
$\Delta T_{phys}=30\pm16\,\mathrm{K}$\tabularnewline
(5) & Amorphous water ice & $731$ & -\tabularnewline
(6) & Amorphous water ice (heat-treated low density) & $522$ and $789$ & \tabularnewline
 & Amorphous water ice (vapor-deposited low density) & $453$ and $778$ & Binding energy distribution described by two values\tabularnewline
 & Amorphous water ice (high density) & $615$ and $801$ & \tabularnewline
\hline 
\end{tabular*}
\par\end{centering}

\smallskip{}

\centering{}(1) : \citet{1999ApJ...522..305K}; (2) : \citealp{2007ApJ...661L.163P};
(3) : \citealp{2010JPCM...22D4012V}; (4) : \citet{1993JChPh..99.2265B};
(5) : \citet{2001ApJ...548L.253M}; (6) : \citet{2002ApJ...581..276R}
\end{table*}

In competition with the conversion process, the H$_{2}$ molecule
can also desorb thermally from the surface. The thermal desorption
rate of ortho- and para-H$_{2}$ is determined by their physisorption
energies. Several studies have measured the adsorption energy of H$_{2}$
on various surfaces, most of which do not distinguish between ortho-
and para-H$_{2}$. As para-H$_{2}$ is insensitive to the anisotropic
part of the adsorption potential while ortho-H$_{2}$ is not, ortho-H$_{2}$
tends to have a higher binding energy than para-H$_{2}$ \citep{Fukutani13}.

For crystalline silicates, \citet{1999ApJ...522..305K} give an adsorption
energy of $27.1\,\mathrm{meV}$ ($314.5\,\mathrm{K}$). For amorphous
silicates, \citealp{2007JPCA..11112611V,2007ApJ...661L.163P} find
that the distribution of binding energies can be described with two
values : $35\,\mathrm{meV}$ ($406\,\mathrm{K}$) and $53\,\mathrm{meV}$
($615\,\mathrm{K}$). \citealp{2010JPCM...22D4012V} determine the
full distribution of binding energies and find a distribution with
a peak at $57\,\mathrm{meV}$ ($662\,\mathrm{K}$), and extending
from $45\,\mathrm{meV}$ ($522\,\mathrm{K}$) to $90\,\mathrm{meV}$
($\mathrm{1044\,K}$). For amorphous carbon, \citet{1999ApJ...522..305K}
give $46.7\,\mathrm{meV}$ ($542\,\mathrm{K}$). For amorphous water
ice (ASW), \citet{2001ApJ...548L.253M} find $63\,\mathrm{meV}$ ($731\,\mathrm{K}$).
\citet{2002ApJ...581..276R} study the binding energy distribution
for various types of ices, describing the distribution by two energy
values. They find $45\,\mathrm{meV}$ ($522\,\mathrm{K}$) and $68\,\mathrm{meV}$
($789\,\mathrm{K}$) for heat-treated low density ASW, $39\,\mathrm{meV}$
($453\,\mathrm{K}$) and $67\,\mathrm{meV}$ ($778\,\mathrm{K}$)
for vapor-deposited low density ASW, and $53\,\mathrm{meV}$ ($615\,\mathrm{K})$
and $69\,\mathrm{meV}$ ($801\,\mathrm{K}$) for high density ice.
Finally, \citet{1993JChPh..99.2265B} give the only measurement of
the binding energy difference between ortho- and para-H$_{2}$ on
an astrophysically relevant surface: amorphous water ice. Experimentally,
they find a lower limit on the adsorbed OPR of 9 at $12\,\mathrm{K}$
before conversion has time to take place, corresponding to a difference
between ortho- and para-binding energies $\Delta T_{\mathrm{phys}}\geq13\,\mathrm{K}.$
They also study numerically the distribution of binding energies for
both ortho and para-H$_{2}$ on amorphous water ice, and find for
para-H2 a binding energy of $646\pm177\,\mathrm{K}$, and a difference
between the ortho- and para- binding energies of $30\pm16\,\mathrm{K}$.

These values of the binding energy are summarized in Table \ref{tab:T_phys_literature}.
Overall, a value of $500-600\,\mathrm{K}$ seems reasonable for both
amorphous carbons and amorphous silicates, a value of $315\,\mathrm{K}$
for crystalline silicates, and a value in the range $500-800\,\mathrm{K}$
for amorphous water ices. We will thus explore a range of binding
energies between $300$ and $800\,\mathrm{K}$. In addition, we will
investigate the effect of an ortho/para difference in binding energies
by testing both without difference and with a difference of $30\,\mathrm{K}$.

The desorption rates of ortho- and para-H$_{2}$ (for one molecule)
are then 
\begin{equation}
k_{\mathrm{des}}^{(i)}(T_{\mathrm{d}})=\nu_{0}^{(i)}\exp\left(-\frac{T_{\mathrm{phys}}^{(i)}}{T_{\mathrm{d}}}\right)
\end{equation}
with $i=\mathrm{o}$ for ortho-H$_{2}$ and $i=\mathrm{p}$ for para-H$_{2}$,
and where the $\nu_{0}$'s are typical vibration frequencies given
by (\citealp{Hasegawa92})
\begin{equation}
\nu_{0}^{(i)}=\frac{1}{\pi}\sqrt{\frac{2\,k_{B}\,T_{\mathrm{phys}}^{(i)}}{d_{0}^{2}\,2\,m_{\mathrm{H}}}}
\end{equation}
with $d_{0}=0.1\,\mathrm{nm}$ the typical width of the potential
well.

\subsection{Photon emission and absorption}

These surface processes are sensitive to the grain temperature. This
temperature is controlled by the absorption and emission of photons
by the grain, which we model as in \citet{Bron14}.

The power received by the grain at a photon energy $U$ is $P_{\mathrm{abs}}(U)=4\pi^{2}\,a^{2}\,Q_{\mathrm{abs}}(U)\,I_{U}(U)$,
with $Q_{\mathrm{abs}}(U)$ the absorption efficiency coefficient
of the grain at photon energy $U$, and $I_{U}(U)$ the radiation
field intensity at photon energy $U$ (in units of $\mathrm{W\cdot m^{-2}\cdot J^{-1}\cdot sr^{-1}}$).
Later, we need transition rates between thermal energy states of the
grain. The rate of photon absorptions at this energy $U$ is
\begin{equation}
R_{\mathrm{abs}}(U)={\displaystyle \frac{P_{\mathrm{abs}}(U)}{U}}.
\end{equation}

We approximate the grain emission by a modified black body law with
a specific intensity $Q_{\mathrm{abs}}(U)\,B_{U}(U,T_{\mathrm{d}})$,
where $B_{U}(U,T)$ is the usual black body specific intensity. The
power emitted at photon energy $U$ is then $P_{\mathrm{em}}(U,T_{\mathrm{d}})=4\pi^{2}\,a^{2}\,Q_{\mathrm{abs}}(U)\,B_{U}(U,T_{\mathrm{d}})$
and the photon emission rate is
\begin{equation}
R_{\mathrm{em}}(U,T_{\mathrm{d}})=\frac{P_{\mathrm{em}}(U,T_{\mathrm{d}})}{U}.
\end{equation}

These events occur randomly as Poisson processes and cause fluctuations
of the grain temperature. The impact of these fluctuations on the
efficiency of ortho-para conversion on dust grains is investigated
in details in the following sections. In the simpler rate equation
treatment that we will use for comparison, we neglect these fluctuations
and use the usual equilibrium temperature $T_{\mathrm{eq}}$ of the
grain, defined by the balance between the instantaneous emitted and
absorbed powers :
\begin{equation}
\int_{0}^{+\infty}dU\,P_{\mathrm{abs}}(U)=\int_{0}^{E_{eq}}dU\,P_{\mathrm{em}}(U,T_{\mathrm{eq}}),\label{eq:T_eq}
\end{equation}
where the upper bound on the right hand side accounts for the finite
total energy of the grain ($E_{\mathrm{eq}}$ is the thermal energy
of the grain at the equilibrium temperature, related to $T_{\mathrm{eq}}$
by $E_{\mathrm{eq}}=\int_{0}^{T_{\mathrm{eq}}}C(T)\,dT$, with $C(T)$
the heat capacity of the grain). 

In the following, we use a standard interstellar radiation field (\citealp{Mathis83})
and apply a scaling factor $\chi$ to the UV component of the field.
We measure the UV intensity of those fields using the usual $G_{0}=\frac{1}{u_{\mathrm{Habing}}}\,\int_{912\text{Å}}^{2400\text{Å}}d\lambda\,u_{\lambda}(\lambda)$,
where $u_{\mathrm{Habing}}=5.3\times10^{-15}\,\mathrm{J}\,\mathrm{m}^{-3}$.
$G_{0}$ is related to $\chi$ as $G_{0}\simeq0.65\,\chi$.

The dust properties ($C(T_{\mathrm{d}})$, $Q_{\mathrm{abs}}(U)$
and $\rho$) are taken from \citet{Compiegne11} and the \texttt{\textcolor{black}{DustEM}}
code\footnote{Available at \texttt{http://www.ias.u-psud.fr/DUSTEM/}.}.
We consider amorphous carbon and silicate dust populations and use
the properties used in this reference (see references therein, in
their Appendix A).

\subsection{Rate equation model at constant temperature\label{sub:Rate-equation-model}}

We will compare the results of our statistical treatment presented
in the next section to a simpler rate equation model that neglects
dust temperature fluctuations, and that we present in this section.
In this simple model, the grain is thus supposed to be at a constant
temperature $T_{\mathrm{d}}$, for which we take the equilibrium temperature
defined by Eq. \ref{eq:T_eq}.

Using the rates of the different processes defined above, we can thus
write a system of rate equations :

\begin{equation}
\left\{ \begin{array}{lr}
{\displaystyle \frac{dn_{\mathrm{o}}}{dt}=} & {\displaystyle k_{\mathrm{ads}}^{(\mathrm{o})}\left(1-\frac{n_{\mathrm{o}}+n_{\mathrm{p}}}{N_{\mathrm{s}}}\right)-n_{\mathrm{o}}\,k_{des}^{(\mathrm{o})}(T_{\mathrm{d}})}\\
 & {\displaystyle -n_{\mathrm{o}}\,k_{\mathrm{o}\rightarrow\mathrm{p}}(T_{\mathrm{d}})+n_{\mathrm{p}}\,k_{\mathrm{p}\rightarrow\mathrm{o}}(T_{\mathrm{d}})}\\
\\
{\displaystyle \frac{dn_{\mathrm{p}}}{dt}=} & {\displaystyle {\displaystyle k_{\mathrm{ads}}^{(\mathrm{p})}\left(1-\frac{n_{\mathrm{o}}+n_{\mathrm{p}}}{N_{\mathrm{s}}}\right)-n_{\mathrm{p}}\,k_{\mathrm{des}}^{(\mathrm{p})}(T_{\mathrm{d}})}}\\
 & {\displaystyle -n_{\mathrm{p}}\,k_{\mathrm{p}\rightarrow\mathrm{o}}(T_{\mathrm{d}})+n_{\mathrm{o}}\,k_{\mathrm{o}\rightarrow\mathrm{p}}(T_{\mathrm{d}})}
\end{array}\right.\label{eq:rate_equations}
\end{equation}

At equilibrium, we find{\footnotesize{}
\begin{equation}
n_{\mathrm{o}}=\frac{k_{\mathrm{ads}}^{(\mathrm{o})}\left(\frac{k_{\mathrm{ads}}^{(\mathrm{p})}}{N_{\mathrm{s}}}+k_{\mathrm{des}}^{(\mathrm{p})}(T_{\mathrm{d}})+k_{\mathrm{p}\rightarrow\mathrm{o}}(T_{\mathrm{d}})\right)+k_{\mathrm{ads}}^{(\mathrm{p})}\left(k_{\mathrm{p}\rightarrow\mathrm{o}}(T_{\mathrm{d}})-\frac{k_{\mathrm{ads}}^{(\mathrm{o})}}{N_{\mathrm{s}}}\right)}{D(T_{\mathrm{d}})}\label{eq:rate_eq_solution_o}
\end{equation}
}and{\footnotesize{}
\begin{equation}
n_{\mathrm{p}}=\frac{k_{\mathrm{ads}}^{(\mathrm{p})}\left(\frac{k_{\mathrm{ads}}^{(\mathrm{o})}}{N_{\mathrm{s}}}+k_{\mathrm{des}}^{(\mathrm{o})}(T_{\mathrm{d}})+k_{\mathrm{o}\rightarrow\mathrm{p}}(T_{\mathrm{d}})\right)+k_{\mathrm{ads}}^{(\mathrm{o})}\left(k_{\mathrm{o}\rightarrow\mathrm{p}}(T_{\mathrm{d}})-\frac{k_{\mathrm{ads}}^{(\mathrm{p})}}{N_{\mathrm{s}}}\right)}{D(T_{\mathrm{d}})}\label{eq:rate_eq_solution_p}
\end{equation}
with 
\begin{multline}
D(T_{\mathrm{d}})=\\
\left(\frac{k_{\mathrm{ads}}^{(\mathrm{o})}}{N_{\mathrm{s}}}+k_{\mathrm{des}}^{(\mathrm{o})}(T_{\mathrm{d}})+k_{\mathrm{o}\rightarrow\mathrm{p}}(T_{\mathrm{d}})\right)\left(\frac{k_{\mathrm{ads}}^{(\mathrm{p})}}{N_{\mathrm{s}}}+k_{\mathrm{des}}^{(\mathrm{p})}(T_{\mathrm{d}})+k_{\mathrm{p}\rightarrow\mathrm{o}}(T_{\mathrm{d}})\right)\\
-\left(k_{\mathrm{o}\rightarrow\mathrm{p}}(T_{\mathrm{d}})-\frac{k_{\mathrm{ads}}^{(\mathrm{p})}}{N_{\mathrm{s}}}\right)\left(k_{\mathrm{p}\rightarrow\mathrm{o}}(T_{\mathrm{d}})-\frac{k_{\mathrm{ads}}^{(\mathrm{o})}}{N_{\mathrm{s}}}\right)
\end{multline}
}{\footnotesize \par}

To evaluate the efficiency of the ortho-para conversion process, we
first define the net ortho-para conversion rate (in $\mathrm{s}^{-1}$)
on the grain surface as 
\begin{equation}
k_{\mathrm{O/P}}^{(\mathrm{grain})}=n_{\mathrm{o}}\,k_{\mathrm{o}\rightarrow\mathrm{p}}-n_{\mathrm{p}}\,k_{\mathrm{p}\rightarrow\mathrm{o}}.
\end{equation}
The net conversion rate is positive when conversion occurs in the
usual ortho-to-para direction, and negative if para-to-ortho conversion
occurs. This latter case could happen if the grains are warmer than
the gas.

We then define the ortho-para conversion efficiency as 
\begin{equation}
\eta_{\mathrm{O/P}}^{(\mathrm{grain)}}=\frac{k_{\mathrm{O/P}}^{(\mathrm{grain})}}{k_{\mathrm{coll}}^{(\mathrm{o})}\,S(T_{\mathrm{gas}})}\label{eq:conversion_efficiency}
\end{equation}
which represents the fraction of the ortho-H$_{2}$ molecules sticking
to the grain that are converted to para-H$_{2}$ (in a stationary
situation). Note that we took the parts that depend on the gas conditions
(density, temperature) out of our definition of the conversion efficiency,
in order to separate the effects of the sticking coefficient from
the surface conversion process. Note also that this definition is
only meaningful if conversion occurs in the ortho-to-para direction,
which will be the case in our applications.

In this constant temperature rate equation model, the results are
not directly affected by the grain size, except through the equilibrium
temperature which is size dependent. We will thus show here a few
results as a function of dust temperature.

As the conversion efficiency is controlled by the competition between
conversion and desorption (an ortho-H$_{2}$ molecule needs to have
time to convert before desorbing), the two main parameters affecting
this efficiency are the binding energy $T_{\mathrm{phys}}$, which
controls the desorption rate, and the conversion timescale $\tau_{\mathrm{conv}}$.

\begin{figure}
\begin{centering}
\includegraphics[width=1\columnwidth]{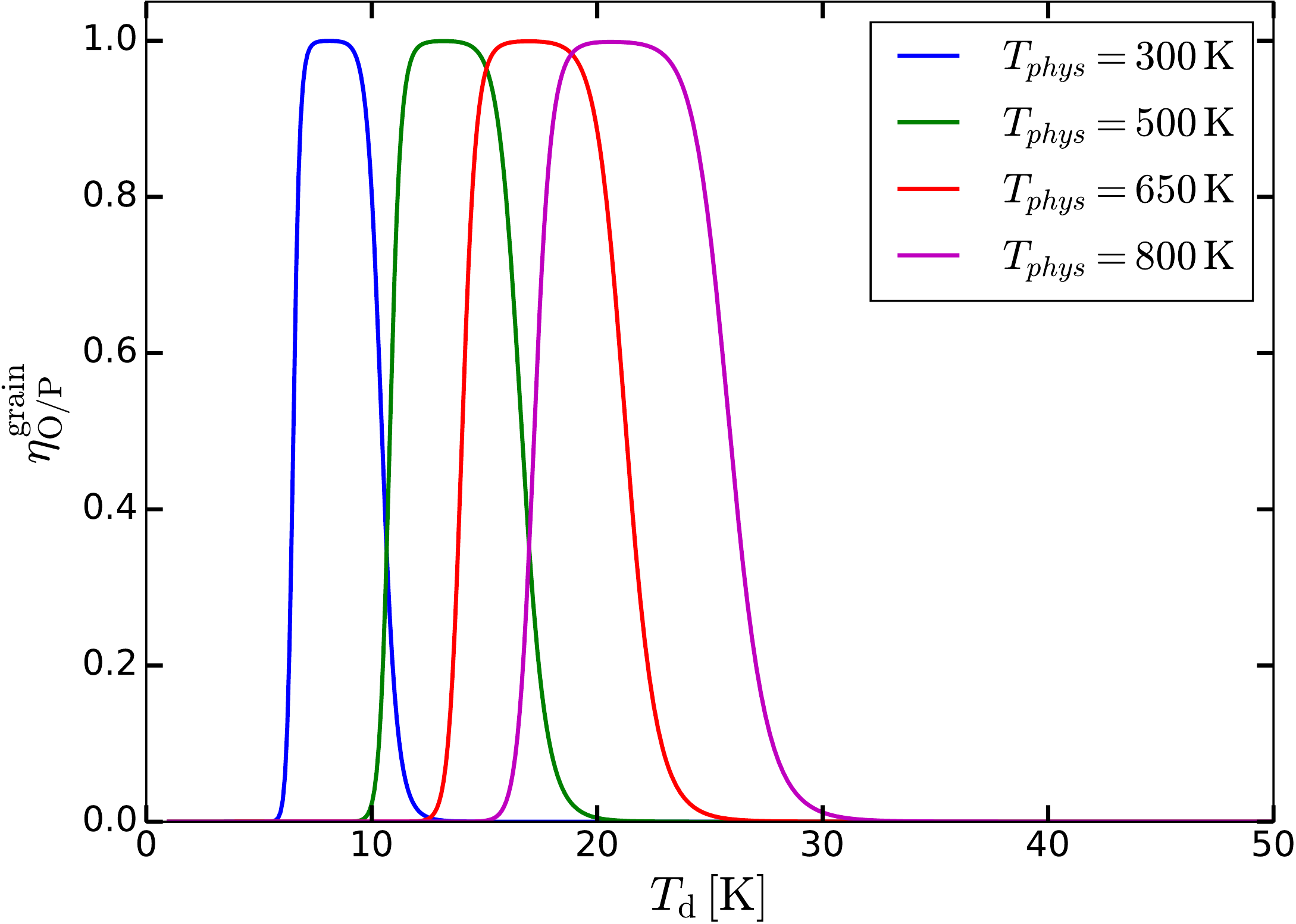}
\par\end{centering}

\protect\caption{Conversion efficiency as a function of the grain temperature for different
para-H$_{2}$ binding energies in the constant temperature rate equation
model. The gas ortho-para ratio is taken as 3, and the conversion
timescale at $10\,\mathrm{s}$.\label{fig:rate_eq_efficiency_binding energy}}

\end{figure}

Fig. \ref{fig:rate_eq_efficiency_binding energy} shows the impact
of the binding energy on the conversion efficiency. In all cases,
the efficiency curves are bell-shaped, with a flat top at full efficiency.
At high temperature, desorption is too fast and the adsorbed H$_{2}$
molecules do not have time to convert from ortho to para. At low temperature,
H$_{2}$ molecules cover all the surface as desorption is very slow,
and most molecules coming from the gas are rejected. The range of
temperature where conversion is efficient is thus very limited. 

The width and position of the efficiency window is strongly affected
by the binding energy, going from $7-10\,\mathrm{K}$ for $T_{\mathrm{phys}}=300\,\mathrm{K}$
to $17-26\,\mathrm{K}$ for $T_{\mathrm{phys}}=800\,\mathrm{K}$.
Consequently, this model predicts that in PDR conditions where the
grains are exposed to a strong UV field and are thus warmer than this
efficiency window (typically $30-50\,\mathrm{K}$ for small grains),
ortho-para conversion on dust grain should be inefficient. We will
see in the following sections that taking into account the dust temperature
fluctuations significantly changes this picture.

\begin{figure}
\begin{centering}
\includegraphics[width=1\columnwidth]{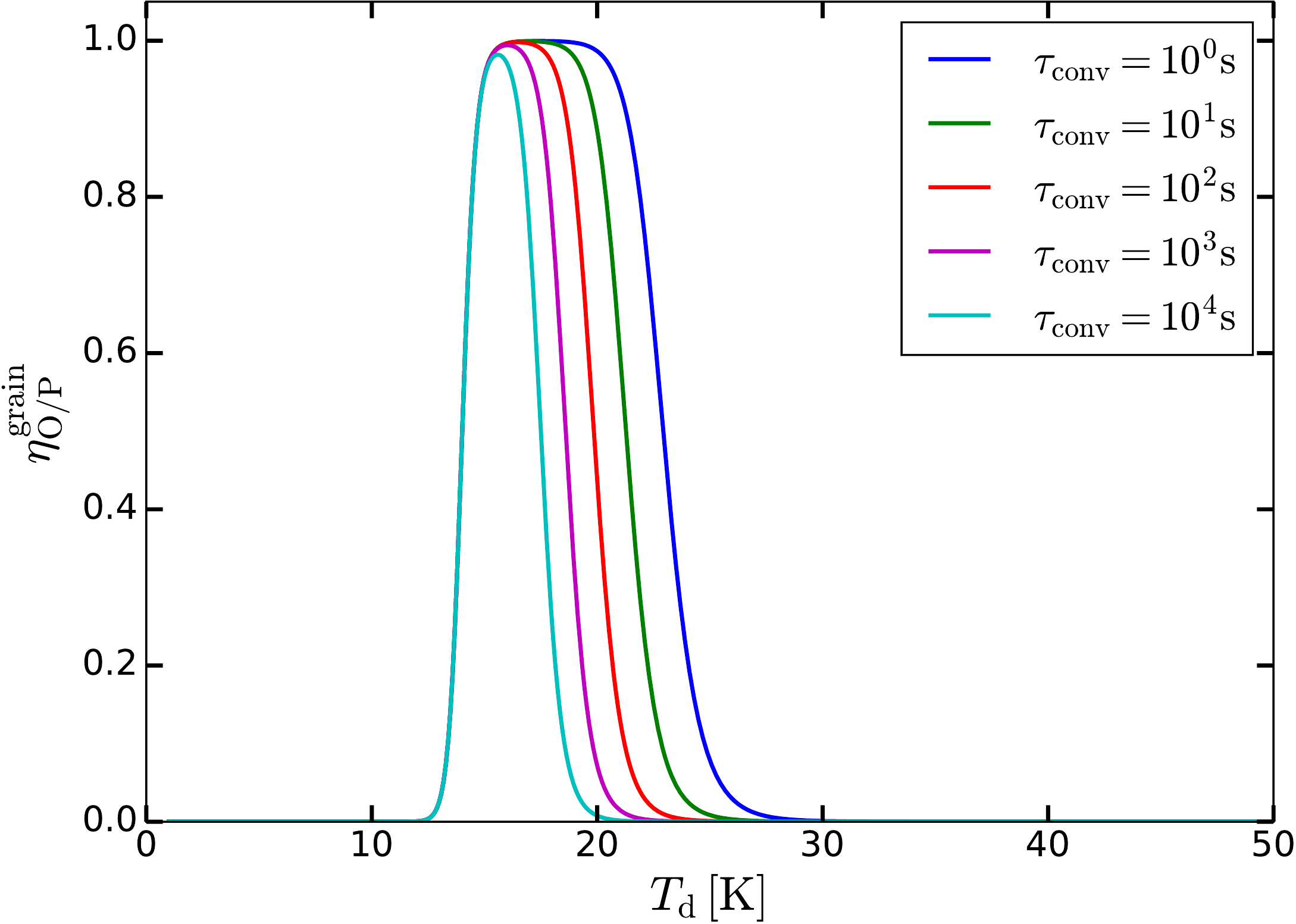}
\par\end{centering}

\protect\caption{Conversion efficiency as a function of the grain temperature for different
conversion timescale in the constant temperature rate equation model.
The gas ortho-para ratio is taken as 3, and the binding energy at
$650\,\mathrm{K}$.\label{fig:rate_eq_efficiency_conversion_timescale}}
\end{figure}

Fig. \ref{fig:rate_eq_efficiency_conversion_timescale} shows the
influence of the conversion timescale. It only affects the upper limit
of the efficiency window as the lower limit is due to rejection and
controlled by the competition between adsorption and desorption. Its
impact is less than that of the binding energy as the conversion rate
is inversely proportional to this timescale, while the desorption
rate depends exponentially on the binding energy. It can still double
the width of the efficiency window over the range of relevant values.

\begin{figure}
\begin{centering}
\includegraphics[width=1\columnwidth]{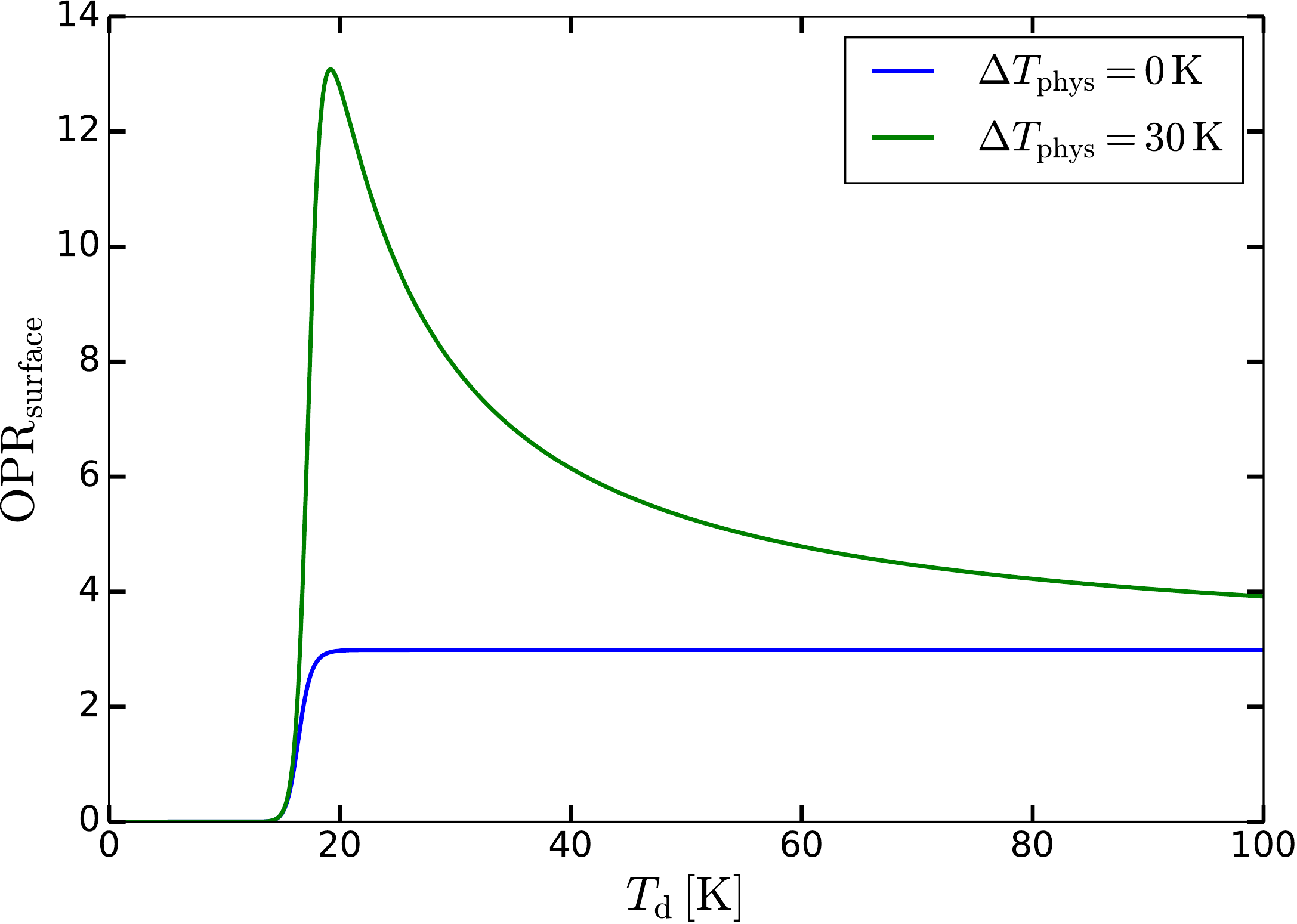}
\par\end{centering}

\protect\caption{Ortho-para ratio of the adsorbed molecules, with and without a difference
of binding energy $\Delta T_{\mathrm{phys}}$ between ortho-H$_{2}$
and para-H$_{2}$. This computation was done with a gas ortho-para
ratio of 3, a binding energy for para-H$_{2}$ of $500\,\mathrm{K}$
and a conversion timescale of $10\,\mathrm{s}$. \label{fig:rate_eq_OPR_surface}}

\end{figure}

Finally, Fig. \ref{fig:rate_eq_OPR_surface} shows how a difference
in binding energy $\Delta T_{\mathrm{phys}}$ between ortho-H$_{2}$
and para-H$_{2}$ affects the OPR of the adsorbed molecules. At low
temperature, conversion is much faster than desorption and the OPR
on the surface is very low. When desorption starts to become significant,
para-H$_{2}$ desorbs faster than ortho-H$_{2}$ and the surface population
is enriched in ortho-H$_{2}$ leading to a surface OPR significantly
higher than 3. This enrichment could play a role for further surface
reactions leading to more complex molecules. A higher binding energy
for ortho-H$_{2}$ than for para-H$_{2}$ also makes the high efficiency
window slightly larger, but the effect is small compared to the effects
of the other parameters described above.

\section{Method for a stochastic model\label{sec:Statistical_method}}

We will now take into account the dust temperature fluctuations. We
are interested in the average conversion efficiency under the effect
of these fluctuations in a statistically stationary situation. It
is thus sufficient to compute the stationary probability density function
(hereafter PDF) of the state of the grain, $f(T_{\mathrm{d}},n_{\mathrm{o}},n_{\mathrm{p}})$,
from which all average quantities of interest can be deduced. In the
following, we will note $T$ the instantaneous temperature of the
grain to simplify the notations. Moreover, we will work with the thermal
energy of the grain $E$ rather than its temperature $T$. The two
are related by 
\begin{equation}
E(T)=\int_{0}^{T}dT'\,C(T')
\end{equation}
where $C(T)$ is the heat capacity of the grain. Conversely, we will
also note $T(E)$ the temperature corresponding to thermal energy
$E$.

\subsection{General master equation}

As the state of the system evolves in time by discrete events (photon
absorption or emission, adsorption, desorption or conversion of a
molecule), the evolution of its PDF is governed by a master equation
of the form
\begin{equation}
\frac{df(X)}{dt}=\int_{\mathrm{states}}dY\,f(Y)\,p_{Y\rightarrow X}-\int_{states}dY\,f(X)\,p_{X\rightarrow Y}\label{eq:Generic_master_equation}
\end{equation}
where $X$ generically describes the state of the system, and $p_{X\rightarrow Y}$
are the transition rates from state $X$ to state $Y$. The integrals
are to be interpreted as integrals over the continuous state variables
(e.g., the thermal energy $E$ of the grain) and sums over the discrete
state variables (e.g., the number $n_{\mathrm{o}}$ of ortho-H$_{2}$
molecules on the surface).

The transition rates of the possible events are 
\begin{equation}
\begin{cases}
E,n_{\mathrm{o}},n_{\mathrm{p}}\rightarrow E'>E,n_{\mathrm{o}},n_{\mathrm{p}}\,: & R_{\mathrm{abs}}(E'-E)\\
E,n_{\mathrm{o}},n_{\mathrm{p}}\rightarrow E'<E,n_{\mathrm{o}},n_{\mathrm{p}}\,: & R_{\mathrm{em}}(E-E',T(E))\\
E,n_{\mathrm{o}},n_{\mathrm{p}}\rightarrow E,n_{\mathrm{o}}+1,n_{\mathrm{p}}\,: & k_{\mathrm{ads}}^{(\mathrm{o})}\left(1-\frac{n_{\mathrm{o}}+n_{\mathrm{p}}}{N_{\mathrm{s}}}\right)\\
E,n_{\mathrm{o}},n_{\mathrm{p}}\rightarrow E,n_{\mathrm{o}},n_{\mathrm{p}}+1\,: & k_{\mathrm{ads}}^{(\mathrm{p})}\left(1-\frac{n_{\mathrm{o}}+n_{\mathrm{p}}}{N_{\mathrm{s}}}\right)\\
E,n_{\mathrm{o}},n_{\mathrm{p}}\rightarrow E,n_{\mathrm{o}}-1,n_{\mathrm{p}}\,: & n_{\mathrm{o}}\,k_{\mathrm{des}}^{(\mathrm{o})}(T_{\mathrm{grain}})\\
E,n_{\mathrm{o}},n_{\mathrm{p}}\rightarrow E,n_{\mathrm{o}},n_{\mathrm{p}}-1\,: & n_{\mathrm{p}}\,k_{\mathrm{des}}^{(\mathrm{p})}(T_{\mathrm{grain}})\\
E,n_{\mathrm{o}},n_{\mathrm{p}}\rightarrow E,n_{\mathrm{o}}-1,n_{\mathrm{p}}+1\,: & n_{\mathrm{o}}\,k_{\mathrm{o}\rightarrow\mathrm{p}}(T_{\mathrm{grain}})\\
E,n_{\mathrm{o}},n_{\mathrm{p}}\rightarrow E,n_{\mathrm{o}}+1,n_{\mathrm{p}}-1\,: & n_{\mathrm{p}}\,k_{\mathrm{p}\rightarrow\mathrm{o}}(T_{\mathrm{grain}})\\
\mathrm{all}\:\mathrm{other}\:\mathrm{cases}\,: & 0
\end{cases}
\end{equation}

Writing explicitly the terms corresponding to the different processes
in Eq. \ref{eq:Generic_master_equation} and considering statistical
equilibrium, we get the stationary master equation governing $f(E,n_{\mathrm{o}},n_{\mathrm{p}})$:
\begin{multline}
\int_{0}^{E}dE'\,R_{\mathrm{abs}}(E-E')\,f(E',n{}_{\mathrm{o}},n{}_{\mathrm{p}})\\
+\int_{E}^{+\infty}dE'\,R_{\mathrm{em}}(E'-E,T(E'))\,f(E',n{}_{\mathrm{o}},n{}_{\mathrm{p}})\\
+k_{\mathrm{ads}}^{(\mathrm{o})}\,\left(1-\frac{n_{\mathrm{o}}+n_{\mathrm{p}}-1}{N_{\mathrm{s}}}\right)\,f(E,n{}_{\mathrm{o}}-1,n{}_{\mathrm{p}})\\
+k_{\mathrm{ads}}^{(\mathrm{p})}\,\left(1-\frac{n_{\mathrm{o}}+n_{\mathrm{p}}-1}{N_{\mathrm{s}}}\right)\,f(E,n{}_{\mathrm{o}},n{}_{\mathrm{p}}-1)\\
+(n_{\mathrm{o}}+1)\,k_{\mathrm{des}}^{(\mathrm{o})}(T(E))\,f(E,n{}_{\mathrm{o}}+1,n{}_{\mathrm{p}})\\
+(n_{\mathrm{p}}+1)\,k_{\mathrm{des}}^{(\mathrm{p})}(T(E))\,f(E,n{}_{\mathrm{o}},n{}_{\mathrm{p}}+1)\\
+(n_{\mathrm{o}}+1)\,k_{\mathrm{o}\rightarrow\mathrm{p}}(T(E))\,f(E,n\mathrm{_{o}}+1,n{}_{\mathrm{p}}-1)\\
+(n_{\mathrm{p}}+1)\,k_{\mathrm{p}\rightarrow\mathrm{o}}(T(E))\,f(E,n{}_{\mathrm{o}}-1,n{}_{\mathrm{p}}+1)=\\
f(E,n{}_{\mathrm{o}},n{}_{\mathrm{p}})M(E,n_{\mathrm{o}},n_{\mathrm{p}})\label{eq:Master_general}
\end{multline}
where
\begin{multline}
M(E,n_{\mathrm{o}},n_{\mathrm{p}})=\int_{0}^{E}dE'R_{\mathrm{em}}(E-E',T(E))+\int_{E}^{+\infty}dE'R_{\mathrm{abs}}(E'-E)\\
+k_{\mathrm{ads}}^{(\mathrm{o})}\,\left(1-\frac{n_{\mathrm{o}}+n_{\mathrm{p}}}{N_{\mathrm{s}}}\right)+k_{\mathrm{ads}}^{(\mathrm{p})}\,\left(1-\frac{n_{\mathrm{o}}+n_{\mathrm{p}}}{N_{\mathrm{s}}}\right)+n_{\mathrm{o}}\,k_{\mathrm{des}}^{(\mathrm{o})}(T(E))\\
+n_{\mathrm{p}}\,k_{\mathrm{des}}^{(\mathrm{p})}(T(E))+n_{\mathrm{o}}\,k_{\mathrm{o}\rightarrow\mathrm{p}}(T(E))+n_{\mathrm{p}}\,k_{\mathrm{p}\rightarrow\mathrm{o}}(T(E))\label{eq:loss_general}
\end{multline}
As boundary conditions, $f(E,n_{\mathrm{o}},n_{\mathrm{p}})=0$ if
$n_{o}<0$, $n_{\mathrm{p}}<0$, $E<0$ or $n_{o}+n_{p}>N_{\mathrm{s}}$.

Directly solving this equation numerically would be extremely time-consuming
as the unknown is a function of three variables. We can however simplify
the problem by noting that all the average quantities of interest
can be expressed in terms of the marginal thermal energy PDF 
\begin{equation}
f_{E}(E)=\sum_{n_{\mathrm{o}},n_{\mathrm{p}}}f(E,n_{\mathrm{o}},n_{\mathrm{p}})
\end{equation}
and of the conditional expectation for the ortho and para populations
at a given instantaneous thermal energy
\begin{equation}
\left\langle n_{\mathrm{o}}\left|E\right.\right\rangle =\sum_{n_{\mathrm{o}},n_{\mathrm{p}}}n_{\mathrm{o}}\frac{f(E,n_{\mathrm{o}},n_{\mathrm{p}})}{f_{E}(E)}
\end{equation}
and
\begin{equation}
\left\langle n_{\mathrm{p}}\left|E\right.\right\rangle =\sum_{n_{\mathrm{o}},n_{\mathrm{p}}}n_{\mathrm{p}}\frac{f(E,n_{\mathrm{o}},n_{\mathrm{p}})}{f_{E}(E)}
\end{equation}
For instance, the average conversion efficiency (using the definition
by Eq. \ref{eq:conversion_efficiency})
\begin{multline}
\left\langle \eta_{\mathrm{O/P}}^{(\mathrm{grain})}\right\rangle =\\
\int_{0}^{+\infty}dE\sum_{n_{\mathrm{o}},n_{\mathrm{p}}}\frac{n_{\mathrm{o}}k_{\mathrm{o}\rightarrow\mathrm{p}}(T(E))-n_{\mathrm{p}}k_{\mathrm{p}\rightarrow\mathrm{o}}(T(E))}{k_{\mathrm{coll}}^{(\mathrm{o})}\,S(T_{\mathrm{gas}})}f(E,n_{\mathrm{o}},n_{\mathrm{p}})
\end{multline}
can also be expressed as
\begin{multline}
\left\langle \eta_{\mathrm{O/P}}^{(\mathrm{grain})}\right\rangle =\\
\int_{0}^{+\infty}dE\frac{\left\langle n_{\mathrm{o}}\left|E\right.\right\rangle k_{\mathrm{o}\rightarrow\mathrm{p}}(T(E))-\left\langle n_{\mathrm{p}}\left|E\right.\right\rangle k_{\mathrm{p}\rightarrow\mathrm{o}}(T(E))}{k_{\mathrm{coll}}^{(\mathrm{o})}\,S(T_{\mathrm{gas}})}f_{E}(E)\label{eq:average_efficiency}
\end{multline}
We will thus compute three single-variable functions rather than one
3-variable function. We now deduce the equations governing these three
functions.

\subsection{Marginal temperature equation}

We start by deriving the equation governing the marginal thermal energy
PDF. This equation can be deduced from the main master equation Eq.
\ref{eq:Generic_master_equation}.

By applying ${\displaystyle \sum_{n_{\mathrm{o}}=0,n_{\mathrm{p}}=0}^{n_{\mathrm{o}}+n_{\mathrm{p}}\leq N_{s}}}$
to Eq. \ref{eq:Generic_master_equation}, we get 
\begin{multline}
P(E)\,f_{E}(E)=\int_{0}^{E}dE'\,R_{\mathrm{abs}}(E-E')\,f_{E}(E')\\
+\int_{E}^{+\infty}dE'\,R_{\mathrm{em}}(E'-E,T(E'))\,f_{E}(E')\label{eq:Temperature_marginal_equation}
\end{multline}
with 
\begin{equation}
P(E)=\int_{E}^{+\infty}dE'\,R_{\mathrm{abs}}(E'-E)+\int_{0}^{E}dE'\,R_{\mathrm{em}}(E-E',T(E))
\end{equation}

This equation governs the marginal PDF $f_{E}(E)$ of the thermal
energy of the grain, which describes the stationary statistics of
the temperature fluctuations. It was already encountered in \citet{Bron14}
when studying the impact of dust temperature fluctuations on H$_{2}$
formation on grain surfaces. We use here the same numerical resolution
method, consisting in iteratively applying the operator 
\begin{multline}
\mathcal{L}[f](E)=\\
\frac{\int_{0}^{E}dE'\,R_{\mathrm{abs}}(E-E')\,f_{E}(E')+\int_{E}^{+\infty}dE'\,R_{\mathrm{em}}(E'-E,T(E'))\,f_{E}(E')}{P(E)}
\end{multline}
to an initial guess until some convergence criterium is met.

The thermal energy PDF $f_{E}$ can be converted into the temperature
PDF $f_{T}$ using the usual variable change rule for PDFs:
\begin{equation}
f_{T}(T)=f_{E}(E)\,C(T)
\end{equation}

\subsection{Marginal population equations}

We now deduce the equations governing the conditional expectation
of the ortho and para populations. Applying the operators ${\displaystyle \sum_{n_{\mathrm{o}}=0,n_{\mathrm{p}}=0}^{n_{\mathrm{o}}+n_{\mathrm{p}}\leq N_{s}}}n_{\mathrm{o}}\times\cdot$
and ${\displaystyle \sum_{n_{\mathrm{o}}=0,n_{\mathrm{p}}=0}^{n_{\mathrm{o}}+n_{\mathrm{p}}\leq N_{s}}}n_{p}\times\cdot$
to Eq. \ref{eq:Generic_master_equation} yields the system of equations
governing $\left\langle n_{\mathrm{o}}\left|E\right.\right\rangle $
and $\left\langle n_{\mathrm{p}}\left|E\right.\right\rangle $: 
\begin{multline}
\int_{0}^{E}dE'\,R_{\mathrm{abs}}(E-E')\,\frac{f_{E}(E')}{f_{E}(E)}\,\left\langle \left.n_{\mathrm{o}}\right|E'\right\rangle \\
+\int_{E}^{+\infty}dE'\,R_{\mathrm{em}}(E'-E,T(E'))\,\frac{f_{E}(E')}{f_{E}(E)}\,\left\langle \left.n_{\mathrm{o}}\right|E'\right\rangle =\\
-k_{\mathrm{ads}}^{(\mathrm{o})}+T_{\mathrm{op}}(E)\,\left\langle \left.n_{\mathrm{p}}\right|E\right\rangle +T_{\mathrm{oo}}(E)\,\left\langle \left.n_{\mathrm{o}}\right|E\right\rangle \label{eq:ortho_marginal_equation}
\end{multline}
and
\begin{multline}
\int_{0}^{E}dE'\,R_{\mathrm{abs}}(E-E')\,\frac{f_{E}(E')}{f_{E}(E)}\,\left\langle \left.n_{\mathrm{p}}\right|E'\right\rangle \\
+\int_{E}^{+\infty}dE'\,R_{\mathrm{em}}(E'-E,T(E'))\,\frac{f_{E}(E')}{f_{E}(E)}\,\left\langle \left.n_{\mathrm{p}}\right|E'\right\rangle =\\
-k_{\mathrm{ads}}^{(\mathrm{p})}+T_{\mathrm{po}}(E)\,\left\langle \left.n_{\mathrm{o}}\right|E\right\rangle +T_{\mathrm{pp}}(E)\,\left\langle \left.n_{\mathrm{p}}\right|E\right\rangle \label{eq:para_marginal_equation}
\end{multline}

where
\begin{equation}
T_{\mathrm{op}}(E)=\frac{k_{\mathrm{ads}}^{(\mathrm{o})}}{N_{\mathrm{s}}}-k_{\mathrm{p}\rightarrow\mathrm{o}}(T(E))
\end{equation}
\begin{equation}
T_{\mathrm{oo}}(E)=P(E)+k_{\mathrm{o}\rightarrow\mathrm{p}}(T(E))+\frac{k_{\mathrm{ads}}^{(\mathrm{o})}}{N_{\mathrm{s}}}+k_{\mathrm{des}}^{(\mathrm{o})}(T(E))
\end{equation}
\begin{equation}
T_{\mathrm{po}}(E)=\frac{k_{\mathrm{ads}}^{(\mathrm{p})}}{N_{\mathrm{s}}}-k_{\mathrm{o}\rightarrow\mathrm{p}}(T(E))
\end{equation}
\begin{equation}
T_{\mathrm{pp}}(E)=P(E)+k_{\mathrm{p}\rightarrow\mathrm{o}}(T(E))+\frac{k_{\mathrm{ads}}^{(\mathrm{p})}}{N_{\mathrm{s}}}+k_{\mathrm{des}}^{(\mathrm{p})}(T(E))
\end{equation}

This is a system of coupled inhomogeneous second-kind Fredholm equations.
Similarly to the situation encountered in \citet{Bron14} with a similar
equation, the discretized version of this system gives a linear system
that converges exponentially fast towards a singular system when the
grain size $a$ grows. To avoid numerical problems, we use the same
trick as in \citet{Bron14} and eliminate the constant terms $-k_{\mathrm{ads}}^{(\mathrm{o})}$
and $-k_{\mathrm{ads}}^{(\mathrm{p})}$ to get a system of homogeneous
second-kind Fredholm equation that can be solved by the same iterative
method as the marginal temperature equation in the previous section.

By multiplying equations \ref{eq:ortho_marginal_equation} and \ref{eq:para_marginal_equation}
by $f_{E}(E)$ and integrating over $E$, we obtain
\begin{multline}
k_{\mathrm{ads}}^{(\mathrm{o})}=\int_{0}^{+\infty}dE'\,T_{\mathrm{op}}(E')\,f_{E}(E')\,\left\langle \left.n_{\mathrm{p}}\right|E'\right\rangle \\
+\int_{0}^{+\infty}dE'\,T_{\mathrm{oo}}'(E')\,f_{E}(E')\,\left\langle \left.n_{\mathrm{o}}\right|E'\right\rangle \label{eq:constant_elimination_ortho}
\end{multline}
and
\begin{multline}
k_{\mathrm{ads}}^{(\mathrm{p})}=\int_{0}^{+\infty}dE'\,T_{\mathrm{po}}(E')\,f_{E}(E')\,\left\langle \left.n_{\mathrm{o}}\right|E'\right\rangle \\
+\int_{0}^{+\infty}dE'\,T_{\mathrm{pp}}'(E')\,f_{E}(E')\,\left\langle \left.n_{\mathrm{p}}\right|E'\right\rangle \label{eq:constant_elimination_para}
\end{multline}
with $T_{\mathrm{oo}}'(E)=T_{\mathrm{oo}}(E)-P(E)$ and $T_{\mathrm{pp}}'(E)=T_{\mathrm{pp}}(E)-P(E)$.

Injecting equations \ref{eq:constant_elimination_ortho} and \ref{eq:constant_elimination_para}
into equations \ref{eq:ortho_marginal_equation} and \ref{eq:para_marginal_equation},
we get
\begin{multline}
\int_{0}^{+\infty}dE'\left[\frac{G(E,E')}{f(E)}+T_{\mathrm{oo}}'(E')\right]f_{E}(E')\,\left\langle \left.n_{\mathrm{o}}\right|E'\right\rangle \\
+\int_{0}^{+\infty}dE'\,T_{\mathrm{op}}(E')\,f_{E}(E')\,\left\langle \left.n_{\mathrm{p}}\right|E'\right\rangle =\\
T_{\mathrm{op}}(E)\,\left\langle \left.n_{\mathrm{p}}\right|E\right\rangle +T_{\mathrm{oo}}(E)\,\left\langle \left.n_{\mathrm{o}}\right|E\right\rangle \label{eq:pop_equations_interm_o}
\end{multline}
and
\begin{multline}
\int_{0}^{+\infty}dE'\left[\frac{G(E,E')}{f_{E}(E)}+T_{\mathrm{pp}}'(E')\right]f_{E}(E')\,\left\langle \left.n_{\mathrm{p}}\right|E'\right\rangle \\
+\int_{0}^{+\infty}dE'\,T_{\mathrm{po}}(E')\,f_{E}(E')\,\left\langle \left.n_{\mathrm{o}}\right|E'\right\rangle =\\
T_{\mathrm{po}}(E)\,\left\langle \left.n_{\mathrm{o}}\right|E\right\rangle +T_{\mathrm{pp}}(E)\,\left\langle \left.n_{\mathrm{p}}\right|E\right\rangle \label{eq:pop_equation_interm_p}
\end{multline}
with 
\begin{equation}
G(E,E')=\begin{cases}
R_{\mathrm{abs}}(E-E') & \mathrm{if}\quad E'<E\\
R_{\mathrm{em}}(E'-E,T(E')) & \mathrm{if}\quad E'>E
\end{cases}.
\end{equation}
Equations \ref{eq:pop_equations_interm_o} and \ref{eq:pop_equation_interm_p}
are now homogeneous.

Finally, we can rewrite this system in vector form as {\tiny{}
\begin{multline}
\left(\begin{array}{cc}
T_{\mathrm{oo}}(E) & T_{\mathrm{op}}(E)\\
T_{\mathrm{po}}(E) & T_{\mathrm{pp}}(E)
\end{array}\right)^{-1}\times\int_{0}^{+\infty}dE'\,\Biggl[\\
\left(\begin{array}{cc}
\left[\frac{G(E,E')}{f(E)}+T_{\mathrm{oo}}'(E')\right]f_{E}(E') & T_{\mathrm{op}}(E')\,f_{E}(E')\\
T_{\mathrm{po}}(E')\,f_{E}(E') & \left[\frac{G(E,E')}{f(E)}+T_{\mathrm{pp}}'(E')\right]f_{E}(E')
\end{array}\right)\left(\begin{array}{c}
\left\langle \left.n_{\mathrm{o}}\right|E'\right\rangle \\
\left\langle \left.n_{\mathrm{p}}\right|E'\right\rangle 
\end{array}\right)\\
\Biggr]=\left(\begin{array}{c}
\left\langle \left.n_{\mathrm{o}}\right|E\right\rangle \\
\left\langle \left.n_{\mathrm{p}}\right|E\right\rangle 
\end{array}\right)\label{eq:population_marginal_equation_vector_form_final}
\end{multline}
}where the integral sign is to be applied to each component of the
vector. The first matrix is always invertible, as can be easily verified
by expressing its determinant.

This equation shows that the solution is an eigenvector associated
with eigenvalue 1 for the linear integral operator defined by the
left-hand side. After discretization of the problem, we numerically
compute such an eigenvector, and we then normalize it using equations
\ref{eq:constant_elimination_ortho} and \ref{eq:constant_elimination_para}.
Once $f_{E}(E)$, $\left\langle n_{\mathrm{o}}\left|E\right.\right\rangle $
and $\left\langle n_{\mathrm{p}}\left|E\right.\right\rangle $ are
computed, we can compute all the average quantities of interest.

\section{Results\label{sec:Results}}

We now present the results of this computation of the ortho-para conversion
efficiency taking into account the dust temperature fluctuations.
We first discuss how the temperature fluctuations change significantly
the efficiency in PDR-like conditions, before investigating the influence
of the microphysical parameters of the model. When not specified otherwise,
our standard model consists of amorphous carbon grains with a binding
energy of H$_{2}$ $T_{\mathrm{phys}}=550\,\mathrm{K}$ and a conversion
timescale $\tau_{\mathrm{conv}}=10\,\mathrm{s}$.

\subsection{Efficiency of the conversion process in PDR-like conditions}

\begin{figure}
\begin{centering}
\includegraphics[width=1\columnwidth]{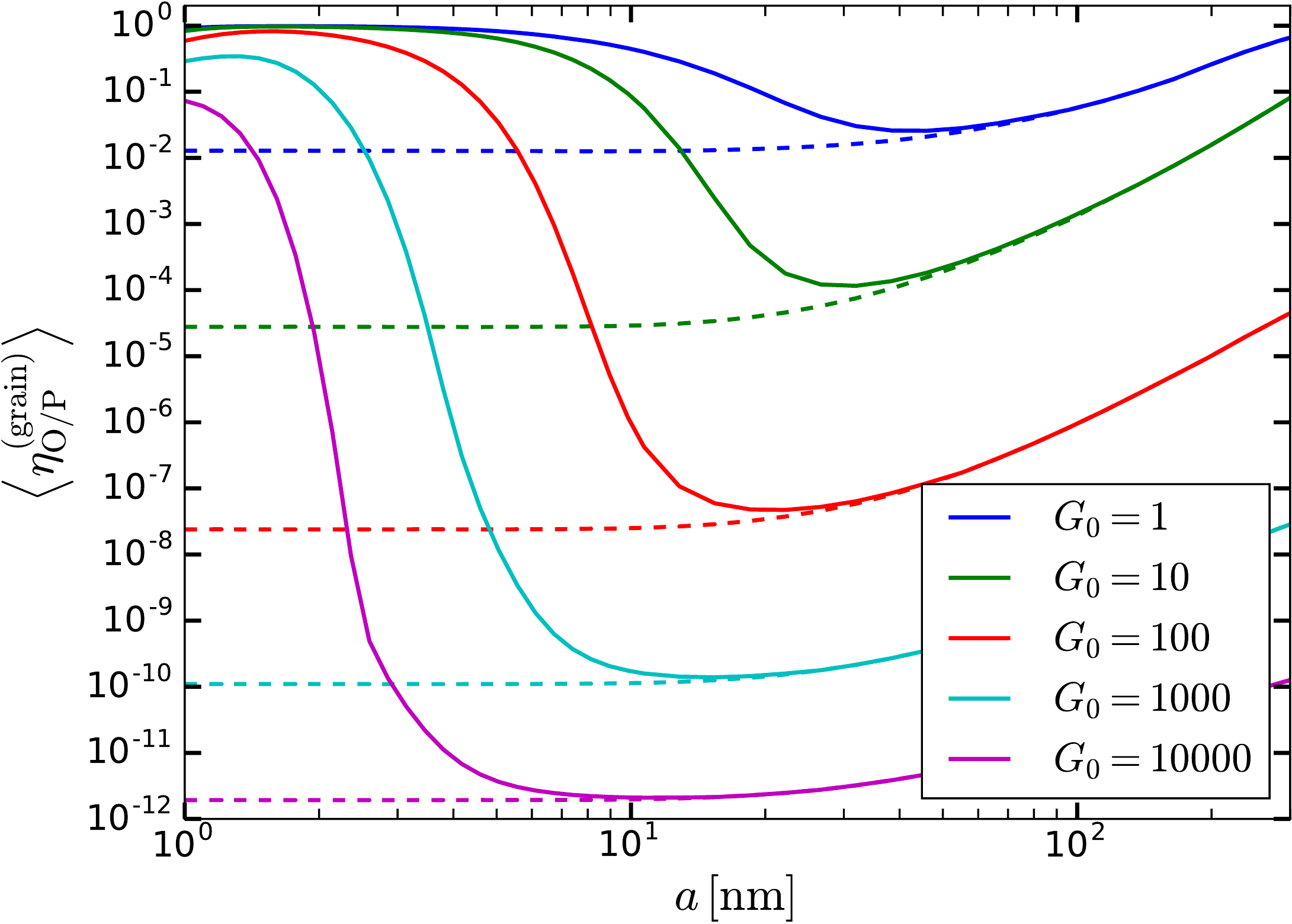}
\par\end{centering}

\protect\caption{Conversion efficiency on one grain as a function of grain size for
different radiation field intensities $G_{0}$. Solid lines : Full
statistical model, dashed lines : rate equation model without dust
temperature fluctuations.\label{fig:eta_vs_a_khis}}
\end{figure}

We will present the results in terms of the average conversion efficiency
defined by Eq. \ref{eq:average_efficiency}. Fig. \ref{fig:eta_vs_a_khis}
shows this average conversion efficiency for a single grain as a function
of grain size, and for different ambient UV radiation field intensities
$G_{0}$. The results of the full statistical computation (solid lines)
are compared to those of the simpler rate equation model which neglects
the dust temperature fluctuations (dashed lines). 

In the rate equation model, the grains are assumed to be at constant
temperature at their equilibrium temperature. As discussed in Sect.
\ref{sub:Rate-equation-model}, the ortho-para conversion process
is then only efficient when the grains are sufficiently cold. The
rate equation model thus only approaches full efficiency for the largest
grains, and gives efficiencies that decrease sharply with the radiation
field intensities for all sizes. Under $G_{0}=1$, the efficiently
is of the order of a few percents for most sizes.

The full statistical model gives significantly different results.
For large grains, UV photons (limited to the Lyman limit in PDRs)
do not have enough energy to cause significant fluctuations of the
dust temperature. The full model and the rate equation model thus
give the same efficiencies for large grains. For small grains, the
full model finds efficiencies orders of magnitude higher than the
rate equation model. There seems to be a critical size below which
conversion occurs at almost full efficiency. This critical size distinguishes
small grains for which the temperature PDF is wide enough to overlap
with the high efficiency window found in Sect. \ref{sub:Rate-equation-model}
from large grains whose narrow temperature PDF falls entirely outside
this window. Small grains thus spend a very large fraction of their
time at low temperatures, where conversion is efficient, between very
short high temperature spikes during which desorption becomes dominant.

\begin{figure}
\begin{centering}
\includegraphics[width=1\columnwidth]{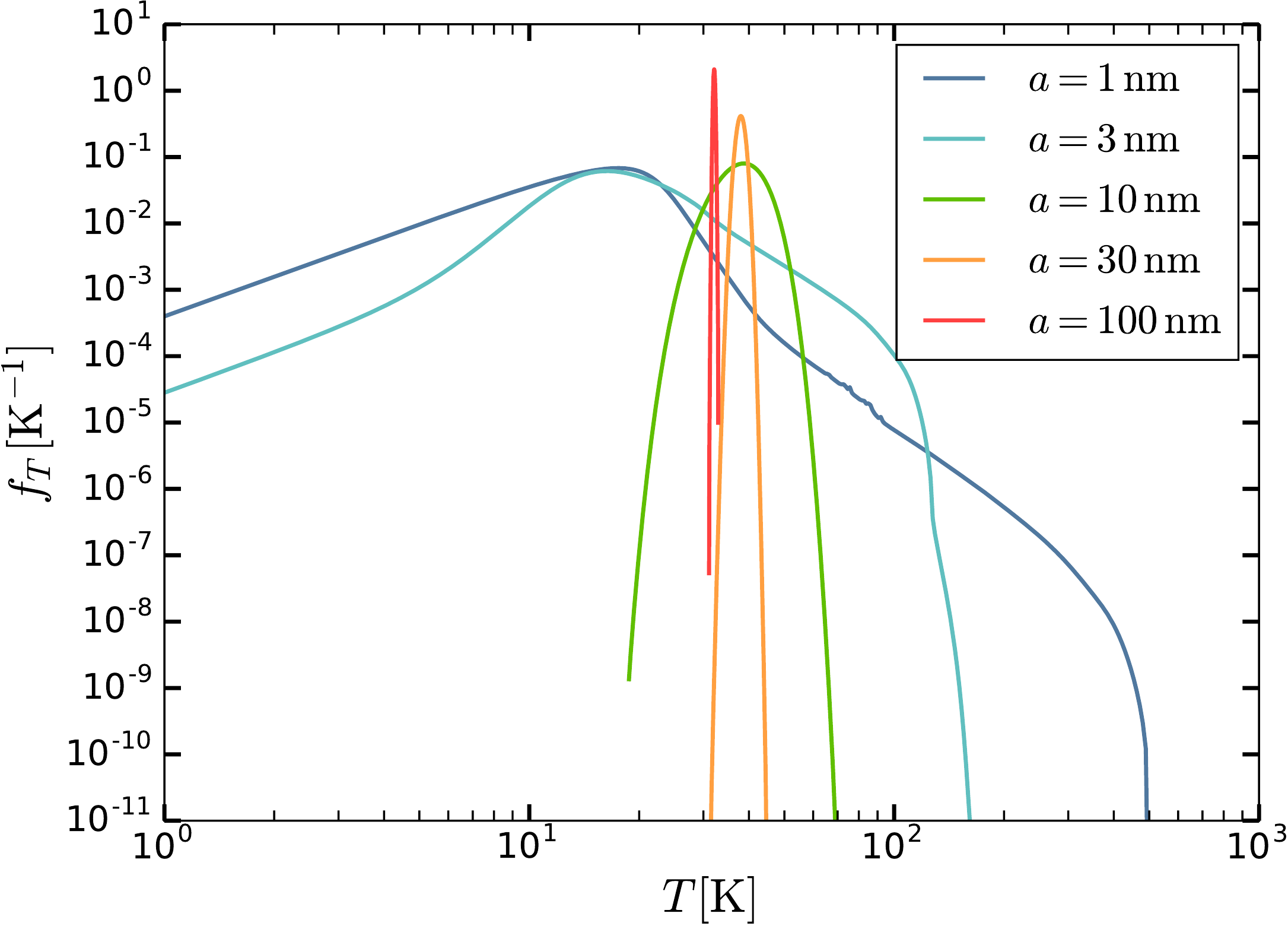}
\par\end{centering}

\protect\caption{Temperature PDF of amorphous carbon grains of various sizes under
an external UV field with $G_{0}=100$. \label{fig:Temperature-PDF}}
\end{figure}

These temperature PDFs can be seen in Fig. \ref{fig:Temperature-PDF}
in the case of an external radiation field with $G_{0}=100$ (corresponding
to the red curve on Fig. \ref{fig:eta_vs_a_khis}). Grains of size
$1\,\mathrm{nm}$ and $3\,\mathrm{nm}$ have very wide temperature
PDFs with a large probability around $10-20\,\mathrm{K}$, corresponding
to the cold state of grains between high-temperature spikes caused
by UV-photon absorption (the high-temperature spikes correspond here
to the long low-probability high-temperature tails for these sizes).
This temperature range also corresponds to the high conversion efficiency
window found in Sect. \ref{sub:Rate-equation-model}, which explains
why we find high conversion efficiency for these grain sizes in Fig.
\ref{fig:eta_vs_a_khis}. In comparison, the temperature PDFs for
sizes of $10$, $30$ and $100\,\mathrm{nm}$ are not wide enough
to cover the high efficiency window and these grain sizes have much
lower conversion efficiencies.

In order to evaluate the resulting overall efficiency, we can then
integrate the average net conversion rate 
\begin{multline}
\left\langle k_{\mathrm{O/P}}^{(\mathrm{grain})}\right\rangle =\\
\int dE\,\left(\left\langle n_{\mathrm{o}}\left|E\right.\right\rangle k_{\mathrm{o}\rightarrow\mathrm{p}}(T(E))-\left\langle n_{\mathrm{p}}\left|E\right.\right\rangle k_{\mathrm{p}\rightarrow\mathrm{o}}(T(E))\right)f_{E}(E)
\end{multline}
 over the full dust size distribution to obtain the overall net conversion
rate
\begin{equation}
k_{\mathrm{O/P}}^{(\mathrm{tot.)}}=\int da\,f_{a}(a)\,\left\langle k_{\mathrm{O/P}}^{(\mathrm{grain})}(a)\right\rangle 
\end{equation}
where $f_{a}(a)$ is the dust size distribution. We then define the
overall conversion efficiency as 
\begin{equation}
\eta_{\mathrm{O/P}}^{\mathrm{tot.}}=\frac{k_{\mathrm{O/P}}^{(\mathrm{tot.)}}}{k_{\mathrm{coll}}^{(\mathrm{o})\,\mathrm{tot.}}S(T_{\mathrm{gas}})}
\end{equation}
where 
\begin{equation}
k_{\mathrm{coll}}^{(\mathrm{o})\,\mathrm{tot.}}=\int da\,f_{a}(a)\,k_{\mathrm{coll}}^{(\mathrm{o})}(a)
\end{equation}
 is the total collision rate of ortho-H$_{2}$ on grains.

In this section, we use a simplified dust population comprising only
amorphous carbon grains from $1\,\mathrm{nm}$ to $0.3\,\mu\mathrm{m}$
with a MRN-like \citep{Mathis77} power-law size distribution with
exponent $-3.5$. As we assume that conversion is inefficient on PAHs,
not including them in our dust population does not affect the results.

\begin{figure}
\begin{centering}
\includegraphics[width=1\columnwidth]{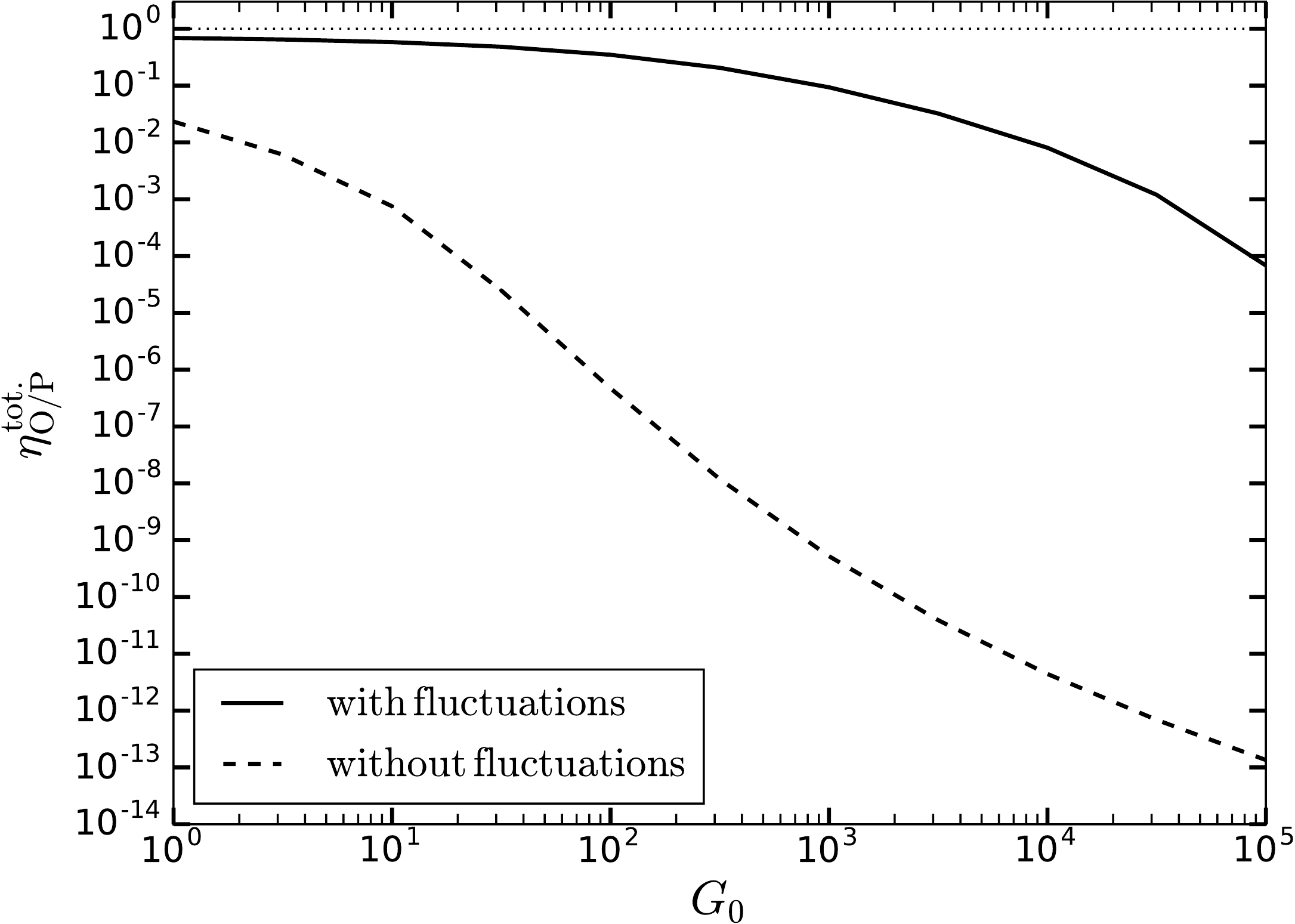}
\par\end{centering}

\protect\caption{Global conversion efficiency for a full dust size distribution as
a function of the UV intensity $G_{0}$. \label{fig:eta_full_vs_G0_plain}}
\end{figure}

Fig. \ref{fig:eta_full_vs_G0_plain} shows the resulting total conversion
efficiency for the full dust distribution as a function of UV intensity
$G_{0}$. The results of the full statistical computation (solid line)
are again compared to those of the rate equation model without fluctuations
(dashed lines). In the rate equation model, we saw on Fig. \ref{fig:eta_vs_a_khis}
that all sizes excepted the largest had low efficiencies that were
sharply decreasing when increasing the UV intensity. As small grains
represent most of the dust surface in our dust population, the overall
conversion efficiency in the rate equation model is never higher than
a few percents and decreases quickly when $G_{0}$ is increased. In
the full model, the smallest grains had their efficiencies almost
unaffected by the UV intensity. As they dominate the total dust surface
available for conversion, the overall conversion efficiency is much
less affected by $G_{0}$ and remains above $10\%$ up to $G_{0}=1000$.

Dust temperature fluctuations thus make ortho-para conversion on grains
efficient in most PDR conditions, because small grains, which dominate
the total surface and undergo large temperature fluctuations, spend
a large fraction of their time at low temperature between the temperature
spikes caused by photon absorption events. Note that this efficiency
does not include the sticking efficiency.

\subsection{Influence of the microphysical parameters}

We now investigate the impact of the uncertainties in the microphysical
parameters. The two most important parameters are the binding energy
$T_{\mathrm{phys}}$, which controls the desorption rate, and the
conversion timescale $\tau_{\mathrm{conv}}$, which controls the conversion
rate.

\begin{figure}
\begin{centering}
\includegraphics[width=1\columnwidth]{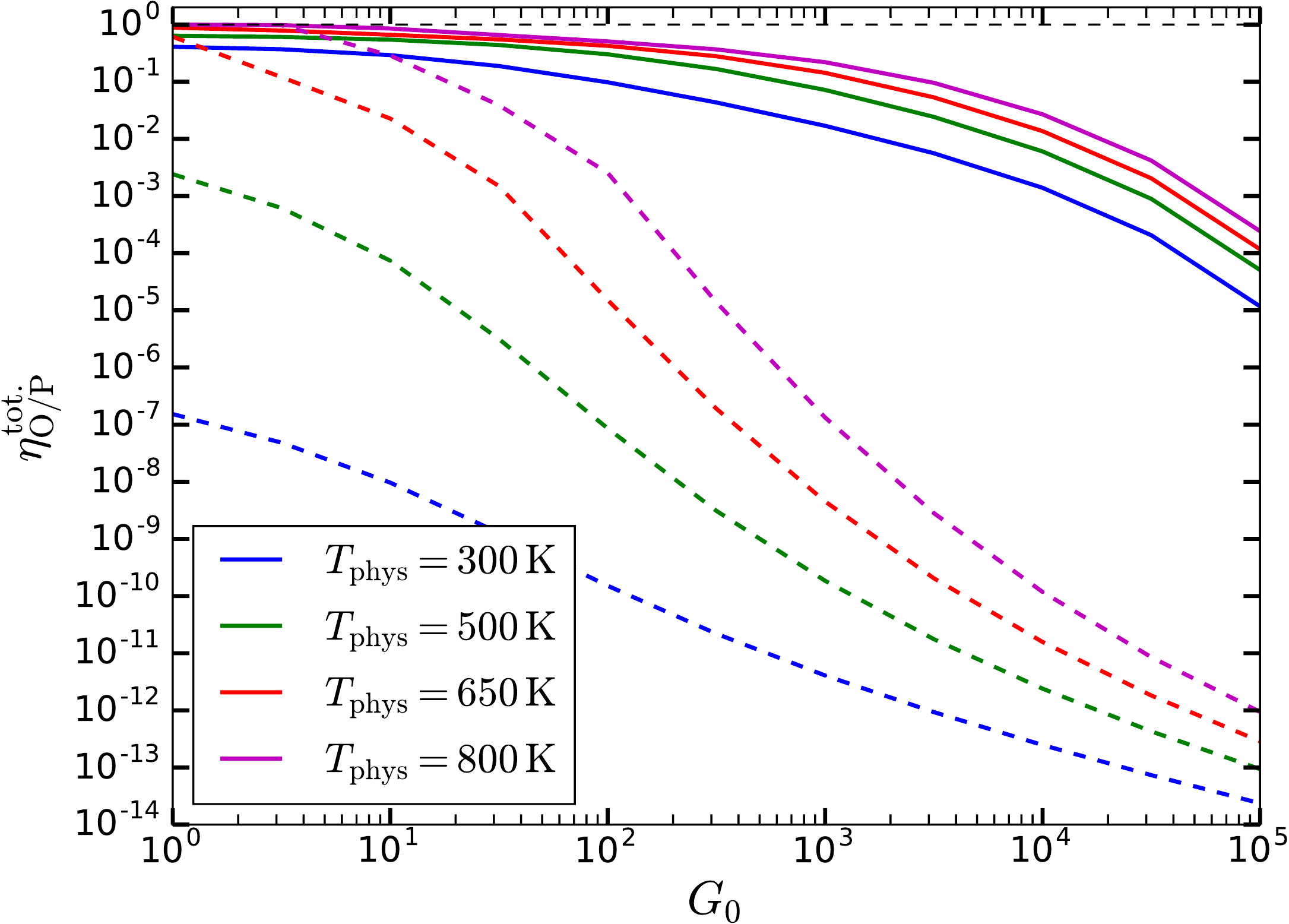}
\par\end{centering}

\protect\caption{Global conversion efficiency as a function $G_{0}$ for different
values of the binding energy of H$_{2}$, $T_{\mathrm{phys}}$.\label{fig:eta_full_vs_G0_T_phys}}

\end{figure}

Fig. \ref{fig:eta_full_vs_G0_T_phys} shows the effect of the binding
energy $T_{\mathrm{phys}}$ on both the rate equation model (dashed
lines) and the full statistical model. While the rate equation model
is strongly sensitive to the value of the binding energy (seven order
of magnitude difference at low $G_{0}$ between $T_{\mathrm{phys}}=300\,\mathrm{K}$
and $800\,\mathrm{K}$), the effect on the full statistical model
is much smaller. As $T_{\mathrm{phys}}$ is decreased, the high efficiency
window described in Sect. \ref{sub:Rate-equation-model} is shifted
to lower temperature. In the rate equation model where a grain has
a single constant temperature, this temperature will at some point
fall out of the efficiency window, resulting in a sharply decreasing
efficiency. In contrast, small grains in the full statistical model
have a wide temperature PDF. As the efficiency window is shifted,
it will still be covered by the tails of the PDF, resulting in a much
smoother decrease of the efficiency. The uncertainties on the binding
energies thus only cause uncertainties on the efficiency of at most
slightly more than one order of magnitude, instead of the seven orders
of magnitude uncertainty when neglecting the fluctuations.

\begin{figure}
\begin{centering}
\includegraphics[width=1\columnwidth]{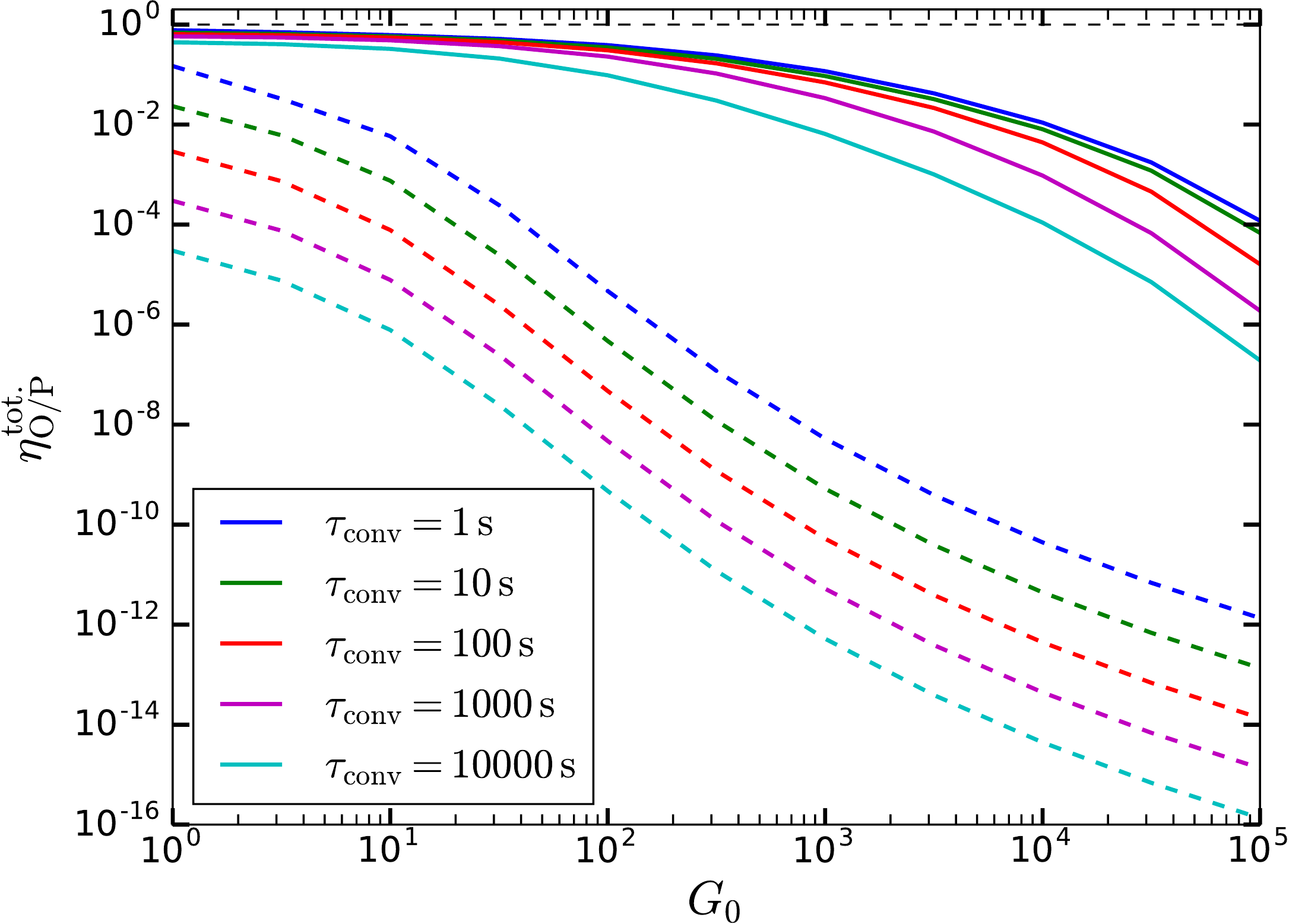}
\par\end{centering}

\protect\caption{Global conversion efficiency as a function $G_{0}$ for different
values of the conversion timescale $\tau_{\mathrm{conv}}$.\label{fig:eta_full_vs_G0_tau}}

\end{figure}

Fig. \ref{fig:eta_full_vs_G0_tau} shows similarly the impact of the
uncertainties on $\tau_{\mathrm{conv}}$. A similar effect is observed,
although less dramatic. While in the rate equation models, the uncertainties
on $\tau_{\mathrm{conv}}$ cause a four orders of magnitude difference
in the efficiency, the full model is almost unaffected up to $G_{0}\simeq100$,
and gives differences of slightly more than two orders of magnitude
at most for the strongest UV fields.

The dust temperature fluctuations thus significantly reduce the impact
of the uncertainties about the microphysical parameters. Uncertainties
of typically one or two orders of magnitude still remain on the final
conversion efficiency.

\section{PDR models and PDR observations\label{sec:PDR_models_and_obs}}

We now investigate the effects of this new computation of the ortho-para
conversion rate on grains in full PDR models. The code developed here
to perform the statistical calculation of the conversion rate has
been coupled to the Meudon PDR Code \citep{LePetit06,Goicoechea07,GonzalezGarcia08,LeBourlot12}.

The Meudon PDR Code\footnote{The Meudon PDR Code can be downloaded by following the instructions
given at http://ism.obspm.fr} solves the stationary state of a one-dimensional PDR by computing
self-consistently the chemical balance (147 species and 2835 reactions
here), the thermal balance between heating (photoelectric effect,
cosmic rays, exothermic reactions, as well as H$_{2}$ collisional
deexcitation and dust-gas collisions that can act as heating or cooling
terms) and cooling (by the lines of 28 species for which level populations
are computed from statistical balance), and the radiative transfer
in the continuum (from radio to UV wavelength, taking into account
dust absorption and scattering as well as continuum absorption by
the ionization of species such as C and S) and in the lines of the
species for which the level populations are computed.

At each position in the cloud, the local radiation field, gas density,
gas-phase H$_{2}$ density and OPR computed by the Meudon PDR Code
are used for the statistical computation of the ortho-para conversion
rate, which is sent back to the Meudon PDR Code for a new iteration.
The dust population comprises a mixture of carbonaceous and silicate
grains following a power-law size distribution with exponent -3.5
\citep{Mathis77} from $1\,\mathrm{nm}$ to $0.3\,\mu\mathrm{m}$.

In addition, the Meudon PDR code includes the other processes affecting
the ortho-para ratio : formation on grains is assumed to occur with
an OPR of 3, photodissociation and shielding are computed level by
level and thus naturally include preferential shielding of ortho-H$_{2}$,
and gas-phase ortho-para conversion includes reactive collisions with
H \citep[and references therein]{LeBourlot99}, H$^{+}$ \citep{Gerlich90}
and H$^{3+}$ (assuming identical rates as with H$^{+}$).

In this section, we first present a detailed study of an example PDR
model, discuss the local OPR in the regions emitting the rotational
lines of H$_{2}$ and show the impact of the conversion process on
grains on the local OPR and on the intensities of the rotational lines.
We then compare for a full grid of models the OPR values that can
be deduced from the predicted line intensities to actual PDR observations.
Finally, we discuss the influence of the microphysical parameters
$\tau_{\mathrm{conv}}$ and $T_{\mathrm{phys}}$ on these results.

\subsection{A typical PDR}

We consider here as a typical PDR example an isobaric model with $P=10^{7}\,\mathrm{K\cdot cm}^{-3}$
illuminated by the standard ISRF scaled by a factor $\chi=10^{3}$.
In such a PDR, the first rotational lines of H$_{2}$ are mainly emitted
in the warm molecular layer that follows the H/H$_{2}$ transitions,
and will thus provide informations about the local OPR in this region.

\begin{figure}
\begin{centering}
\includegraphics[width=1\columnwidth]{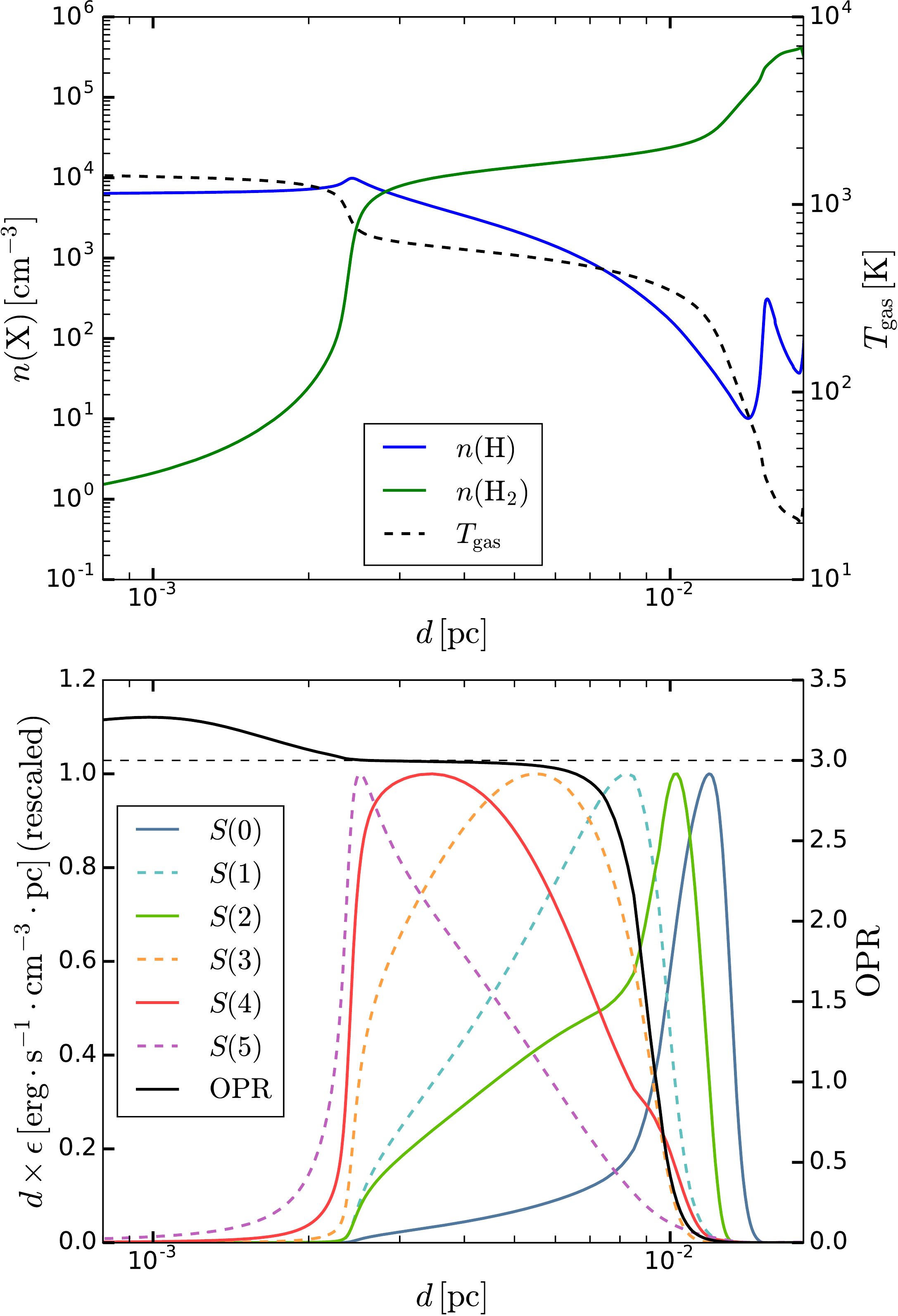}
\par\end{centering}

\protect\caption{Upper panel: density profiles of H (blue) and H$_{2}$ (green) and
gas temperature profile (red) in the PDR ($P=10^{7}\,\mathrm{K\cdot cm}^{-3}$,
$\chi=10^{3}$). Lower panel: local emissivity profiles (color lines)
of the first rotational lines of H$_{2}$ (solid lines for para transitions
and dashed lines for ortho transitions) compared to the local OPR
profile (black). Local emissivities are multiplied by $d$ so that
the contribution of a given region to the total intensity is proportional
to its area under the curve despite the logarithmic axis. Each emissivity
curve has been scaled so that its maximum is 1 for ease of comparison.\label{fig:H2_emission_regions}}
\end{figure}

Due to the large energy differences between the first rotational levels
and the temperature gradient present in this layer, the successive
lines are emitted in overlapping but relatively separated layers,
as can be seen on Fig. \ref{fig:H2_emission_regions}. This figure
shows the local emissivity profiles as a function of position, compared
to the local OPR (lower panel), for a model that includes our treatment
of surface conversion. We multiply the local emissivity $\epsilon$
by the distance from the edge so that the contribution of a given
region to the total line intensity can be evaluated visually as the
area under the curve in this region despite the logarithmic scale
used for the distance axis. The emissivity profile of each line has
also been scaled so that its maximum is 1. For reference, the density
profiles of H and H$_{2}$ and the temperature profile are shown in
the upper panel.

The local OPR shows a bump above 3 just before the H/H$_{2}$ transition,
due to preferential self-shielding of ortho-H$_{2}$ compared to para-H$_{2}$
as H$_{2}$ is assumed to form with an OPR of 3 (as discussed in \citealt{Abgrall92,Sternberg99}).
The OPR is then at 3 immediately after the H/H$_{2}$ transition before
decreasing sharply between $8\times10^{-3}$ and $10^{-2}\,\mathrm{pc}$.
The position of this sharp decrease is controlled by the efficiency
of the ortho-para conversion process on grains, as we show later.

We see that the emission regions of the H$_{2}$ lines, starting at
the H/H$_{2}$ transition for the higher lines, extend further than
the OPR drop for the lower lines. The first five rotational lines
usually observed in PDR are thus emitted in a region with a strongly
varying OPR. We can thus expect that the OPR values deduced from the
line intensities will vary from 3 for the highest lines to a significantly
lower value for the lowest lines. As described in the next section,
we will thus compute observational OPR values based on each successive
triplet of lines in our comparison between models and observations.

We can also note that the sharp decrease of the OPR causes an increase
of the para lines relative to the ortho lines, resulting in an inversion
of the emission peaks of the S(1) and S(2) lines, the S(1) emission
peak occurring before (and thus in warmer gas than) the S(2) peak
which occurs after the OPR drop. This inversion is highly dependent
on the efficiency of ortho-para conversion on grains. Resolving this
inversion at the distance of a PDR such as NGC~7023 would for instance
require a $\sim1"$ spatial resolution for the rotational lines of
H$_{2}$, and will become possible with the James Webb Space Telescope.

\begin{figure}
\begin{centering}
\includegraphics[width=1\columnwidth]{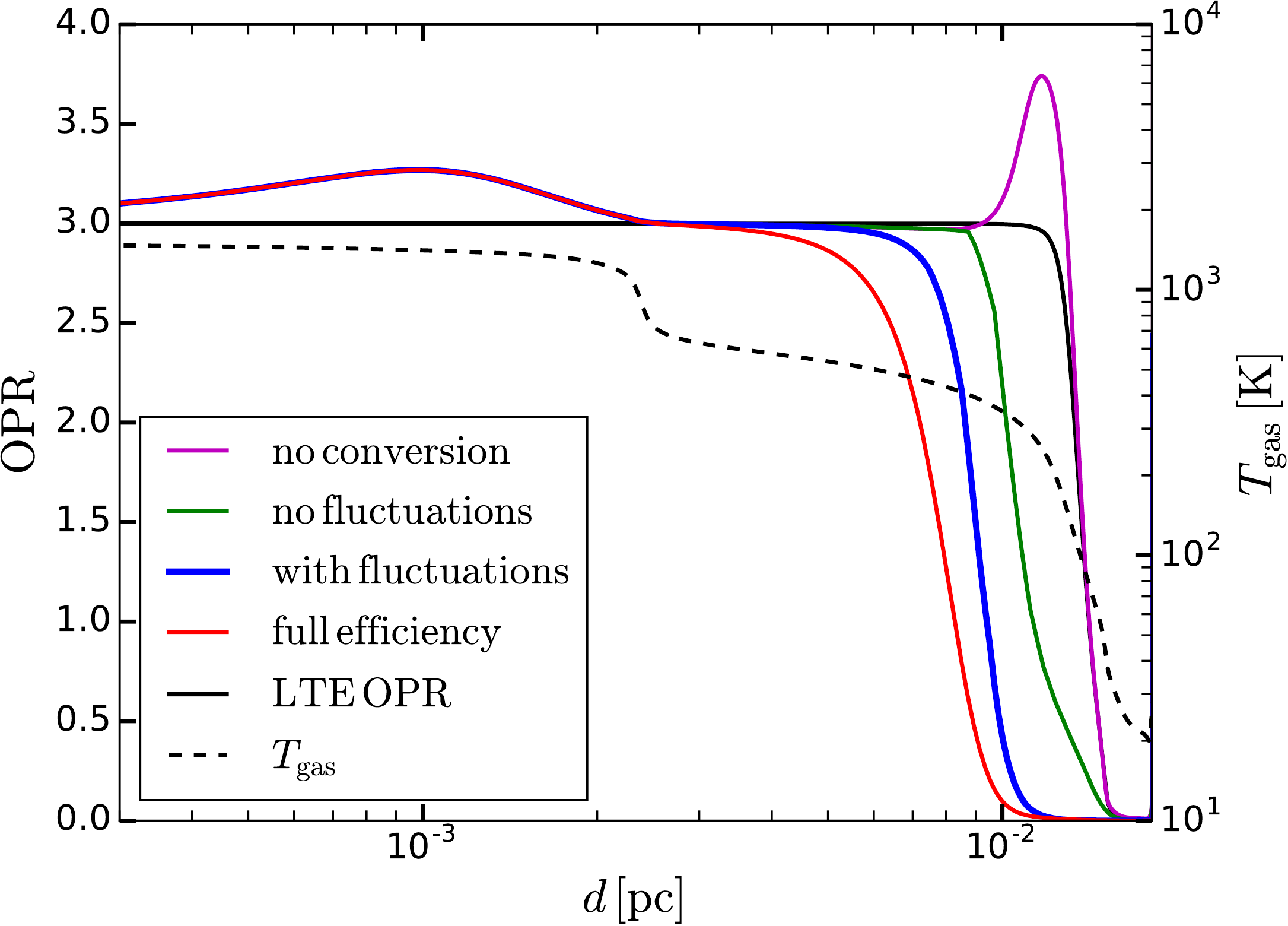}
\par\end{centering}

\protect\caption{Comparison of the local OPR profiles in the PDR in models with $P=10^{7}\,\mathrm{K\cdot cm}^{-3}$
and $\chi=10^{3}$ for the four prescriptions for ortho-para conversion
on grains : no conversion (purple), rate equation approach neglecting
the fluctuations (green), statistical approach taking the fluctuations
into account (blue), and full conversion for H$_{2}$ molecules sticking
to grains (red). The temperature profile of the PDR is also shown
for reference (dotted black line), along with the corresponding LTE
OPR profile (solid black line). \label{fig:OPR_profile}}
\end{figure}

Figure \ref{fig:OPR_profile} shows the influence of the surface conversion
efficiency on the position of the OPR drop. We compare the OPR profile
corresponding to LTE with the local temperature (black) to the OPR
profiles obtained in PDR models implementing four different prescriptions
for ortho-para conversion on grains : no surface conversion (purple),
the rate equation treatment presented in Sec. \ref{sub:Rate-equation-model}
which neglects dust temperature fluctuations (green), the full statistical
treatment described in Sec. \ref{sec:Statistical_method} which takes
fluctuations into account (blue), and a model assuming that all H$_{2}$
molecules sticking to grains are converted (red). As expected, the
more efficient surface conversion is, the closer to the H/H$_{2}$
transition the OPR drop occurs. We also see that the results of the
statistical computation are distinctly different from those of the
simpler approximations and cannot be approximated by a simpler formalism.
In the absence of surface conversion, the OPR drop starts when the
temperature falls below $200\,\mathrm{K}$ as expected for a LTE OPR.
Just before the drop in the case without surface conversion, the OPR
peaks at $\sim3.7$ due to photodissociation-formation cycling becoming
the dominant mechanism again, as reactive collisions with H collapse
proportionally to the H density after the H/H$_{2}$ transition. This
only affects the local OPR in the absence of efficient surface conversion,
and no such effect is present in our model including the statistical
treatment of surface conversion.

\begin{figure}
\begin{centering}
\includegraphics[width=1\columnwidth]{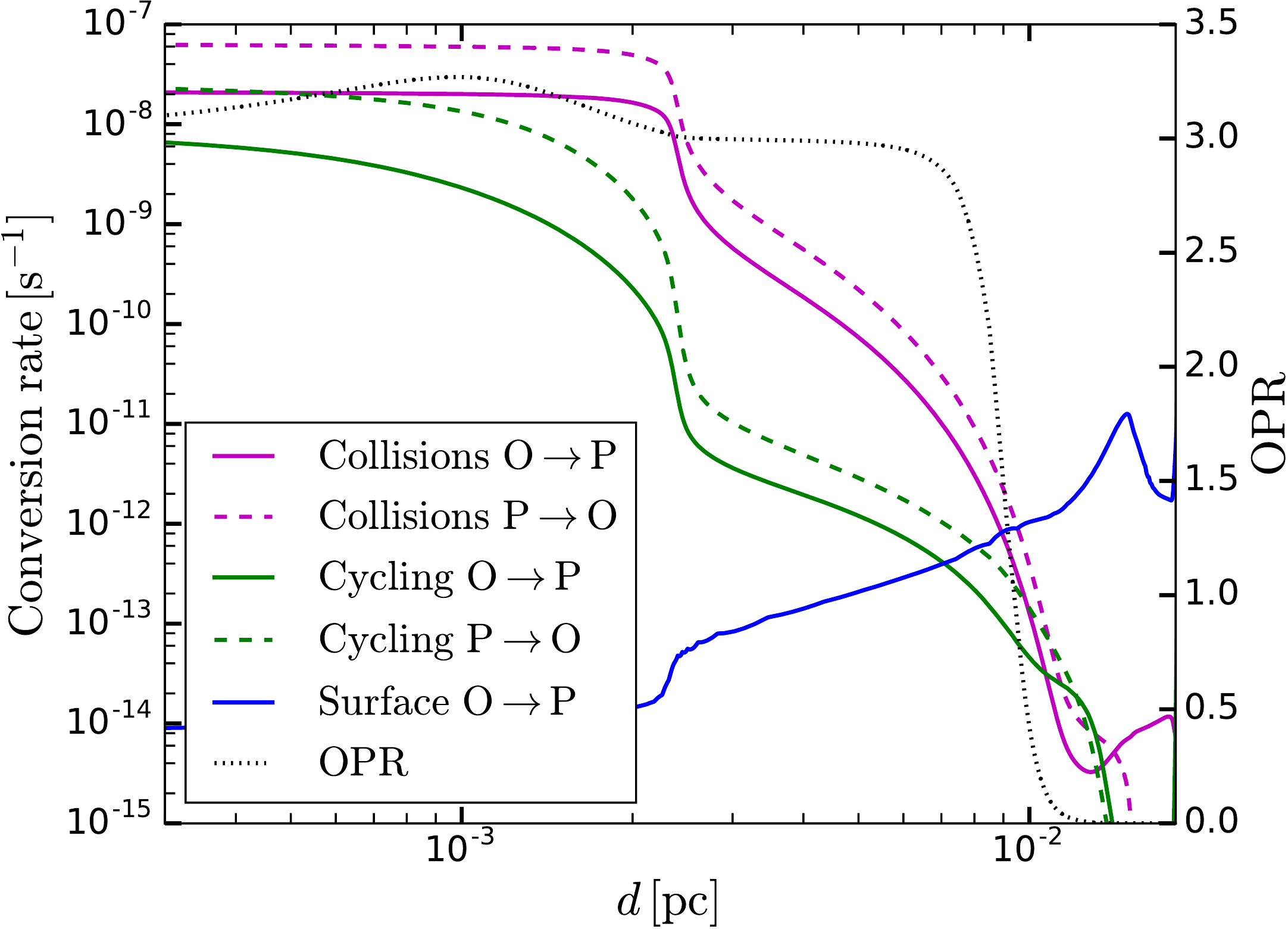}
\par\end{centering}

\protect\caption{Conversion rates by the different processes : gas-phase reactive collisions
(magenta), destruction-formation cycling (green), and dust surface
conversion (blue). \label{fig:Conversion-rates}}

\end{figure}

These OPR profiles can be better understood by considering the rates
of the different conversion processes that control the local OPR.
Figure \ref{fig:Conversion-rates} shows the conversion rates per
H$_{2}$ molecules by gas-phase reactive collisions (magenta), by
destruction-formation cycling (green) and by dust surface conversion
(blue), for a model that includes our statistical treatment of surface
conversion. Both ortho-para (solid lines) and para-ortho (dashed lines)
rates are shown. In this representation (assuming chemical balance),
the local OPR is equal to the ratio of total para-ortho conversion
rate to total ortho-para conversion. Before the H/H$_{2}$ transition,
conversion is mainly due to reactive collisions (with H), but photodissociation
is non-negligible and causes the OPR bump (OPR>3) due to preferential
shielding of ortho-H$_{2}$. After the transition, reactive collisions
dominate until conversion on grain surface becomes comparable, at
which points the local OPR decreases sharply to low values. This transition
of the local OPR from 3 to very low values is thus due to surface
conversion becoming dominant over reactive collisions. This explains
why the position of this transition appeared to be controlled by the
efficiency of surface conversion on Fig. \ref{fig:OPR_profile}.

Previous studies of the OPR in PDR observations (e.g. \citealp{Fleming10})
have advocated advection flows bringing cold (low-OPR) gas from the
molecular cloud through the PDR front to explain the low OPR values
that are observed. We will see in the next section that the observed
values can be explained by our model without advection flows. We can
however estimate here the minimum velocity required for an advection
flow to affect H$_{2}$ rotational lines intensities. Such a flow
would need to bring cold (low-OPR) gas into the region where we predict
an OPR of three. The OPR transition occurs on a width of $\sim3\times10^{-3}\,\mathrm{pc}$,
and the para-ortho conversion rate at the start of this transition
is of the order of $10^{-11}\,\mathrm{s}^{-1}$. A minimum advection
velocity of $\sim1\,\mathrm{km}\cdot\mathrm{s}^{-1}$ would thus be
required to affect the observable OPR in rotational intensities. This
value corresponds to the maximum advection velocity through the dissociation
front computed by \citet{Storzer98} for PDRs with an advancing photo-ionization
front. We thus estimate that advection could only affect the observable
OPR of the rotational lines in the most dynamical PDRs with fast photo-ionization
fronts.

\begin{figure}
\begin{centering}
\includegraphics[width=1\columnwidth]{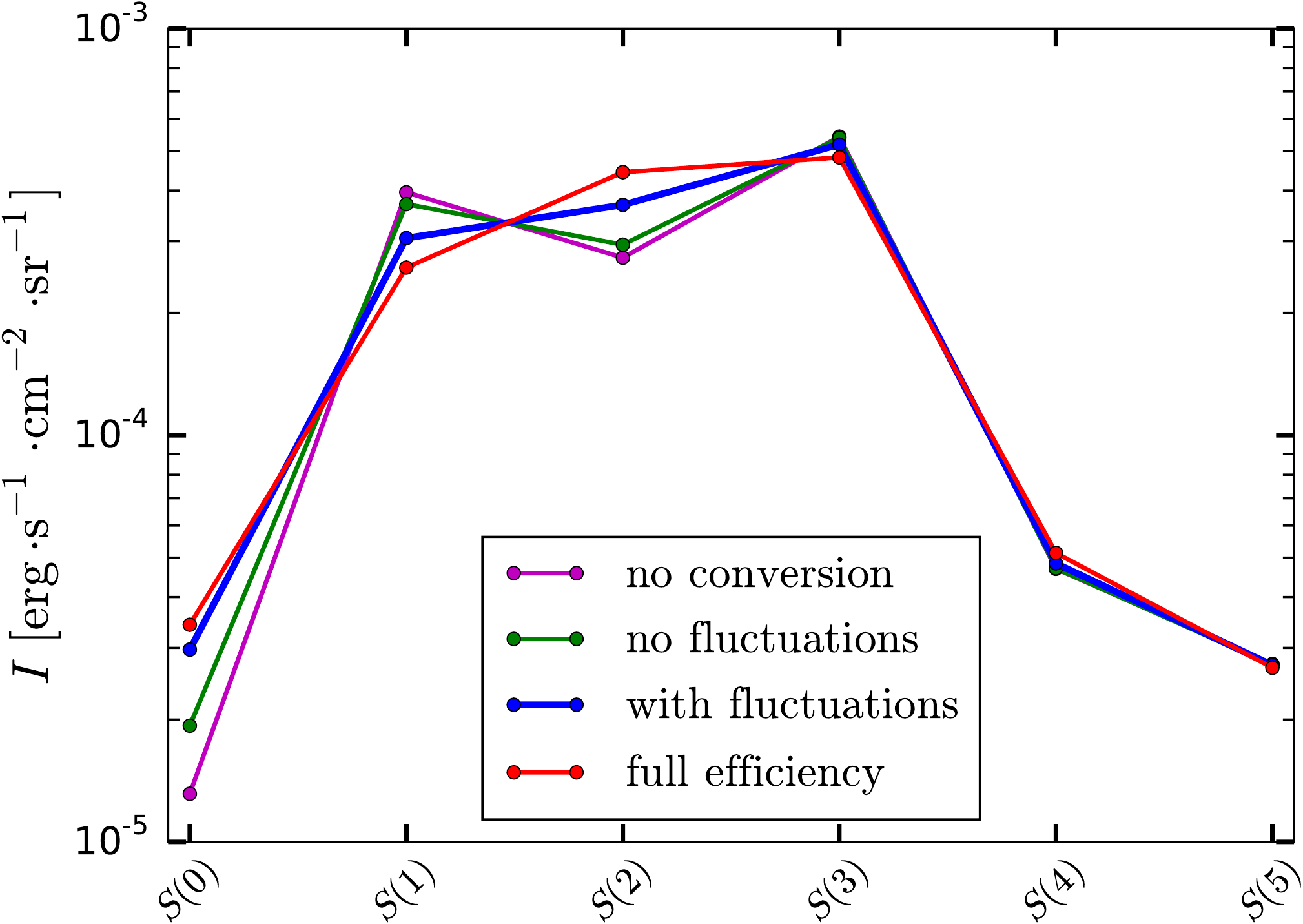}
\par\end{centering}

\protect\caption{Intensities of the first rotational lines of H$_{2}$ predicted by
the same PDR model ($P=10^{7}\,\mathrm{K\cdot cm}^{-3}$ and $\chi=10^{3}$)
for the four prescriptions for ortho-para conversion on grains: no
conversion (purple), rate equation approach neglecting the fluctuations
(green), statistical approach taking the fluctuations into account
(blue), and full conversion for H$_{2}$ molecules sticking to grains
(red).\label{fig:H2_intensities}}
\end{figure}

Finally, we show the impact of these differences on the line intensities
of the rotational lines of H$_{2}$ on Fig. \ref{fig:H2_intensities}.
The lines S(3), S(4) and S(5), which are mainly emitted before the
OPR drop in all cases are almost unaffected. The lines S(0), S(1)
and S(2) are strongly affected with a S(1)/S(2) ratio varying by a
factor of 4, and a S(3)/S(2) ratio varying by a factor of almost 3.
The impact tends to be stronger for lower pressure models, as will
be seen in the next section.

\subsection{Comparison of a grid of models to PDR observations}

\begin{table*}[t]
\begin{centering}
\protect\caption{Sample of PDR observations of lines $S(0)$ to $S(3)$ with the derived
values of $T_{24}$, OPR$_{234}$, $T_{35}$ and OPR$_{345}$. \label{tab:OPR_observations}}
\smallskip{}
\begin{tabular}{lllllll}
\hline 
\hline Object & Ref. & Note & $T_{24}$ & OPR$_{234}$ & $T_{35}$ & OPR$_{345}$\tabularnewline
 &  &  & (K) &  & (K) & \tabularnewline
\hline 
L1721 & (1) & upper limit for $S(3)$ & $206\pm13$ & $0.99\pm0.22$ & $<308$ & $<3.00$\tabularnewline
California & (1) & - & $208\pm19$ & $0.82\pm0.15$ & $303\pm31$ & $2.32\pm1.18$\tabularnewline
NGC~7023 East & (1) & - & $269\pm9$ & $0.86\pm0.07$ & $328\pm37$ & $1.35\pm0.21$\tabularnewline
Horsehead & (1) & - & $263\pm20$ & $0.65\pm0.21$ & $399\pm33$ & $1.54\pm0.50$\tabularnewline
$\rho$ Oph. & (1) & - & $251\pm16$ & $0.75\pm0.23$ & $291\pm14$ & $1.08\pm0.32$\tabularnewline
NGC~2023 North & (1) & - & $222\pm18$ & $0.68\pm0.21$ & $302\pm33$ & $1.54\pm0.65$\tabularnewline
NGC~7023 South-West & (2) & upper limit for $S(0)$ & $>242$ & - & $452\pm14$ & $1.43\pm0.25$\tabularnewline
NGC~7023 North-West & (3) & - & $293\pm24$ & $0.94\pm0.28$ & $438\pm6$ & $2.01\pm0.28$\tabularnewline
$\rho$ Oph. ``pos. 1'' & (4) & - & $282\pm23$ & $1.05\pm0.11$ & $322\pm8$ & $1.41\pm0.47$\tabularnewline
$\rho$ Oph. ``pos. 2'' & (4) & - & $300\pm21$ & $0.74\pm0.22$ & $313\pm14$ & $0.82\pm0.20$\tabularnewline
$\rho$ Oph. ``pos. 3'' & (4) & - & $255\pm7$ & $0.58\pm0.06$ & $285\pm7$ & $0.77\pm0.08$\tabularnewline
S140 & (5) & missing $S(0)$ & - & - & $354\pm59$ & $1.57\pm1.11$\tabularnewline
Orion Bar & (6) & - & $258\pm19$ & $1.35\pm0.53$ & $361\pm17$ & $2.85\pm0.57$\tabularnewline
\hline 
\end{tabular}\smallskip{}

\par\end{centering}

\centering{}(1) : \citet{Habart11}, (2) : \citet{Fuente00}, (3)
: \citet{Fuente99}, (4) : \citet{Habart03}, (5) : \citet{Timmermann96},
(6) : \citet{Joblin_in_prep.}
\end{table*}

As discussed in the previous section, the different rotational lines
are emitted in regions with large differences in local OPR values.
Interpreting observed rotational diagrams with a single OPR value
is thus insufficient. We try here to measure the OPR as locally as
possible by computing separate OPR values for each triplet of successive
lines. We note OPR$_{j-1,j,j+1}$ the OPR computed from the column
densities of levels $J=j-1$, $j$ and $j+1$ (derived from the line
intensities of lines $S(j-3)$, $S(j-2)$ and $S(j-1)$). This OPR
value is computed from the misalignment of level $j$ with respect
to the line defined by levels $j-1$ and $j+1$ in the rotational
diagram as{\footnotesize{}
\begin{equation}
\mathrm{OPR}_{j-1,j,j+1}=\begin{cases}
\mathrm{OPR_{LTE}}(T_{j-1,j+1})\frac{N_{j}/g_{j}}{N_{j-1}/g_{j-1}}\exp\left(\frac{E_{j}-E_{j-1}}{k_{B}T_{j-1,j+1}}\right) & \mathrm{if}\;\mathrm{j}\:\mathrm{odd}\\
\\
\mathrm{OPR_{LTE}}(T_{j-1,j+1})\frac{N_{j-1}/g_{j-1}}{N_{j}/g_{j}}\exp\left(-\frac{E_{j}-E_{j-1}}{k_{B}T_{j-1,j+1}}\right) & \mathrm{if}\;\mathrm{j}\:\mathrm{even}
\end{cases}
\end{equation}
}where $N_{j}$ is the column density of level $j$ (deduced from
the intensity of line $S(j-2)$), $g_{j}$ and $E_{j}$ the degeneracies
and energies of level $j$ and $T_{j-1,j+1}$ is the excitation temperature
computed from levels $j-1$ and $j+1$ (thus unaffected by OPR effects):
\begin{equation}
T_{j-1,j+1}=-\frac{1}{k_{B}}\frac{E_{j+1}-E_{j-1}}{\log\left(\frac{N_{j+1}/g_{j+1}}{N_{j-1}/g_{j-1}}\right)}
\end{equation}
and OPR$_{\mathrm{LTE}}(T)$ is the thermal equilibrium value of the
OPR at temperature $T$.

Note that this measure of the OPR can be affected by the presence
of a curvature in the excitation diagram, corresponding for instance
to a strong temperature gradient in the region of emission. A positive
curvature will lead to an overestimation of OPR$_{j-1,j,j+1}$ for
even values of $j$ and an underestimation for odd values of $j$,
as it will create a misalignment caused not by an actual out-of-equilibrium
OPR value but by a varying excitation temperature for increasing values
of $J$. We will thus compare the OPR values derived from the observations
to OPR values similarly derived from the line intensities predicted
by the models rather than to the actual local OPR value. We will see
that lower-than-three OPR values are found for both odd and even values
of $j$ in the observations, indicating a physical OPR lower than
3.

We will focus on the lines that are the most affected by surface conversion
on grains, the lines $S(0)$ to $S(3)$, which are also observed in
a larger sample of PDR observations. The observation sample gathered
from the literature is presented in Table \ref{tab:OPR_observations},
with the values of $T_{24}$, OPR$_{234}$, $T_{35}$ and OPR$_{345}$
derived from the observed line intensities. 

As shown by \citet{Joblin_in_prep.}, the pressure in the dense structures
of PDRs seems to be related to the intensity of the UV radiation field,
with a roughly constant $P/G_{0}$ ratio. The typical value for the
ratio $P/G_{0}$ in PDRs appears to be $\sim2\times10^{4}$ with a
scatter of a factor of 2-3 above and below. In our PDR models, we
recall that $G_{0}$ is related to the scaling factor $\chi$ by the
relation $G_{0}\sim0.65\chi$. The observed $P-G_{0}$ relation thus
corresponds to $P/\chi\sim10^{4}$. We thus use a grid of models covering
a range of $P/\chi$ from $2\times10^{3}$ to $5\times10^{4}$ (a
factor 5 above and below the observed value), and a range of pressures
from $P=10^{5}$ to $10^{9}\,\mathrm{K\cdot cm}^{-3}$.

In the region were lines $S(0)$ to $S(3)$ are emitted, the local
OPR is controlled by the balance between gas phase reactive collisions,
which tends to thermalize the OPR to 3 and whose efficiency is dependent
on the gas temperature, and surface conversion, which tends to thermalize
the OPR to a lower value corresponding to dust temperature. The resulting
observable OPR is thus dependent both on the surface conversion efficiency
(which we wish to constrain) and on the gas temperature (which is
controlled by the photoelectric efficiency in this region). Rather
than simply comparing the measured OPR values, we thus investigate
the OPR-$T_{\mathrm{gas}}$ relation in both observations and model.
When considering OPR$_{j-1,j,j+1}$, we thus estimate the gas temperature
of the corresponding layer of the PDR through the excitation temperature
$T_{j-1,j+1}$. 

\begin{figure}
\begin{centering}
\includegraphics[width=1\columnwidth]{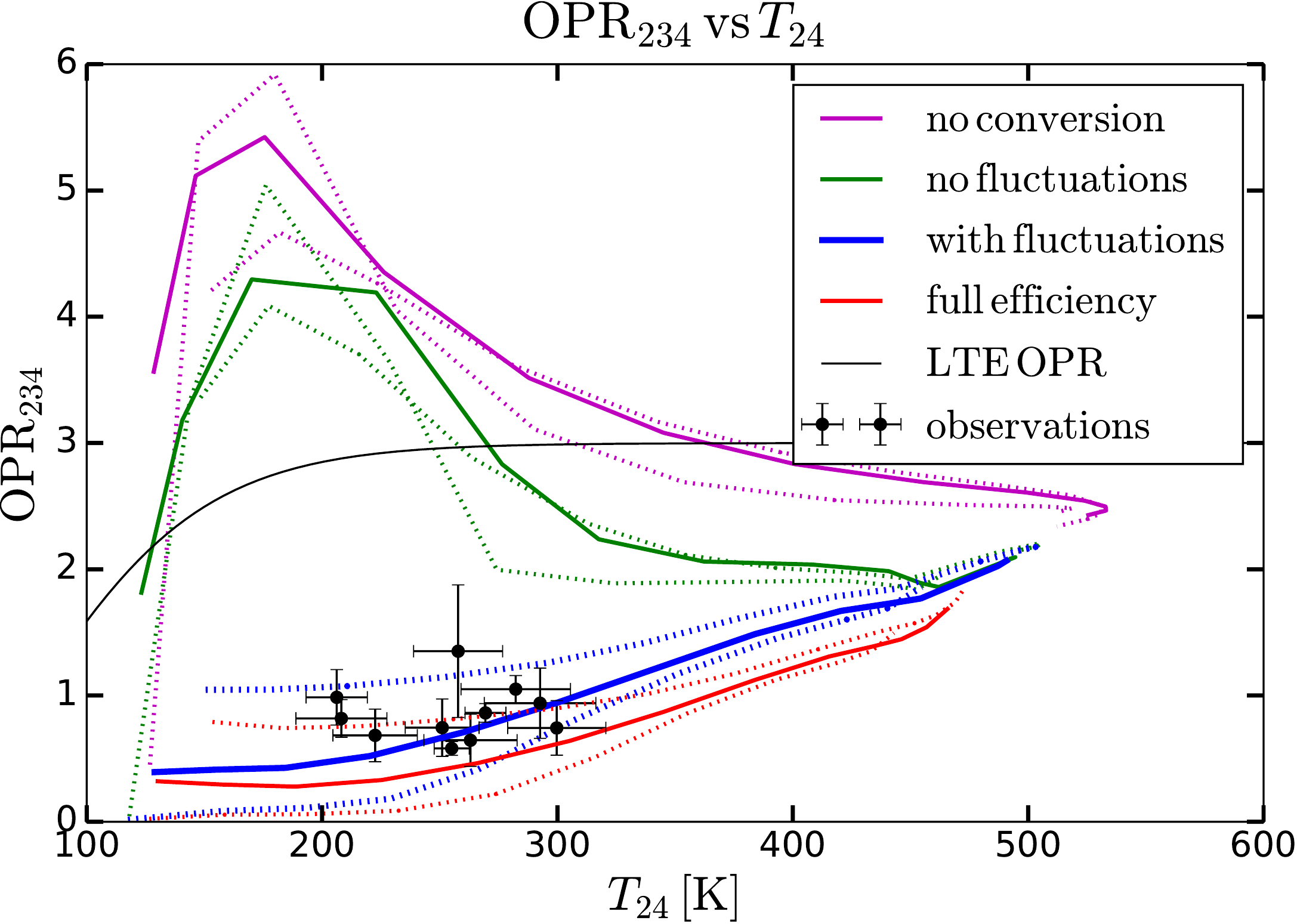}
\par\end{centering}

\protect\caption{OPR computed from levels $J=2,\,3\:\mathrm{and}\:4$ as a function
of the excitation temperature $T_{24}$, comparing PDR observations
(symbols) to for PDR models (lines). PDR models implement four prescriptions
for ortho-para conversion on grains: no conversion (purple), rate
equation approach without fluctuations (green), statistical approach
with fluctuations (blue), and full conversion (red). Solid lines correspond
to models with $P/\chi=10^{4}$, while models with $P/\chi$ a factor
of 5 above and below this value are shown as dotted lines.\label{fig:Compa_obs_OPR_234}}
\end{figure}

\begin{figure}
\begin{centering}
\includegraphics[width=1\columnwidth]{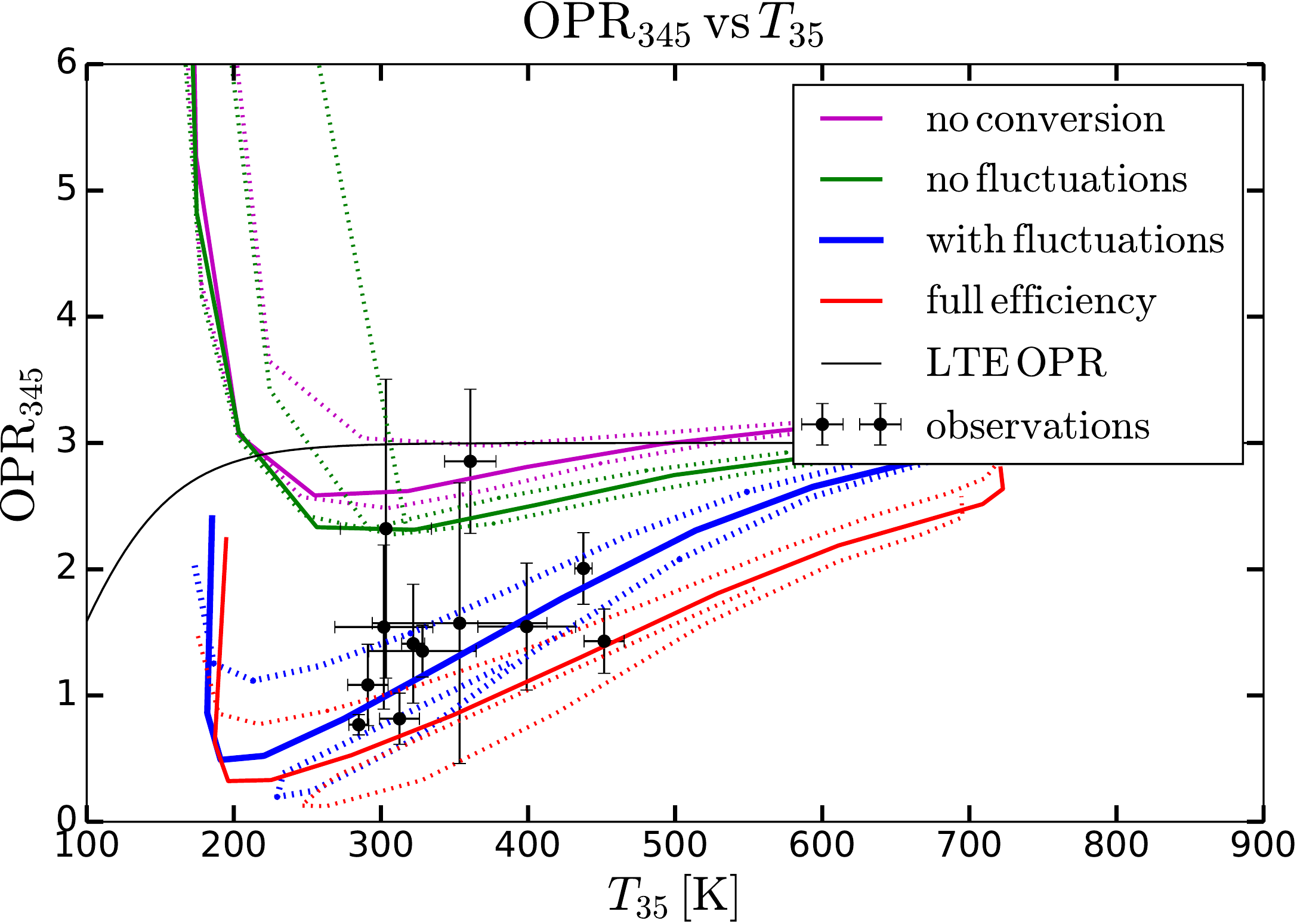}
\par\end{centering}

\protect\caption{Same as Fig. \ref{fig:Compa_obs_OPR_234} for the OPR computed from
levels $J=3$, $4$ and $5$, shown as a function of the excitation
temperature $T_{35}$.\label{fig:Compa_obs_OPR_345}}
\end{figure}

We thus present our comparison between models and observations on
two figures presenting the OPR$_{234}$-$T_{24}$ relationship (Fig.
\ref{fig:Compa_obs_OPR_234}) and the OPR$_{345}$-$T_{35}$ relationship
(Fig. \ref{fig:Compa_obs_OPR_345}). On both figures, we compare the
observed values (symbols with error-bars) to model results. As previously,
we compare models with four different prescriptions for ortho-para
conversion on grains : no surface conversion (purple), the rate equation
treatment presented in Sec. \ref{sub:Rate-equation-model} which neglects
dust temperature fluctuations (green), the full statistical treatment
described in Sec. \ref{sec:Statistical_method} which takes fluctuations
into account (blue), and a model assuming that all H$_{2}$ molecules
sticking to grains are converted (red). The solid lines show results
of models with $P/\chi=10^{4}$, while the dotted lines corresponds
to model with $P/\chi$ ratios a factor of 5 above and below this
value. Finally, the LTE value of the OPR as a function of the local
temperature is shown as a black line.

On Fig. \ref{fig:Compa_obs_OPR_234}, corresponding to lines $S(0)$,
$S(1)$ and $S(2)$, we see a clear separation between high efficiency
models (statistical approach and full efficiency hypothesis) and the
models without conversion on dust. The observational data points all
fall close to the curve of these high efficiency models and are clearly
incompatible with the no-conversion models. The observations thus
indicate that high-efficiency conversion on grains is indeed occurring
in PDRs. We also see that the models with the rate equation treatment
of surface conversion, which neglects dust temperature fluctuations,
give results that are close to models with no conversion. The statistical
effect of fluctuations is thus necessary to explain the high conversion
efficiency that the observations seem to indicate. Finally, the observations
seem to favor the full statistical treatment over a simpler full efficiency
hypothesis, but the difference is smaller and other uncertainties
(for instance on the total dust surface available) could induce similar
differences. However, the statistical treatment of fluctuations is
again the only way to physically explain a high conversion efficiency.

On Fig. \ref{fig:Compa_obs_OPR_345}, corresponding to lines $S(1)$,
$S(2)$ and $S(3)$, the observations are more scattered, but most
datapoint again fall close to the results of models with the statistical
treatment of fluctuations. The two outliers are the California PDR,
whose large error-bars still make it compatible with the models, and
the Orion Bar PDR which seems to have a clearly different behavior
from all other PDRs. Again, the observations indicate a high conversion
efficiency on grains which cannot be explained when neglecting the
fluctuations, and slightly favor our statistical model over a full
efficiency hypothesis.

On both figures, higher excitation temperatures correspond to higher
pressure models. We see that the low-conversion-efficiency models
(no conversion and rate equations neglecting the fluctuations) sometimes
exhibit OPR values higher than 3. For OPR$_{345}$, this is due to
the strong curvature of the rotational diagram for low pressure models.
For OPR$_{234}$, curvature effects play in the opposite direction,
and only explain the drop of the OPR at the lowest temperature. The
values above 3 are actually caused by the local peak of the OPR, seen
in Fig. \ref{fig:OPR_profile} after the H/H$_{2}$ transition for
models without efficient surface conversion, and which is due to preferential
photodissociation of para-H$_{2}$ in a fully molecular region where
reactive collisions with H are very rare. Moderate curvature effects
are also seen in the fact that the OPR values of models at high temperatures
seem to converge towards a value lower than 3 for OPR$_{234}$ and
higher than 3 for OPR$_{345}$. These measures of the OPR are thus
slightly biased due to the curvature of the rotational diagrams. Observations
and models are however similarly biased, and these figures present
equivalent information as $S(1)/S(0)$ vs $S(2)/S(0)$ and $S(2)/S(1)$
vs $S(3)/S(1)$ graphs, in a more physically meaningful form.

Finally, we can note that the range of excitation temperatures found
by the models is significantly more extended than the range of observed
excitation temperatures, despite the fact that the observed objects
cover a wide range of conditions. This probably indicates that the
temperature profile (and possibly the density profile) in the region
that emits H$_{2}$ rotational lines is not adequate in the models.
It could come from an incorrect estimation of photoelectric heating
or from the dynamics of the photodissociation front affecting the
density profile of the PDR. The observations seem to indicate that
PDRs with very different excitation conditions still have relatively
similar temperatures ($200-300\,\mathrm{K}$) in the region where
the first rotational lines of H$_{2}$ ($S(0)$ to $S(2)$) are emitted.

\subsection{Influence of the microphysical parameters}

Our microphysical model of ortho-para conversion on dust grains depends
on two poorly constrained parameters : the surface conversion timescale
$\tau_{\mathrm{conv}}$ and the physisorption binding energy $T_{\mathrm{phys}}$
(cf. Sec. \ref{sub:conversion} and \ref{sub:desorption}). We now
investigate the impact of these uncertainties on the previously presented
results.

\begin{figure}
\begin{centering}
\includegraphics[width=1\columnwidth]{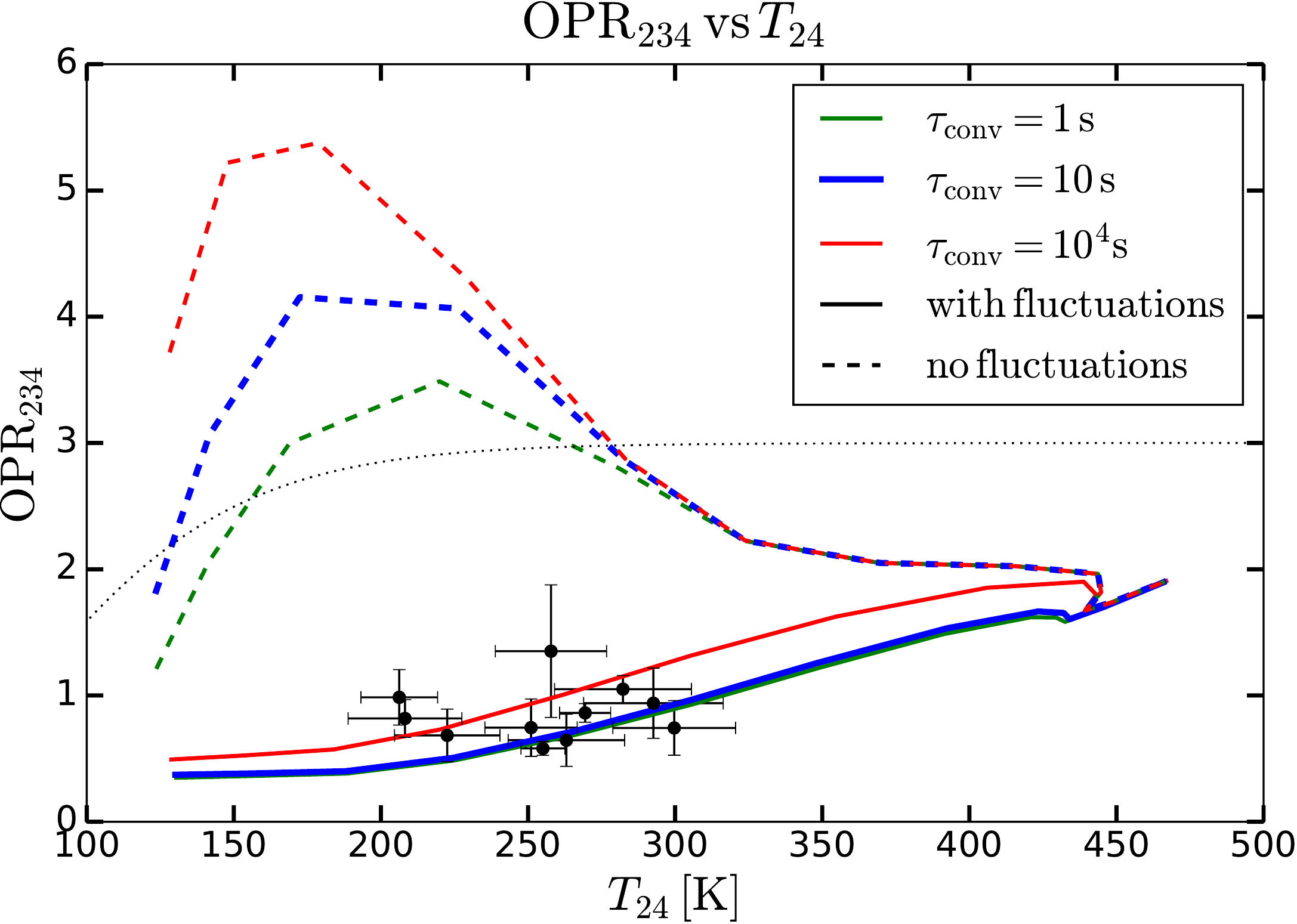}
\par\end{centering}

\protect\caption{Influence of the surface conversion timescale $\tau_{\mathrm{conv}}$
on the observable OPR$_{234}$ as a function of $T_{24}$. We compare
models results for three values of $\tau_{\mathrm{conv}}$ ($1\,\mathrm{s}$
in red, $10\,\mathrm{s}$ in blue and $10^{4}\mathrm{s}$ in green)
and for two different prescriptions of the ortho-para conversion on
grains (rate equation neglecting fluctuations in dashed lines, full
statistical treatment of fluctuations in solid lines) to the PDR observations.\label{fig:tau_effect_OPR234}}
\end{figure}

\begin{figure}
\begin{centering}
\includegraphics[width=1\columnwidth]{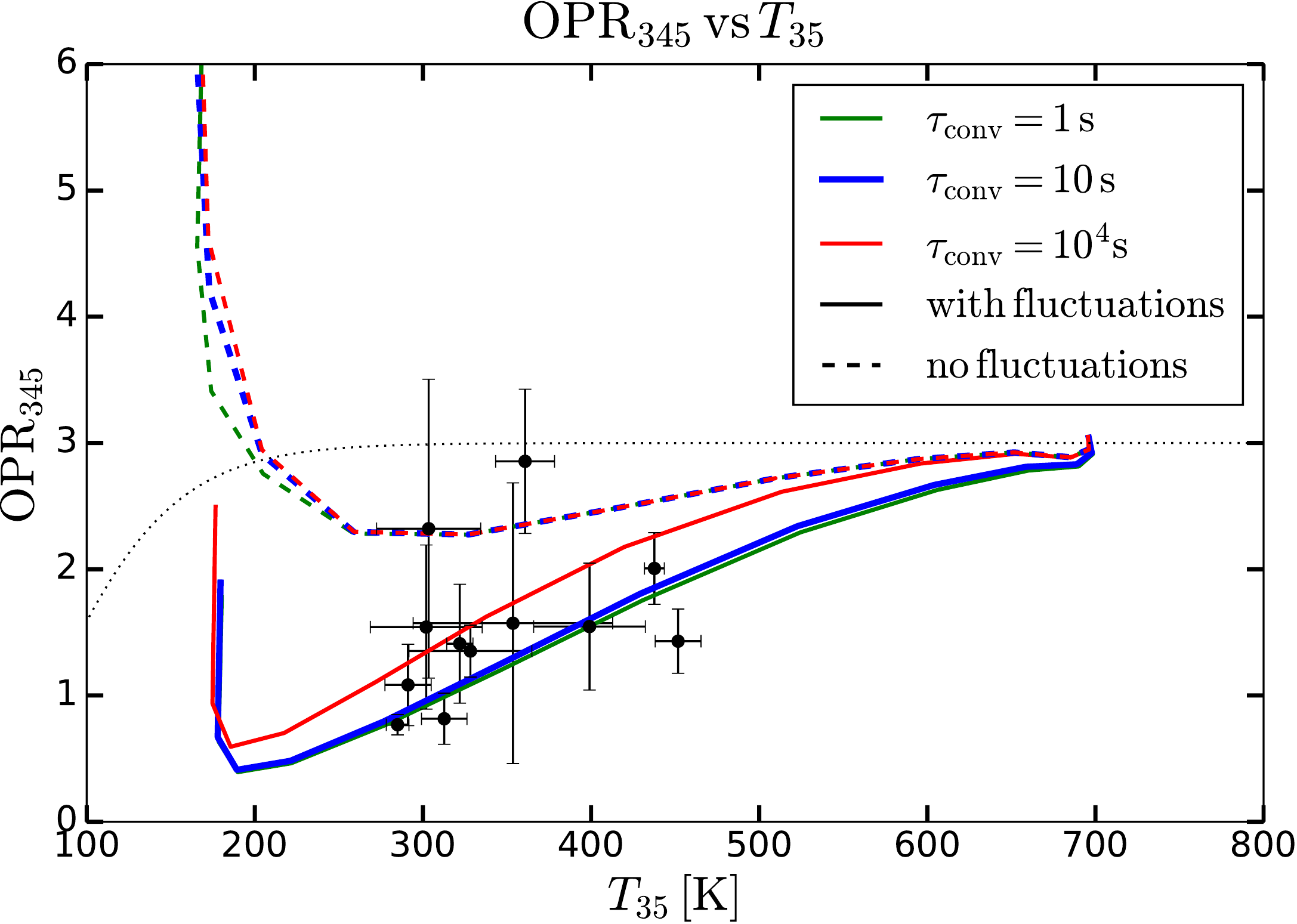}
\par\end{centering}

\protect\caption{Same as Fig. \ref{fig:tau_effect_OPR234} for OPR$_{345}$ as a function
of $T_{35}$. \label{fig:tau_effect_OPR345}}
\end{figure}

We first study the impact of the conversion timescale $\tau_{\mathrm{conv}}$.
We took $\tau_{\mathrm{conv}}=10\,\mathrm{s}$ as our standard value,
while the possible values found in the literature range from $1\,\mathrm{s}$
to $10^{4}\,\mathrm{s}$. Fig. \ref{fig:tau_effect_OPR234} and \ref{fig:tau_effect_OPR345}
thus compare the results of PDR models using these three values of
$\tau_{\mathrm{conv}}$. Only the models with the statistical treatment
of fluctuations (solid lines) and with the rate equation treatment
without fluctuations (dashed lines) are shown, in comparison to the
observations. On both figures, the impact of the variations of $\tau_{\mathrm{conv}}$
on the models with the statistical treatment is limited and the model
results remain compatible with the observations. Models implementing
the rate equation treatment are more strongly affected (for OPR$_{234}$)
but remain incompatible with the observations.

\begin{figure}
\begin{centering}
\includegraphics[width=1\columnwidth]{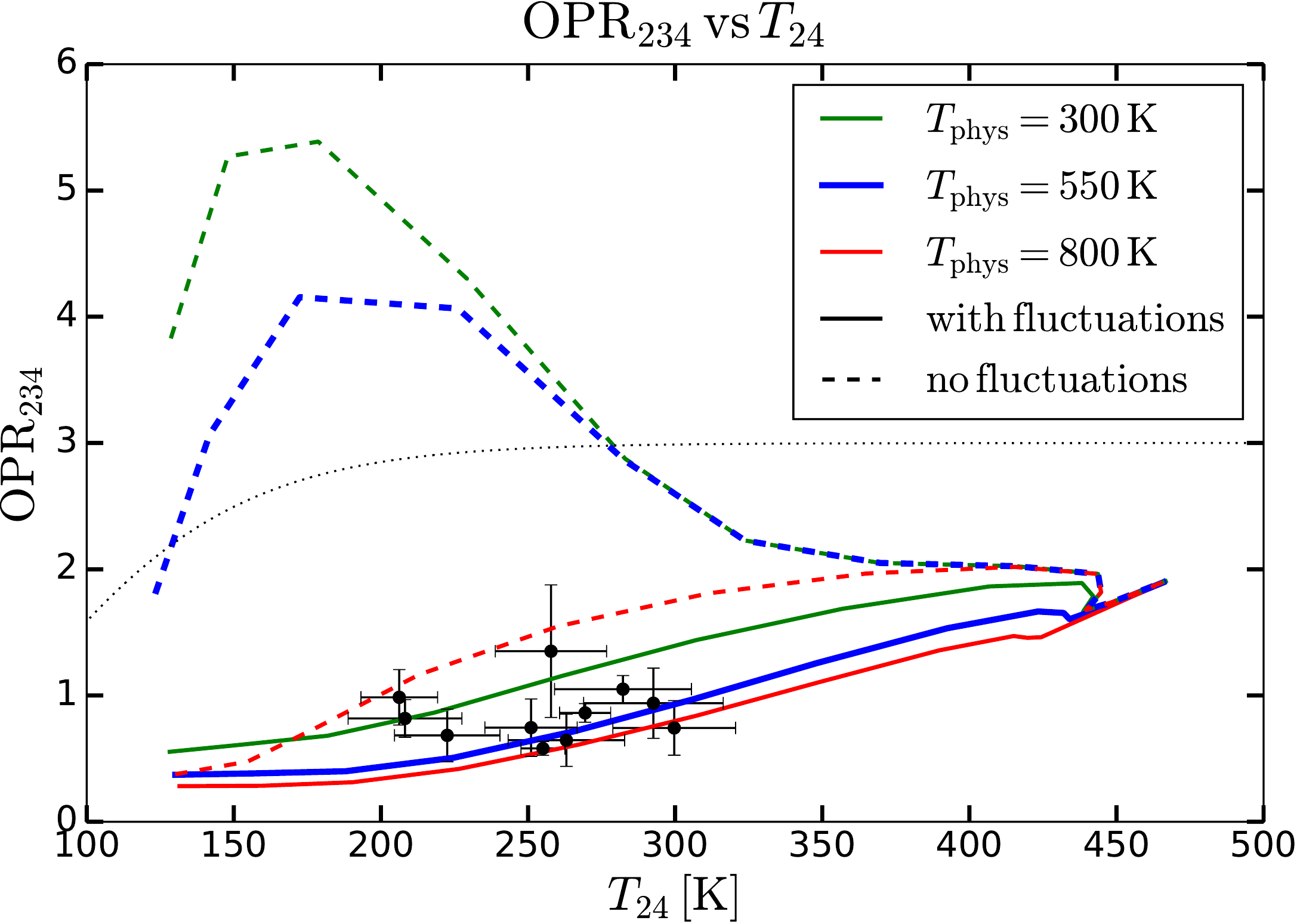}
\par\end{centering}

\protect\caption{Influence of the physisorption binding energy $T_{\mathrm{phys}}$
on the observable OPR$_{234}$ as a function of $T_{24}$. We compare
models results for three values of $T_{\mathrm{phys}}$ ($300\,\mathrm{K}$
in red, $550\,\mathrm{K}$ in blue and $800\,\mathrm{K}$ in green)
and for two different prescriptions of the ortho-para conversion on
grains (rate equation neglecting fluctuations in dashed lines, full
statistical treatment of fluctuations in solid lines) to the PDR observations.\label{fig:Tphys_effect_OPR234}}
\end{figure}

\begin{figure}
\begin{centering}
\includegraphics[width=1\columnwidth]{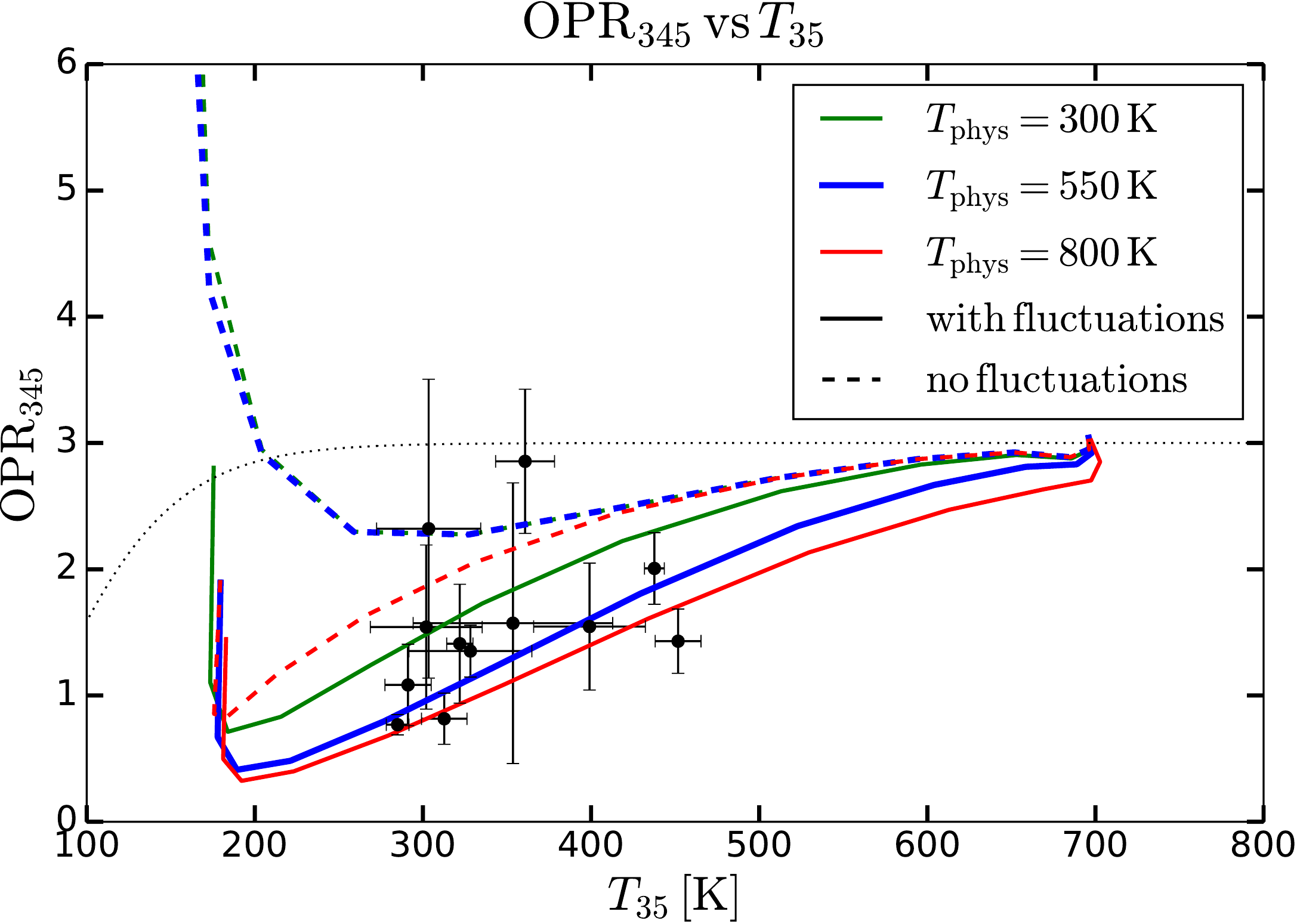}
\par\end{centering}

\protect\caption{Same as Fig. \ref{fig:Tphys_effect_OPR234} for OPR$_{345}$ as a
function of $T_{35}$. \label{fig:Tphys_effect_OPR345}}
\end{figure}

For the physisorption binding energy $T_{\mathrm{phys}}$, we took
a standard value of $550\,\mathrm{K}$. The values found in the literature
range from $300\,\mathrm{K}$ to $800\,\mathrm{K}$. Fig. \ref{fig:Tphys_effect_OPR234}
and \ref{fig:Tphys_effect_OPR345} show the model results for these
three values. Again only the rate equation treatment (dashed lines)
and the statistical treatment (solid lines) are shown. The binding
energy $T_{\mathrm{phys}}$ has a dramatic impact on the results when
neglecting fluctuations, while its impact when taking the fluctuations
into account is small. At $T_{\mathrm{phys}}=800\,\mathrm{K}$, the
rate equation results start to approach the observations as desorption
becomes slower and efficient conversion can happen, but the results
remain less fitting than the results of the statistical treatment.

The conclusions of the previous section are thus unaffected by the
uncertainties on the microphysical parameters. The dust temperature
fluctuations make the conversion rate much less dependent on the detail
of the microphysics. Dust temperature indeed explore a large range
of temperatures during the fluctuations, and the average efficiency
is thus controlled by the fraction of the grains whose temperature
fall in the range where the instantaneous rate is high (in other terms,
the portion of the temperature PDF that falls in this range). Varying
the microphysical parameters, and thus the extend of the temperature
range where the instantaneous rate is high, only changes this fraction
slowly while it can change dramatically the instantaneous rate at
a given temperature (for instance the equilibrium temperature used
when neglecting fluctuations). A similar effect of the dust temperature
fluctuations was found in \citet{Bron14} for H$_{2}$ formation.

There is however one source of uncertainties that is not reduced by
the effects of temperature fluctuations: as discussed in Sect. \ref{sub:sticking}
the sticking function on bare grains is not well known and the conversion
efficiency is directly proportional to the sticking probability. In
the region where surface conversion affects the local OPR and H$_{2}$
rotational emission lines, the gas temperature is in the range $100-400\,\mathrm{K}$.
The sticking function that we used \citep{2010JChPh.133j4507M} gives
sticking probabilities in the range $0.1-0.4$ for this temperature
range. A sticking function significantly lower than these values would
thus reduce the impact of surface conversion on H$_{2}$ emission.
Finally, the total available dust surface and especially the dust
surface corresponding to small grains would also affect the total
surface conversion rate. The PDR observations to which we compared
our results seem to be in agreement with the prescriptions used for
the sticking function and the dust population.

\section{Conclusions\label{sec:Conclusions}}

We have built a model of ortho-para conversion of H$_{2}$ on dust
grains based on the latest experimental and theoretical results. When
neglecting dust temperature fluctuations, conversion is found to be
strongly suppressed by the presence of a UV radiation field.

We developed a statistical calculation of the conversion rate, based
on a master equation approach (similar to the method used in \citealt{Bron14}
for H$_{2}$ formation), that takes into account the statistical effect
of dust temperature fluctuations. Conversion on grains is found to
stay efficient under much higher UV radiation fields when fluctuations
are taken into account. Small grains, despite being too warm on average,
spend a sufficiently large fraction of their time between temperature
spikes at colder temperatures where conversion is efficient.

This conversion process on grains is found to play an important role
in PDRs, affecting the rotational lines intensities. The local OPR
falls from 3 to a low value inside the region where H$_{2}$ rotational
lines are emitted, and the position of this transition is controlled
by the conversion efficiency on dust grains. As a result, the OPR
determined from the line intensities of the first few rotational lines
is a signature of the efficiency of this process.

The comparison of our models to a sample of PDR observations of rotational
H$_{2}$ lines indicates a high conversion efficiency on dust grains.
Models implementing the exact statistical treatment with dust temperature
fluctuations give results that are consistent with the observations.
Models that neglect the fluctuations cannot account for the high conversion
efficiency indicated by the observations. Ortho-para conversion on
dust grains is thus an efficient and important process in PDRs, which
can only be accurately described by a statistical treatment of the
impact of dust temperature fluctuations. This process is responsible
for the OPR values lower than 3 derived from rotational lines observations
of PDRs.

We also found that this efficiency induced by temperature fluctuations
is much less sensitive to the microphysical parameters of the model
(binding energy, surface conversion timescale) than the efficiency
at a single fixed temperature (e.g. the equilibrium temperature).
As a consequence, the results obtained here are robust despite large
uncertainties on the microphysical parameters (binding energy in the
range $300-800\,\mathrm{K}$, conversion timescale in the range $1-10^{4}\,\mathrm{s}$).
A similar effect was found in \citet{Bron14} for H$_{2}$ formation
on grains, and it seems to be a general consequence of having a distribution
of dust temperatures rather than a single dust temperature.

In this study, the statistical formalism used to take temperature
fluctuations into account, developed in the case of a single chemical
variable for H$_{2}$ formation in \citet{Bron14}, was extended to
a case with two chemical variables, demonstrating that this method
could be generalized to larger chemical networks. For instance, it
could be used to study the impact of cosmic-rays-induced fluctuations
on ice chemistry.
\begin{acknowledgements}
This work was supported by the French CNRS national program PCMI.
We thank Evelyne Roueff and the anonymous referee for their comments
on the paper.
\end{acknowledgements}

\bibliographystyle{aa}
\bibliography{OSP}

\begin{thebibliography}{95}
\expandafter\ifx\csname natexlab\endcsname\relax\def\natexlab#1{#1}\fi

\bibitem[{{Abgrall} {et~al.}(1992){Abgrall}, {Le Bourlot}, {Pineau Des Forets},
  {Roueff}, {Flower}, \& {Heck}}]{Abgrall92}
{Abgrall}, H., {Le Bourlot}, J., {Pineau Des Forets}, G., {et~al.} 1992, \aap,
  253, 525

\bibitem[{{Acharyya}(2014)}]{2014MNRAS.443.1301A}
{Acharyya}, K. 2014, \mnras, 443, 1301

\bibitem[{{Allers} {et~al.}(2005){Allers}, {Jaffe}, {Lacy}, {Draine}, \&
  {Richter}}]{Allers05}
{Allers}, K.~N., {Jaffe}, D.~T., {Lacy}, J.~H., {Draine}, B.~T., \& {Richter},
  M.~J. 2005, \apj, 630, 368

\bibitem[{{Amiaud} {et~al.}(2008){Amiaud}, {Momeni}, {Dulieu}, {Fillion},
  {Matar}, \& {Lemaire}}]{2008PhRvL.100e6101A}
{Amiaud}, L., {Momeni}, A., {Dulieu}, F., {et~al.} 2008, Physical Review
  Letters, 100, 056101

\bibitem[{{Bron}(2014)}]{These}
{Bron}, E. 2014, PhD thesis, {Université Paris Diderot}

\bibitem[{{Bron} {et~al.}(2014){Bron}, {Le Bourlot}, \& {Le Petit}}]{Bron14}
{Bron}, E., {Le Bourlot}, J., \& {Le Petit}, F. 2014, \aap, 569, A100

\bibitem[{{Buch} {et~al.}(1993){Buch}, {Silva}, \&
  {Devlin}}]{1993JChPh..99.2265B}
{Buch}, V., {Silva}, S.~C., \& {Devlin}, J.~P. 1993, \jcp, 99, 2265

\bibitem[{{Burton} {et~al.}(1992){Burton}, {Hollenbach}, \&
  {Tielens}}]{Burton92}
{Burton}, M.~G., {Hollenbach}, D.~J., \& {Tielens}, A.~G.~G. 1992, \apj, 399,
  563

\bibitem[{{Carmona-Novillo} {et~al.}(2007){Carmona-Novillo}, {Bartolomei},
  {Hern{\'a}ndez}, \& {Campos-Mart{\'{\i}}nez}}]{Carmona-Novillo07}
{Carmona-Novillo}, E., {Bartolomei}, M., {Hern{\'a}ndez}, M.~I., \&
  {Campos-Mart{\'{\i}}nez}, J. 2007, \jcp, 126, 124315

\bibitem[{{Chehrouri} {et~al.}(2011){Chehrouri}, {Fillion}, {Chaabouni},
  {Mokrane}, {Congiu}, {Dulieu}, {Matar}, {Michaut}, \&
  {Lemaire}}]{2011PCCP...13.2172C}
{Chehrouri}, M., {Fillion}, J.-H., {Chaabouni}, H., {et~al.} 2011, Physical
  Chemistry Chemical Physics (Incorporating Faraday Transactions), 13, 2172

\bibitem[{{Compi{\`e}gne} {et~al.}(2011){Compi{\`e}gne}, {Verstraete}, {Jones},
  {Bernard}, {Boulanger}, {Flagey}, {Le Bourlot}, {Paradis}, \&
  {Ysard}}]{Compiegne11}
{Compi{\`e}gne}, M., {Verstraete}, L., {Jones}, A., {et~al.} 2011, \aap, 525,
  A103

\bibitem[{{Dislaire} {et~al.}(2012){Dislaire}, {Hily-Blant}, {Faure}, {Maret},
  {Bacmann}, \& {Pineau Des For{\^e}ts}}]{Dislaire12}
{Dislaire}, V., {Hily-Blant}, P., {Faure}, A., {et~al.} 2012, \aap, 537, A20

\bibitem[{{Falgarone} {et~al.}(2005){Falgarone}, {Verstraete}, {Pineau Des
  For{\^e}ts}, \& {Hily-Blant}}]{Falgarone05}
{Falgarone}, E., {Verstraete}, L., {Pineau Des For{\^e}ts}, G., \&
  {Hily-Blant}, P. 2005, \aap, 433, 997

\bibitem[{{Faure} {et~al.}(2013){Faure}, {Hily-Blant}, {Le Gal}, {Rist}, \&
  {Pineau des For{\^e}ts}}]{Faure13}
{Faure}, A., {Hily-Blant}, P., {Le Gal}, R., {Rist}, C., \& {Pineau des
  For{\^e}ts}, G. 2013, \apjl, 770, L2

\bibitem[{{Fleming} {et~al.}(2010){Fleming}, {France}, {Lupu}, \&
  {McCandliss}}]{Fleming10}
{Fleming}, B., {France}, K., {Lupu}, R.~E., \& {McCandliss}, S.~R. 2010, \apj,
  725, 159

\bibitem[{{Flower} {et~al.}(2006){Flower}, {Pineau Des For{\^e}ts}, \&
  {Walmsley}}]{Flower06}
{Flower}, D.~R., {Pineau Des For{\^e}ts}, G., \& {Walmsley}, C.~M. 2006, \aap,
  449, 621

\bibitem[{{Fuente} {et~al.}(2000){Fuente}, {Martin-Pintado},
  {Rodriguez-Fern{\'a}ndez}, {Cernicharo}, \& {Gerin}}]{Fuente00}
{Fuente}, A., {Martin-Pintado}, J., {Rodriguez-Fern{\'a}ndez}, N.~J.,
  {Cernicharo}, J., \& {Gerin}, M. 2000, \aap, 354, 1053

\bibitem[{{Fuente} {et~al.}(1999){Fuente}, {Mart{\'{\i}}n-Pintado},
  {Rodr{\'{\i}}guez-Fern{\'a}ndez}, {Rodr{\'{\i}}guez-Franco}, {de Vicente}, \&
  {Kunze}}]{Fuente99}
{Fuente}, A., {Mart{\'{\i}}n-Pintado}, J., {Rodr{\'{\i}}guez-Fern{\'a}ndez},
  N.~J., {et~al.} 1999, \apjl, 518, L45

\bibitem[{{Fukutani} \& {Sugimoto}(2013)}]{Fukutani13}
{Fukutani}, K. \& {Sugimoto}, T. 2013, Progress In Surface Science, 88, 279

\bibitem[{{Gavilan} {et~al.}(2012){Gavilan}, {Vidali}, {Lemaire}, {Chehrouri},
  {Dulieu}, {Fillion}, {Congiu}, \& {Chaabouni}}]{Gavilan12}
{Gavilan}, L., {Vidali}, G., {Lemaire}, J.~L., {et~al.} 2012, \apj, 760, 35

\bibitem[{{Gerlich}(1990)}]{Gerlich90}
{Gerlich}, D. 1990, \jcp, 92, 2377

\bibitem[{{Gillmon} {et~al.}(2006){Gillmon}, {Shull}, {Tumlinson}, \&
  {Danforth}}]{Gillmon06}
{Gillmon}, K., {Shull}, J.~M., {Tumlinson}, J., \& {Danforth}, C. 2006, \apj,
  636, 891

\bibitem[{{Godard} {et~al.}(2014){Godard}, {Falgarone}, \& {Pineau des
  For{\^e}ts}}]{Godard14}
{Godard}, B., {Falgarone}, E., \& {Pineau des For{\^e}ts}, G. 2014, \aap, 570,
  A27

\bibitem[{{Goicoechea} \& {Le Bourlot}(2007)}]{Goicoechea07}
{Goicoechea}, J.~R. \& {Le Bourlot}, J. 2007, \aap, 467, 1

\bibitem[{{G{\'o}mez-Carrasco} {et~al.}(2012){G{\'o}mez-Carrasco},
  {Gonz{\'a}lez-S{\'a}nchez}, {Aguado}, {Sanz-Sanz}, {Zanchet}, \&
  {Roncero}}]{Gomez12}
{G{\'o}mez-Carrasco}, S., {Gonz{\'a}lez-S{\'a}nchez}, L., {Aguado}, A.,
  {et~al.} 2012, \jcp, 137, 094303

\bibitem[{{Gonzalez Garcia} {et~al.}(2008){Gonzalez Garcia}, {Le Bourlot}, {Le
  Petit}, \& {Roueff}}]{GonzalezGarcia08}
{Gonzalez Garcia}, M., {Le Bourlot}, J., {Le Petit}, F., \& {Roueff}, E. 2008,
  \aap, 485, 127

\bibitem[{{Gredel} {et~al.}(2002){Gredel}, {Pineau des For{\^e}ts}, \&
  {Federman}}]{Gredel02}
{Gredel}, R., {Pineau des For{\^e}ts}, G., \& {Federman}, S.~R. 2002, \aap,
  389, 993

\bibitem[{{Gry} {et~al.}(2002){Gry}, {Boulanger}, {Nehm{\'e}}, {Pineau des
  For{\^e}ts}, {Habart}, \& {Falgarone}}]{Gry02}
{Gry}, C., {Boulanger}, F., {Nehm{\'e}}, C., {et~al.} 2002, \aap, 391, 675

\bibitem[{{Habart} {et~al.}(2011){Habart}, {Abergel}, {Boulanger}, {Joblin},
  {Verstraete}, {Compi{\`e}gne}, {Pineau Des For{\^e}ts}, \& {Le
  Bourlot}}]{Habart11}
{Habart}, E., {Abergel}, A., {Boulanger}, F., {et~al.} 2011, \aap, 527, A122

\bibitem[{{Habart} {et~al.}(2003){Habart}, {Boulanger}, {Verstraete}, {Pineau
  des For{\^e}ts}, {Falgarone}, \& {Abergel}}]{Habart03}
{Habart}, E., {Boulanger}, F., {Verstraete}, L., {et~al.} 2003, \aap, 397, 623

\bibitem[{{Habart} {et~al.}(2004){Habart}, {Boulanger}, {Verstraete},
  {Walmsley}, \& {Pineau des For{\^e}ts}}]{Habart04}
{Habart}, E., {Boulanger}, F., {Verstraete}, L., {Walmsley}, C.~M., \& {Pineau
  des For{\^e}ts}, G. 2004, \aap, 414, 531

\bibitem[{{Hasegawa} {et~al.}(1992){Hasegawa}, {Herbst}, \&
  {Leung}}]{Hasegawa92}
{Hasegawa}, T.~I., {Herbst}, E., \& {Leung}, C.~M. 1992, \apjs, 82, 167

\bibitem[{{Higdon} {et~al.}(2006){Higdon}, {Armus}, {Higdon}, {Soifer}, \&
  {Spoon}}]{Higdon06}
{Higdon}, S.~J.~U., {Armus}, L., {Higdon}, J.~L., {Soifer}, B.~T., \& {Spoon},
  H.~W.~W. 2006, \apj, 648, 323

\bibitem[{{Hixson} {et~al.}(1992){Hixson}, {Wojcik}, {Devlin}, {Devlin}, \&
  {Buch}}]{1992JChPh..97..753H}
{Hixson}, H.~G., {Wojcik}, M.~J., {Devlin}, M.~S., {Devlin}, J.~P., \& {Buch},
  V. 1992, \jcp, 97, 753

\bibitem[{{Honvault} {et~al.}(2011){Honvault}, {Jorfi}, {Gonz{\'a}lez-Lezana},
  {Faure}, \& {Pagani}}]{Honvault11}
{Honvault}, P., {Jorfi}, M., {Gonz{\'a}lez-Lezana}, T., {Faure}, A., \&
  {Pagani}, L. 2011, Physical Review Letters, 107, 023201

\bibitem[{{Ilisca} \& {Ghiglieno}(2014)}]{Ilisca14}
{Ilisca}, E. \& {Ghiglieno}, F. 2014, European Physical Journal B, 87, 235

\bibitem[{{Joblin} {et~al.}(in prep.){Joblin}, {Bron}, {Pinto}, {Pilleri},
  {Gerin}, {Le Bourlot}, {Le Petit}, {Fuente}, {Berne}, {Goicoechea}, {Habart},
  {K\"ohler}, {Teyssier}, {Nagy}, {Montillaud}, {Vastel}, {Cernicharo},
  {R\"ollig}, {V. Ossenkopf (WADI team)}, \& {E. A. Bergin (HEXOS
  team)}}]{Joblin_in_prep.}
{Joblin}, C., {Bron}, E., {Pinto}, C., {et~al.} in prep., to be submitted in
  A\&A

\bibitem[{{Katz} {et~al.}(1999){Katz}, {Furman}, {Biham}, {Pirronello}, \&
  {Vidali}}]{1999ApJ...522..305K}
{Katz}, N., {Furman}, I., {Biham}, O., {Pirronello}, V., \& {Vidali}, G. 1999,
  \apj, 522, 305

\bibitem[{{Kubik} {et~al.}(1985){Kubik}, {Hardy}, \&
  {Glattli}}]{1985CaJPh..63..605K}
{Kubik}, P.~R., {Hardy}, W.~N., \& {Glattli}, H. 1985, Canadian Journal of
  Physics, 63, 605

\bibitem[{{Lacour} {et~al.}(2005){Lacour}, {Ziskin}, {H{\'e}brard}, {Oliveira},
  {Andr{\'e}}, {Ferlet}, \& {Vidal-Madjar}}]{Lacour05}
{Lacour}, S., {Ziskin}, V., {H{\'e}brard}, G., {et~al.} 2005, \apj, 627, 251

\bibitem[{{Le Bourlot}(2000)}]{LeBourlot00}
{Le Bourlot}, J. 2000, \aap, 360, 656

\bibitem[{{Le Bourlot} {et~al.}(2012){Le Bourlot}, {Le Petit}, {Pinto},
  {Roueff}, \& {Roy}}]{LeBourlot12}
{Le Bourlot}, J., {Le Petit}, F., {Pinto}, C., {Roueff}, E., \& {Roy}, F. 2012,
  \aap, 541, A76

\bibitem[{{Le Bourlot} {et~al.}(1999){Le Bourlot}, {Pineau des For{\^e}ts}, \&
  {Flower}}]{LeBourlot99}
{Le Bourlot}, J., {Pineau des For{\^e}ts}, G., \& {Flower}, D.~R. 1999, \mnras,
  305, 802

\bibitem[{{Le Petit} {et~al.}(2006){Le Petit}, {Nehm{\'e}}, {Le Bourlot}, \&
  {Roueff}}]{LePetit06}
{Le Petit}, F., {Nehm{\'e}}, C., {Le Bourlot}, J., \& {Roueff}, E. 2006, \apjs,
  164, 506

\bibitem[{{Ledoux} {et~al.}(2003){Ledoux}, {Petitjean}, \&
  {Srianand}}]{Ledoux03}
{Ledoux}, C., {Petitjean}, P., \& {Srianand}, R. 2003, \mnras, 346, 209

\bibitem[{{Leitch-Devlin} \& {Williams}(1985)}]{1985MNRAS.213..295L}
{Leitch-Devlin}, M.~A. \& {Williams}, D.~A. 1985, \mnras, 213, 295

\bibitem[{{Lique} {et~al.}(2012){Lique}, {Honvault}, \& {Faure}}]{Lique12}
{Lique}, F., {Honvault}, P., \& {Faure}, A. 2012, \jcp, 137, 154303

\bibitem[{{Mandy} \& {Martin}(1992)}]{Mandy92}
{Mandy}, M.~E. \& {Martin}, P.~G. 1992, \jcp, 97, 265

\bibitem[{{Mandy} \& {Martin}(1993)}]{Mandy93}
{Mandy}, M.~E. \& {Martin}, P.~G. 1993, \apjs, 86, 199

\bibitem[{{Manic{\`o}} {et~al.}(2001){Manic{\`o}}, {Ragun{\`i}}, {Pirronello},
  {Roser}, \& {Vidali}}]{2001ApJ...548L.253M}
{Manic{\`o}}, G., {Ragun{\`i}}, G., {Pirronello}, V., {Roser}, J.~E., \&
  {Vidali}, G. 2001, \apjl, 548, L253

\bibitem[{{Maret} \& {Bergin}(2007)}]{Maret07}
{Maret}, S. \& {Bergin}, E.~A. 2007, \apj, 664, 956

\bibitem[{{Matar} {et~al.}(2010){Matar}, {Bergeron}, {Dulieu}, {Chaabouni},
  {Accolla}, \& {Lemaire}}]{2010JChPh.133j4507M}
{Matar}, E., {Bergeron}, H., {Dulieu}, F., {et~al.} 2010, \jcp, 133, 104507

\bibitem[{{Mathis} {et~al.}(1983){Mathis}, {Mezger}, \& {Panagia}}]{Mathis83}
{Mathis}, J.~S., {Mezger}, P.~G., \& {Panagia}, N. 1983, \aap, 128, 212

\bibitem[{{Mathis} {et~al.}(1977){Mathis}, {Rumpl}, \& {Nordsieck}}]{Mathis77}
{Mathis}, J.~S., {Rumpl}, W., \& {Nordsieck}, K.~H. 1977, \apj, 217, 425

\bibitem[{{Moutou} {et~al.}(1999){Moutou}, {Verstraete}, {Sellgren}, \&
  {Leger}}]{Moutou99}
{Moutou}, C., {Verstraete}, L., {Sellgren}, K., \& {Leger}, A. 1999, in ESA
  Special Publication, Vol. 427, The Universe as Seen by ISO, ed. P.~{Cox} \&
  M.~{Kessler}, 727

\bibitem[{{Muzahid} {et~al.}(2015){Muzahid}, {Srianand}, \&
  {Charlton}}]{Muzahid15}
{Muzahid}, S., {Srianand}, R., \& {Charlton}, J. 2015, \mnras, 448, 2840

\bibitem[{{Naslim} {et~al.}(2015){Naslim}, {Kemper}, {Madden}, {Hony}, {Chu},
  {Galliano}, {Bot}, {Yang}, {Seok}, {Oliveira}, {van Loon}, {Meixner}, {Li},
  {Hughes}, {Gordon}, {Otsuka}, {Hirashita}, {Morata}, {Lebouteiller},
  {Indebetouw}, {Srinivasan}, {Bernard}, \& {Reach}}]{Naslim15}
{Naslim}, N., {Kemper}, F., {Madden}, S.~C., {et~al.} 2015, \mnras, 446, 2490

\bibitem[{{Neufeld} {et~al.}(2006){Neufeld}, {Melnick}, {Sonnentrucker},
  {Bergin}, {Green}, {Kim}, {Watson}, {Forrest}, \& {Pipher}}]{Neufeld06}
{Neufeld}, D.~A., {Melnick}, G.~J., {Sonnentrucker}, P., {et~al.} 2006, \apj,
  649, 816

\bibitem[{{Noterdaeme} {et~al.}(2007){Noterdaeme}, {Ledoux}, {Petitjean}, {Le
  Petit}, {Srianand}, \& {Smette}}]{Noterdaeme07}
{Noterdaeme}, P., {Ledoux}, C., {Petitjean}, P., {et~al.} 2007, \aap, 474, 393

\bibitem[{{Pachucki} \& {Komasa}(2008)}]{Pachucki08}
{Pachucki}, K. \& {Komasa}, J. 2008, \pra, 77, 030501

\bibitem[{{Pagani} {et~al.}(2013){Pagani}, {Lesaffre}, {Jorfi}, {Honvault},
  {Gonz{\'a}lez-Lezana}, \& {Faure}}]{Pagani13}
{Pagani}, L., {Lesaffre}, P., {Jorfi}, M., {et~al.} 2013, \aap, 551, A38

\bibitem[{{Pagani} {et~al.}(2011){Pagani}, {Roueff}, \& {Lesaffre}}]{Pagani11}
{Pagani}, L., {Roueff}, E., \& {Lesaffre}, P. 2011, \apjl, 739, L35

\bibitem[{{Pagani} {et~al.}(2009){Pagani}, {Vastel}, {Hugo}, {Kokoouline},
  {Greene}, {Bacmann}, {Bayet}, {Ceccarelli}, {Peng}, \&
  {Schlemmer}}]{Pagani09}
{Pagani}, L., {Vastel}, C., {Hugo}, E., {et~al.} 2009, \aap, 494, 623

\bibitem[{{Palmer} \& {Willis}(1987)}]{1987SurSc.179L...1P}
{Palmer}, R.~E. \& {Willis}, R.~F. 1987, Surface Science, 179, L1

\bibitem[{{Pereira-Santaella} {et~al.}(2014){Pereira-Santaella}, {Spinoglio},
  {van der Werf}, \& {Piqueras L{\'o}pez}}]{Pereira14}
{Pereira-Santaella}, M., {Spinoglio}, L., {van der Werf}, P.~P., \& {Piqueras
  L{\'o}pez}, J. 2014, \aap, 566, A49

\bibitem[{{Perets} {et~al.}(2007){Perets}, {Lederhendler}, {Biham}, {Vidali},
  {Li}, {Swords}, {Congiu}, {Roser}, {Manic{\'o}}, {Brucato}, \&
  {Pirronello}}]{2007ApJ...661L.163P}
{Perets}, H.~B., {Lederhendler}, A., {Biham}, O., {et~al.} 2007, \apjl, 661,
  L163

\bibitem[{{Rachford} {et~al.}(2009){Rachford}, {Snow}, {Destree}, {Ross},
  {Ferlet}, {Friedman}, {Gry}, {Jenkins}, {Morton}, {Savage}, {Shull},
  {Sonnentrucker}, {Tumlinson}, {Vidal-Madjar}, {Welty}, \&
  {York}}]{Rachford09}
{Rachford}, B.~L., {Snow}, T.~P., {Destree}, J.~D., {et~al.} 2009, \apjs, 180,
  125

\bibitem[{{Rachford} {et~al.}(2002){Rachford}, {Snow}, {Tumlinson}, {Shull},
  {Blair}, {Ferlet}, {Friedman}, {Gry}, {Jenkins}, {Morton}, {Savage},
  {Sonnentrucker}, {Vidal-Madjar}, {Welty}, \& {York}}]{Rachford02}
{Rachford}, B.~L., {Snow}, T.~P., {Tumlinson}, J., {et~al.} 2002, \apj, 577,
  221

\bibitem[{{Richter} {et~al.}(2003){Richter}, {Wakker}, {Savage}, \&
  {Sembach}}]{Richter03}
{Richter}, P., {Wakker}, B.~P., {Savage}, B.~D., \& {Sembach}, K.~R. 2003,
  \apj, 586, 230

\bibitem[{{Rigopoulou} {et~al.}(2002){Rigopoulou}, {Kunze}, {Lutz}, {Genzel},
  \& {Moorwood}}]{Rigopoulou02}
{Rigopoulou}, D., {Kunze}, D., {Lutz}, D., {Genzel}, R., \& {Moorwood},
  A.~F.~M. 2002, \aap, 389, 374

\bibitem[{{Roser} {et~al.}(2002){Roser}, {Manic{\`o}}, {Pirronello}, \&
  {Vidali}}]{2002ApJ...581..276R}
{Roser}, J.~E., {Manic{\`o}}, G., {Pirronello}, V., \& {Vidali}, G. 2002, \apj,
  581, 276

\bibitem[{{Roussel} {et~al.}(2007){Roussel}, {Helou}, {Hollenbach}, {Draine},
  {Smith}, {Armus}, {Schinnerer}, {Walter}, {Engelbracht}, {Thornley},
  {Kennicutt}, {Calzetti}, {Dale}, {Murphy}, \& {Bot}}]{Roussel07}
{Roussel}, H., {Helou}, G., {Hollenbach}, D.~J., {et~al.} 2007, \apj, 669, 959

\bibitem[{{Roy} {et~al.}(2006){Roy}, {Chengalur}, \& {Srianand}}]{Roy06}
{Roy}, N., {Chengalur}, J.~N., \& {Srianand}, R. 2006, \mnras, 365, L1

\bibitem[{{Sandler}(1954)}]{1954JPhysChem..58..54}
{Sandler}, Y.~L. 1954, Journal of Physical Chemistry, 58, 54

\bibitem[{{Savage} {et~al.}(1977){Savage}, {Bohlin}, {Drake}, \&
  {Budich}}]{Savage77}
{Savage}, B.~D., {Bohlin}, R.~C., {Drake}, J.~F., \& {Budich}, W. 1977, \apj,
  216, 291

\bibitem[{{Schulz} \& {Le Roy}(1965)}]{Schulz65}
{Schulz}, W.~R. \& {Le Roy}, D.~J. 1965, \jcp, 42, 3869

\bibitem[{{Sheffer} {et~al.}(2011){Sheffer}, {Wolfire}, {Hollenbach},
  {Kaufman}, \& {Cordier}}]{Sheffer11}
{Sheffer}, Y., {Wolfire}, M.~G., {Hollenbach}, D.~J., {Kaufman}, M.~J., \&
  {Cordier}, M. 2011, \apj, 741, 45

\bibitem[{{Srianand} {et~al.}(2005){Srianand}, {Petitjean}, {Ledoux},
  {Ferland}, \& {Shaw}}]{Srianand05}
{Srianand}, R., {Petitjean}, P., {Ledoux}, C., {Ferland}, G., \& {Shaw}, G.
  2005, \mnras, 362, 549

\bibitem[{{Sternberg} \& {Neufeld}(1999)}]{Sternberg99}
{Sternberg}, A. \& {Neufeld}, D.~A. 1999, \apj, 516, 371

\bibitem[{{St{\"o}rzer} \& {Hollenbach}(1998)}]{Storzer98}
{St{\"o}rzer}, H. \& {Hollenbach}, D. 1998, \apj, 495, 853

\bibitem[{{Sugimoto} \& {Fukutani}(2011)}]{2011NatPh...7..307S}
{Sugimoto}, T. \& {Fukutani}, K. 2011, Nature Physics, 7, 307

\bibitem[{{Sun} \& {Dalgarno}(1994)}]{Sun94}
{Sun}, Y. \& {Dalgarno}, A. 1994, \apj, 427, 1053

\bibitem[{{Takahashi}(2001)}]{Takahashi01}
{Takahashi}, J. 2001, \apj, 561, 254

\bibitem[{{Thi} {et~al.}(2009){Thi}, {van Dishoeck}, {Bell}, {Viti}, \&
  {Black}}]{Thi09}
{Thi}, W.-F., {van Dishoeck}, E.~F., {Bell}, T., {Viti}, S., \& {Black}, J.
  2009, \mnras, 400, 622

\bibitem[{{Timmermann} {et~al.}(1996){Timmermann}, {Bertoldi}, {Wright},
  {Drapatz}, {Draine}, {Haser}, \& {Sternberg}}]{Timmermann96}
{Timmermann}, R., {Bertoldi}, F., {Wright}, C.~M., {et~al.} 1996, \aap, 315,
  L281

\bibitem[{{Troscompt} {et~al.}(2009){Troscompt}, {Faure}, {Maret},
  {Ceccarelli}, {Hily-Blant}, \& {Wiesenfeld}}]{Troscompt09}
{Troscompt}, N., {Faure}, A., {Maret}, S., {et~al.} 2009, \aap, 506, 1243

\bibitem[{{Truhlar}(1976)}]{Truhlar76}
{Truhlar}, D.~G. 1976, \jcp, 65, 1008

\bibitem[{{Tumlinson} {et~al.}(2002){Tumlinson}, {Shull}, {Rachford},
  {Browning}, {Snow}, {Fullerton}, {Jenkins}, {Savage}, {Crowther}, {Moos},
  {Sembach}, {Sonneborn}, \& {York}}]{Tumlinson02}
{Tumlinson}, J., {Shull}, J.~M., {Rachford}, B.~L., {et~al.} 2002, \apj, 566,
  857

\bibitem[{{Vaytet} {et~al.}(2014){Vaytet}, {Tomida}, \& {Chabrier}}]{Vaytet14}
{Vaytet}, N., {Tomida}, K., \& {Chabrier}, G. 2014, \aap, 563, A85

\bibitem[{{Vidali} \& {Li}(2010)}]{2010JPCM...22D4012V}
{Vidali}, G. \& {Li}, L. 2010, Journal of Physics Condensed Matter, 22, D4012

\bibitem[{{Vidali} {et~al.}(2007){Vidali}, {Pirronello}, {Li}, {Roser},
  {Manico}, {Mehl}, {Lederhendler}, {Perets}, {Brucato}, \&
  {Biham}}]{2007JPCA..11112611V}
{Vidali}, G., {Pirronello}, V., {Li}, L., {et~al.} 2007, Journal of Physical
  Chemistry A, 111, 12611

\bibitem[{{Watanabe} {et~al.}(2010){Watanabe}, {Kimura}, {Kouchi}, {Chigai},
  {Hama}, \& {Pirronello}}]{2010ApJ...714L.233W}
{Watanabe}, N., {Kimura}, Y., {Kouchi}, A., {et~al.} 2010, \apjl, 714, L233

\bibitem[{{Yabushita} {et~al.}(2008){Yabushita}, {Hama}, {Iida}, {Kawanaka},
  {Kawasaki}, {Watanabe}, {Ashfold}, \& {Loock}}]{Yabushita08}
{Yabushita}, A., {Hama}, T., {Iida}, D., {et~al.} 2008, \apjl, 682, L69

\bibitem[{{Yuan} \& {Neufeld}(2011)}]{Yuan11}
{Yuan}, Y. \& {Neufeld}, D.~A. 2011, \apj, 726, 76

\bibitem[{{Yucel} {et~al.}(1990){Yucel}, {Alexander}, \&
  {Honig}}]{1990PhRvB..42..820Y}
{Yucel}, S., {Alexander}, N., \& {Honig}, A. 1990, \prb, 42, 820

\end{thebibliography}

\end{document}